\documentclass[fleqn,usenatbib]{mnras}

\usepackage{newtxtext,newtxmath}

\usepackage[T1]{fontenc}
\usepackage{ae,aecompl}

\usepackage{graphicx}	
\usepackage{amsmath}	
\usepackage{amssymb}	

\usepackage{color}
\usepackage{bm}
\usepackage{bbm}
\usepackage{makecell}
\usepackage{hyperref}
\usepackage{lscape}
\usepackage{rotating}
\usepackage{cancel}

\definecolor{green2}{rgb}{0.13, 0.55, 0.13}
\definecolor{gray2}{rgb}{0.6, 0.6, 0.6}
\definecolor{orange}{rgb}{1.0, 0.49, 0}
\definecolor{blue2}{rgb}{0.25,0.5,1}

\def\mean#1{\left< #1 \right>}

\defcitealias{Mead15}{M15}
\defcitealias{Eifler15}{E15}

\newcommand{\R}{\bm R}
\newcommand{\U}{\mathbf U}
\newcommand{\V}{\mathbf V}
\renewcommand{\t}{\rm t}
\newcommand{\D}{\bm D}
\newcommand{\M}{\bm M}
\newcommand{\B}{\bm B}

\newcommand{\Q}{\bm Q}
\renewcommand{\P}{\mathbf P}
\newcommand{\C}{\mathbf C}
\newcommand{\I}{\mathbf I}
\newcommand{\invL}{\mathbf L^{-1}}
\renewcommand{\L}{\mathbf L}
\newcommand{\pco}{\bm p_{\rm co}}
\newcommand{\pnu}{\bm p_{\rm nu}}
\newcommand{\pcosim}{\bm p_{\rm co, sim}}
\newcommand{\pcofid}{\bm p_{\rm co, fid}}
\newcommand{\pcobest}{\bm p_{\rm co, bestfit}}

\newcommand{\hkpc}{\ h^{-1}\mathrm{kpc}}
\newcommand{\hMpc}{\ h^{-1}\mathrm{Mpc}}
\newcommand{\invhMpc}{\ h\mathrm{Mpc}^{-1}}
\newcommand{\hMsun}{\ h^{-1}\mathrm{M}_{\odot}}
\newcommand{\Msun}{\mathrm{M}_{\odot}}

\newcommand*{\vertbar}{\rule[-1ex]{0.5pt}{2.5ex}}

\newcommand{\Omegam}{\Omega_{\mathrm{m}}}
\newcommand{\Omegab}{\Omega_{\mathrm{b}}}
\newcommand{\Omegal}{\Omega_{\Lambda}}
\newcommand{\ns}{n_{\mathrm{s}}}

\title[Baryonic physics mitigation for lensing]{Modeling baryonic physics in future weak lensing surveys}

\author[Huang et al.]{
Hung-Jin Huang,$^{1}$\thanks{E-mail: hungjinh@andrew.cmu.edu}
Tim Eifler,$^{2,3}$
Rachel Mandelbaum$^{1}$
and Scott Dodelson$^{1}$
\\
$^{1}$McWilliams Center for Cosmology, Department of Physics, Carnegie Mellon University, Pittsburgh, PA 15213, USA\\
$^{2}$Steward Observatory/Department of Astronomy, University of Arizona, 933 North Cherry Avenue, Tucson, AZ 85721, USA \\
$^{3}$Jet Propulsion Laboratory, California Institute of Technology, Pasadena, CA 91109, USA
}

\date{Accepted XXX. Received YYY; in original form ZZZ}

\pubyear{2019}

\begin{document}
\label{firstpage}
\pagerange{\pageref{firstpage}--\pageref{lastpage}}
\maketitle

\begin{abstract} 
Modifications of the matter power spectrum due to baryonic physics are one of the major theoretical uncertainties in cosmological weak lensing measurements.  Developing robust mitigation schemes for this source of systematic uncertainty increases the robustness of cosmological constraints, and may increase their precision if they enable the use of information from smaller scales.
Here we explore the performance of two mitigation schemes for baryonic effects in weak lensing cosmic shear: the PCA method and the halo-model approach in \textsc{HMcode}.
We construct mock tomographic shear power spectra from four hydrodynamical simulations, 
and run simulated likelihood analyses with \textsc{CosmoLike} assuming LSST-like survey statistics.
With an angular scale cut of $\ell_{\rm max}<2000$, both methods successfully remove the biases in cosmological parameters due to the various baryonic physics scenarios, with the PCA method causing less degradation in the parameter constraints than \textsc{HMcode}. 
For a more aggressive $\ell_{\rm max}=$5000, the PCA method performs well for all but one baryonic physics scenario,  requiring additional training simulations to account for the extreme baryonic physics scenario of Illustris; \textsc{HMcode} exhibits tensions in the 2D posterior distributions of cosmological parameters due to lack of freedom in describing the power spectrum for $k > 10\ h^{-1}\mathrm{Mpc}$. 
We investigate variants of the PCA method and improve the bias mitigation through PCA by accounting for the noise properties in the data via Cholesky decomposition of the covariance matrix.
Our improved PCA method allows us to retain more statistical constraining power while effectively mitigating baryonic uncertainties even for a broad range of baryonic physics scenarios.
\end{abstract}



\section{Introduction}

The origin of the accelerated expansion of the Universe has been one of the most profound mysteries in modern cosmology since its discovery 
\citep{Riess98, Perlmutter99}.
The $\Lambda$CDM framework is currently consistent with observations of the expansion history of our Universe from early \citep{Planck16} to late times \citep{DES18}. Ongoing photometry surveys such as \href{http://www.astro-wise.org/projects/KIDS/}{KiDS} (Kilo-Degree Survey\footnote{http://www.astro-wise.org/projects/KIDS/}), \href{http://hsc.mtk.nao.ac.jp/ssp/}{HSC} (Hyper Suprime-Cam\footnote{http://hsc.mtk.nao.ac.jp/ssp/}) and 
\href{www.darkenergysurvey.org/}{DES} (Dark Energy Survey\footnote{www.darkenergysurvey.org/}) or future experiments such as \href{http://www.lsst.org/lsst}{LSST} (Large Synoptic Survey Telescope\footnote{http://www.lsst.org/lsst}), \href{sci.esa.int/euclid/}{Euclid}\footnote{sci.esa.int/euclid/}, and \href{http://wfirst.gsfc.nasa.gov/}{WFIRST} (Wide-Field Infrared Survey Telescope\footnote{http://wfirst.gsfc.nasa.gov/}) experiments aim to constrain cosmological parameters to higher precision and search for deviations from $\Lambda$CDM in order to understand the nature of dark energy and General Relativity. 

Weak gravitational lensing (WL), the deflection of light by the gravitational potential of cosmic structure, is one of the most promising cosmological probes to discriminate between  dark energy models \citep{Weinberg13, Mandelbaum18}. Tomographic WL measurements, in which galaxy shapes are cross-correlated within and across bins in redshift space (e.g. \citealt{Hu04}), are directly sensitive to structure growth, with secondary dependence on the relative distance ratios. 
In order to use tomographic WL measurements to constrain cosmological parameters, an accurate model for matter density power spectrum, $P_\delta(k, z)$, is required. 
It has been estimated that $P_\delta(k, z)$ must be predicted to approximately 1\% accuracy for $k \le k_{\rm max} \sim 10 \hMpc$ in order to avoid biasing cosmological parameter constraints  in the era of LSST \citep{Huterer05,Eifler11,Hearin12}.

In the linear and quasi-linear regime, perturbation theory can be used to calculate the matter power spectra for a set of given cosmological parameters  \citep{Bernardeau02}. On smaller scales, N-body simulations are needed in order to capture the complicated non-linear evolution of structure growth. For example, the \textsc{Halofit} method employs a functional form of $P_\delta(k, z)$ derived from halo models, and calibrates the model parameters from N-body simulations at various cosmological parameters \citep{Smith03, Takahashi12}. Alternatively, the \textsc{Cosmic emu} package emulates $P_\delta(k, z)$ by directly interpolating the N-body simulation results at a range of cosmological models \citep{Heitmann10, Heitmann14, Lawrence17}. 
However, only gravitational physics is included in these dark-matter-only (DMO) simulations, which neglects any modification of the matter distribution due to baryonic physics processes such as star formation, radiative cooling and feedback (e.g. \citealt{Cui14, Velliscig14, Mummery17}). 
These processes can modify $P_\delta(k, z)$ by tens of per cent compared to the DMO power spectra from $k \approx$ 1 to 10 $\hMpc$ at $z=0$  \citep{vanDaalen11}. The changes in the matter power spectrum due to baryonic physics can affect our inferences on dark energy (e.g.\ \citealt{Copeland18}) and neutrino mass parameters (e.g.\ \citealt{HarnoisDeraps15}) as they have similar effects on part of the power spectrum, but the different scale and redshift dependencies can help in breaking some of the degeneracies.


There are several approaches to mitigating the impact of uncertainty in how the baryonic physics modifies the matter power spectrum. 
The simplest approach is to eliminate data points that may be severely affected by this uncertainty, so that limitations in small-scale modeling do not bias the inferred cosmology (e.g., see \citealt{Krause17} for the determination of the redshift-dependent angular scale cuts for the DES-Y1 analysis or see \citealt{Taylor18} for another method relating angular scale cuts to physical ($k$) space). 
This approach results in a loss of cosmological constraining power, especially when the statistical precision of the data increases in the future, resulting in the need for even more conservative scale cuts. 
A more economical way of discarding data is through peak clipping \citep{Simpson11,Simpson13}. By cutting the most extreme peaks in the density fields of both observed and mock data sets, the derived summary statistics become less sensitive to the poorly-modeled non-linear regime, while still allowing the use of a wider range of scales to extract cosmological information \citep{Giblin18}.  
\citet{Eifler15} propose the principal component analysis (PCA) framework \citep[see also][]{Kitching16}, which utilizes suites of hydrodynamical simulations to build a set of principal components (PCs) describing the modification of the observables by baryonic physics. The first few PC modes point toward directions in observable space where deviation from DMO power spectra due to baryons are most dominant.  
One can then efficiently remove the vast majority of baryonic uncertainties by discarding the first $3\sim4$ PC modes. \citet{Mohammed18} point out that the training hydro simulations used to construct PCs have to be sufficiently broad in order to offer flexible degrees of freedom to span the possible baryonic scenarios for our Universe.

Other methods focus on modeling the ratio of power spectra that includes baryons to those that do not, with the goal of finding functional forms to describe the range of possible behavior of $P_{\rm \delta, bary}(k, z)/P_{\rm \delta, DMO}(k, z)$. 
\citet{HarnoisDeraps15} use a parametric form with 15 parameters that is able to describe the power spectrum ratio of several OWLS simulations \citep{vanDaalen11} to within 10\% precision up to $k \approx 20 \hMpc$ and $z < 1.5$. 
\citet{Chisari18} show that the above parametric form is sufficiently flexible to fit the power spectra ratio in the Horizon-AGN \citep{Dubois14} simulation to within 3\% across $z \lesssim 4$ up to $k \approx 30 \hMpc$, but with the downside of involving too many free parameters. The authors propose a more compact model with 4 parameters that is capable of providing a fit to Horizon-AGN to within < 5\%.

Based on the fact that baryonic physics mainly affects the matter power spectrum by altering the structure of dark matter halos, another proposed approach is to model the deviations in the matter power spectrum through the framework of the halo model \citep{Peacock00,Seljak00,Cooray02}. 
\citet{Zentner08, Zentner13} demonstrate that incorporating the halo concentration-mass relation and its redshift evolution into the halo model framework and marginalizing over the associated free parameters can successfully mitigate baryonic bias for Stage III surveys such as DES, but is insufficient for Stage IV experiments. 
In addition to the degree of freedom that governs halo concentration, \citet{Mead15, Mead16} consider a parameter that characterizes the mass dependence of feedback, with publicly available software available for this model  in \href{https://github.com/alexander-mead/HMcode}{\textsc{HMcode}}\footnote{https://github.com/alexander-mead/HMcode}. \citet{Copeland18} further extend \textsc{HMcode},  introducing a core radius parameter to characterize the inner halo structure that is believed to be an outcome of baryonic effects \citep{Martizzi12}. 
There are also approaches that go beyond NFW (Navarro-Frenk-White, \citealt{Navarro96}) halo profiles, focusing on modeling the radial density distributions of stellar, gas, and DM components of halos to capture the main features of baryonic feedback \citep{Semboloni11,Semboloni13,Mohammed14, Schneider15, Schneider19}. The improvement of the halo model approach is an active research area, in particular on constraining the prior range. These halo model approaches potentially enable us to jointly constrain halo structural information and cosmological parameters from data.

Baryonic effects can be mitigated also via a joint analysis through optimized combination of different cosmological probes, as demonstrated in \citet{Osato15}.
Finally, a gradient-based method is proposed recently by \citet{Dai18}. Dark matter particles in N-body simulations are moved along the gradient of estimated thermal pressure to mimic the effect of baryonic feedback. 
This method can be implemented as a post-processing step on N-body simulations to produce fast hydrodynamical-like simulations.

In this paper, we focus on studying two of the above baryonic mitigation methods -- the PCA method and \textsc{HMcode}. 
We test the effectiveness of these baryonic physics mitigation techniques on a broad range of possible baryonic scenarios by applying them to LSST-like mock observables constructed from hydrodynamical simulations of MassiveBlack-II \citep{Khandai15}, Illustris \citep{Vogelsberger14}, Eagle \citep{Schaye15}, and Horizon-AGN \citep{Dubois14}, and comparing their cosmological parameter constraints.  
In addition, for the PCA method, we investigate different ways of constructing the PCs, and provide a modification to the original formalism to improve their efficiency.  

This paper is organized as follows. In \S\ref{sec:simulations}, we give an overview on the hydrodynamical simulations used in this work for the construction of our training and test sets. 
\S\ref{sec:likelihood_sim} describes the setup of our simulated LSST-like likelihood simulations . 
In \S\ref{sec:methods}, we provide the detailed theoretical formalism for the baryonic mitigation techniques from literatures and our improved PCA scheme applied in this work. 
\S\ref{sec:result} presents the main results of the likelihood simulations under various baryonic scenarios and compares the performances of different mitigation methods. 
We summarize our findings in \S\ref{sec:summary}, and discuss the prospects of PCA-based methods for future investigation.

\section{Baryonic Effects in Simulations}
\label{sec:simulations}

In this section, we introduce the hydrodynamical simulations involved in our analysis (summarized in Table~\ref{tb:simulation}), and compare the impact of the baryonic physics considered on the matter distributions.

\begin{table*}
\caption{Basic information for the hydrodynamical simulations used in this work.}
\begin{tabular}{lrrrrrl}
\hline 
\textbf{Simulation}	&	\textbf{Box Length}	&	\textbf{Total }	&	\textbf{DM particle}		&	\textbf{initial gas}	& \textbf{force softening}		&	\textbf{cosmology}	\\
	&		&	\textbf{Particle \#}	&	\textbf{mass}		&	\textbf{particle mass}	& \textbf{length}		&		\\ \hline
OWLS			&	100 $\hMpc$		&	$2\times512^{3}$		&	$4.06 \times 10^8 \hMsun$	&	$8.66 \times 10^7 \hMsun$	& 	0.78 $\hkpc$	 			& 	WMAP3			\\ \hline
MassiveBlack-II		&	100 $\hMpc$		&	$2\times1972^{3}$		&	$1.1   \times 10^7 \hMsun$	&	$2.2   \times 10^6 \hMsun$	& 	1.85 $\hkpc$	 			& 	WMAP7			\\ \hline
Illustris			&	  75 $\hMpc$		&	$2\times1820^{3}$		&	$4.41 \times 10^6 \hMsun$	&	$8.87 \times 10^5 \hMsun$	& 	1.4  $\hkpc$	 			& 	WMAP7			\\ \hline
Eagle			&	67.77 $\hMpc$		&	$2\times1504^{3}$		&	$6.57 \times 10^6 \hMsun$	&	$1.23 \times 10^6 \hMsun$	& 	1.8  $\hkpc$	 			& 	Planck2013		\\ \hline
Horizon-AGN		&	100 $\hMpc$		&	$2\times1024^{3}$		&	$1.1   \times 10^7 \hMsun$	&	$2.2   \times 10^6 \hMsun$	& 	1.85 $\hkpc$	 			& 	WMAP7			\\ \hline

\label{tb:simulation}
\end{tabular}
\end{table*}

\subsection{OWLS Simulation Suite}
The OWLS simulations are a large suite of cosmological hydrodynamical simulations with varying implementations of subgrid physics to  enable investigations of the effects of altering or adding a single physical process on the total matter distribution \citep{Schaye10}.
Here we adopt 9 different baryonic simulations from OWLS. 
We refer readers to \citet{vanDaalen11} for a more detailed description.  
\begin{itemize}
 \item REF: The baseline simulation that contains many of the physical processes known to be important for galaxy formation except for the AGN feedback mechanism. REF includes prescriptions of radiative cooling and heating for 11 different elements, star formation assuming the \citet{Chabrier03} stellar initial mass function (IMF), stellar evolution, mass loss, chemical enrichment, and SN feedback in kinetic form (wind mass loading factor $\eta=2$ and initial wind velocity $v_{\rm w}=600 {\rm\ km\ s^{-1}}$ ; all together $\eta v_{\rm w}^2$ determines the energy injected into the winds per unit stellar mass). 
The other 8 hydro simulations are based on REF, with modifications indicated below.
 \item NOSN: Exclude SN feedback.
 \item NOZCOOL: Exclude metal-line cooling. Only assume primordial abundances when computing cooling rates.
 \item NOSN\_NOZCOOL: Exclude both SN feedback and metal-line cooling. 
 \item WML1V848: Adopt the same SN feedback energy per unit stellar mass as for REF, but reduce the mass loading factor by a factor of 2 ($\eta=1$) and increase the wind velocity by a factor of $\sqrt{2}$ ($v_{\rm w}=848 {\rm\ km\ s^{-1}}$).
 \item WDENS: Adopt the same SN feedback energy per unit stellar mass as that of REF, but let $\eta$ and $v_{\rm w}$ depend on gas density ($v_{\rm w} \propto n_{\rm H}^{1/6}$;  $\eta \propto n_{\rm H}^{-1/3}$).
 \item WML4: Double SN feedback per unit stellar mass by increasing the mass loading factor by a factor of 2 ($\eta=4$).
 \item DBLIMFV1618: Once the gas reaches a certain pressure threshold, 10\% of the star formation activity follows a top-heavy IMF. In this case, more high-mass stars are produced, which leads to higher SN energy feedback.
 \item AGN: In addition to physics included in the REF model, add a subgrid model for BH evolution and AGN feedback following the prescription of \citet{Booth09}. BHs inject 1.5\% of the rest mass energy of the accreted gas into the surrounding matter in the form of heat. 
\end{itemize}

The simulation cube for OWLS is $L = 100 \hMpc$ in comoving scale on a side. The OWLS-DMO simulation contains $512^3$ collisionless DM particles; the 9 hydro simulations contain an additional $512^3$ particles in the form of collisional gas or collisionless stars to capture the baryonic processes. The DM and (initial) gas particle masses are $\approx 4.06 \times 10^8 \hMsun$ and $8.66 \times 10^7 \hMsun$, respectively. The gravitational softening length is $\epsilon \approx 0.78 \hkpc$ in comoving scale, and is limited to a maximum physical scale of $2 \hkpc$. The cosmological parameters used in the simulation are based on WMAP3 results \citep{Spergel07}: $\{\Omegam,\ \Omegab,\ \Omegal,\ \sigma_8,\ \ns,\ h\} = \{0.238, 0.0418, 0.762, 0.74, 0.951, 0.73\}$.

The OWLS simulation sets are not specifically fine-tuned to match with key observables. As indicated in \citet{McCarthy17}, the original OWLS models
underpredict the abundance of $M_* < 10^{11} \Msun$ galaxies at the present day due to overly efficient stellar feedback (see their Fig.~1). The successor BAHAMAS simulation lowers the wind velocity $v_{\rm w}$ from 600 to $300 {\rm\ km\ s^{-1}}$ in order to provide a better fit to the observed abundance of low-to-intermediate mass galaxies.\footnote{Due to the low resolution of BAHAMAS, we are not able to include it as one of the hydrodynamical scenarios in this work (see Appendix~\ref{sec:appendixA2_2} for details of our resolution requirement).}

\subsection{Eagle Simulation}

The Eagle simulation \citep{Schaye15} is conducted in a cubic periodic box of side length $L = 67.77 \hMpc$ (comoving). There are $1504^3$ DM particles in both hydrodynamical and DMO simulations, and an approximately equal number of baryonic particles in the hydrodynamical run. The mass of each DM particle is $6.57 \times 10^6 \hMsun$ and the initial baryonic mass resolution is $1.23 \times 10^6 \hMsun$. The gravitational softening length is $\epsilon = 1.8 \hkpc$ in comoving units \citep{Eagle17}. 
The cosmological parameters used in Eagle are consistent with Planck 2013 results \citep{Planck14}: $\{\Omegam,\ \Omegab,\ \Omegal,\ \sigma_8,\ \ns,\ h\} = \{0.307, 0.04825, 0.693, 0.8288, 0.9611, 0.6777\}$.

The subgrid physics used in Eagle is based on OWLS. The physical models include radiative cooling and photoionization heating; star formation associated with stellar mass loss and energy feedback; BH mergers, gas accretion, and AGN feedback. The most important changes compared to OWLS are: star forming feedback energy changing in terms of thermal form rather than kinetic; accounting for angular momentum during the accretion of gas onto BHs; inclusion of a metallicity-dependence in the star formation law.
In contrast to many hydrodynamical simulations, Eagle employs stellar and AGN feedback only in thermal form, which captures the collective effects of mechanisms such as stellar winds, radiation pressure, SN feedback, radio- and quasar-mode AGN feedback. 
One major improvement in the treatment of thermal feedback is that it can be performed without turning off radiative cooling and hydrodynamical forces. 

The galaxy stellar mass function of Eagle matches extremely well with observations at $z=0.1$, because its stellar and AGN feedback related parameters are specifically calibrated at each resolution to reproduce this observable (see \citealt{Crain15, Schaye15} for details of calibration philosophy).


\subsection{MassiveBlack-II Simulation}

The MassiveBlack-II (hereafter MB2) simulation is a high-resolution $\Lambda$CDM cosmological simulation \citep{Khandai15}. 
Both DMO \citep{Tenneti15} and hydrodynamical MB2 simulations are conducted in a cubic simulation box with sides of length $L = 100 \hMpc$ in comoving scale. There are $1972^3$ DM particles in both the MB2-hydro and MB2-DMO simulations, with an additional $1972^3$ initial number of gas particles in the hydro run. The mass of each DM particle is $1.1 \times 10^7 \hMsun$ and the initial baryonic mass resolution is $2.2 \times 10^6 \hMsun$. The gravitational softening length is $\epsilon = 1.85 \hkpc$ in comoving units. 
The cosmological parameters in MB2 are consistent with WMAP7 results \citep{Komatsu11}: $\{\Omegam,\ \Omegab,\ \Omegal,\ \sigma_8,\ \ns,\ h\} = \{0.275, 0.046, 0.725, 0.816, 0.968, 0.701\}$.

The subgrid models of baryonic physics in MB2 includes a multiphase interstellar medium model with star formation and associated feedback by SN and stellar winds \citep{Springel03}; BH accretion, merger, and associated AGN feedback in quasar-mode \citep{DiMatteo05, Springel05}. 

The AGN feedback efficiency of MB2 is relatively weak compared with other hydrodynamical simulations that have AGN subgrid physics involved in this work. One outcome of this is that MB2 overpredicts the abundance of massive galaxies at low redshift \citep{Khandai15}. 

\subsection{Illustris Simulation}

The Illustris simulation \citep{Vogelsberger14} is carried out in a cubic periodic box with sides of length $L = 75 \hMpc$ (comoving). We download the highest resolution snapshot data from the public release website \citep{Nelson15} to calculate power spectra for both hydrodynamical and DMO runs. There are $1820^3$ DM particles in both hydrodynamical and DMO simulations, and an approximately equal number of baryonic particles in the hydrodynamical run. The mass of each DM particle is $4.41 \times 10^6 \hMsun$ and the initial baryonic mass resolution is $8.87 \times 10^5 \hMsun$. The gravitational softening length is $\epsilon = 1.4 \hkpc$ in comoving units. 
The cosmological parameters adopted in Illustris are consistent with WMAP7 results \citep{Komatsu11}: $\{\Omegam,\ \Omegab,\ \Omegal,\ \sigma_8,\ \ns,\ h\} = \{0.2726, 0.0456, 0.7274, 0.809, 0.963, 0.704\}$. 

Illustris incorporates a broad range of galaxy formation physics \citep{Vogelsberger13}: gas cooling in primordial and metal-lines; stellar evolution associated with chemical enrichment and stellar mass-loss; kinetic stellar feedback driven by SN; BH accretion, merging, and related AGN feedback in terms of quasar- and radio-modes as well as associated radiative electromagnetic feedback.  

Illustris is run using the moving-mesh-based code \textsc{AREPO} \citep{Springel10}, which is more efficient in cooling compared with classical particle-based SPH codes (e.g.\ \citealt{Springel05b}). The energy input from feedback is designed to be strong to avoid efficient stellar mass buildup. Even with this setting, Illustris still overshoots the observed low redshift stellar mass function on both high and low mass ends. The radio-mode AGN feedback is also too violent for the gas component, under predicting the baryon content in  lower-redshift high mass halos where the radio-mode feedback is the dominant heating channel  \citep{Genel14, Haider16}. The successor IllustrisTNG simulation replaces the intense thermal energy dump of radio-mode feedback with kinematic kicks to heat up affected gas particles \citep{Weinberger18}.

\subsection{Horizon-AGN Simulation}

The Horizon-AGN \citep{Dubois14} is carried out in a cubic periodic box of side length $L = 100 \hMpc$ (comoving). There are $1024^3$ DM particles in both the DMO and hydrodynamical runs, with the DM particle mass of $9.9 \times 10^7 \hMsun$ for the DMO run, and $8.3 \times 10^7 \hMsun$ for the hydrodynamical run. The initial gas particle mass is about $1 \times 10^7 \hMsun$. The cosmological parameters used in the simulation are compatible with WMAP7 cosmology \citep{Komatsu11}: $\{\Omegam,\ \Omegab,\ \Omegal,\ \sigma_8,\ \ns,\ h\} = \{0.272, 0.045, 0.728, 0.81, 0.967, 0.704\}$.

Subgrid physics models for a variety of baryonic physics effects are implemented in Horizon-AGN. 
Gas is allowed to cool down to $10^4$ K via transition lines of hydrogen and helium as well as metals using the \citealt{Sutherland93} model. 
When the hydrogen number density exceeds a threshold of $0.1~\rm{H}~\rm{cm}^{-3}$, star formation is triggered following a random Poisson process \citep{Rasera06, Shandarin89}. SN feedback is taken into account assuming an IMF with a low-mass cut-off at 0.1 $\mathrm{M}_{\odot}$ and a high-mass cut-off at 100 $\mathrm{M}_{\odot}$. 
Chemical enrichment happens along with SN explosions and stellar winds. 
The AGN feedback is modeled in a combination of two different modes: the kinematic radio mode when $\dot{M}_{\rm BH} / \dot{M}_{\rm Edd} < 0.01 $ and the thermal quasar mode otherwise \citep{Dubois12}.

Although Horizon-AGN is not specifically tuned to reproduce the galaxy stellar mass function at local Universe, it shows reasonable consistency with  observations, with slight overproduction of galaxies at the low mass end \citep{Kaviraj17}.

\subsection{Comparison of Power Spectra in Hydrodynamical versus DMO Simulations}

From the snapshot data release of Eagle, MB2 and Illustris, we calculate the matter power spectra as detailed in Appendix~\ref{sec:appendixA1}. For OWLS and Horizon-AGN simulations, we use the computed results from \citet{vanDaalen11} and \citet{Chisari18}, respectively. 
Power spectra from DMO simulations, with the same initial condition as their paired hydrodynamical simulations, are also computed in order to perform a fair comparison across simulations with different cosmological parameters and with reduced cosmic variance. For each paired simulation set, only a single realization was available to construct the power spectrum ratio.

\begin{figure}
\begin{center}
\includegraphics[width=0.48\textwidth]{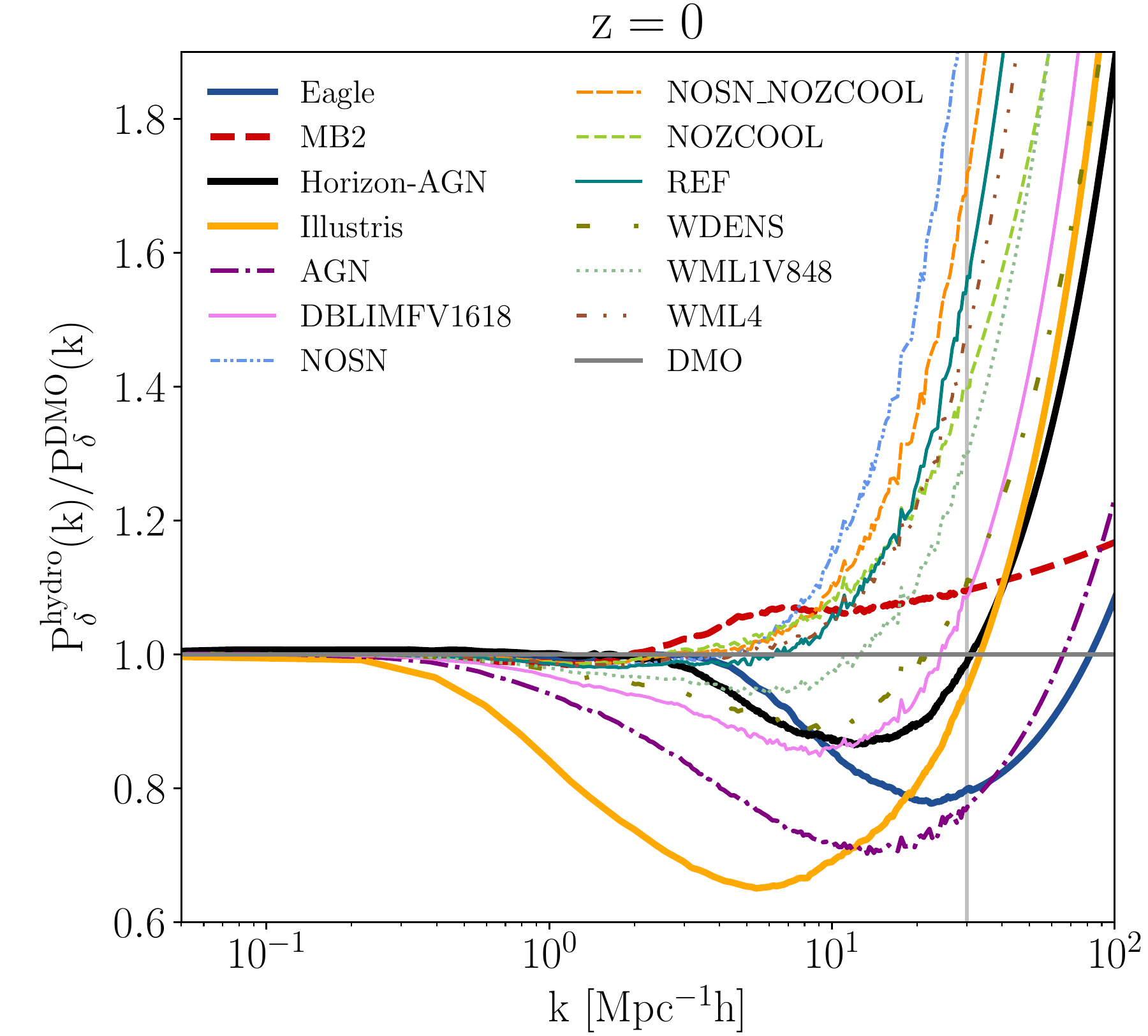}
\caption{The ratios of the matter power spectra in different hydrodynamical simulations with respect to their counterpart DMO simulations at $z=0$. The thick lines show results for the Eagle, MB2 and Illustris simulations, while the thin lines indicate the 9 different baryonic scenarios in OWLS simulation suite. The gray vertical line separates between regions where the data points come from direct measurement ($k \lesssim 30 \hMpc$) and from extrapolation with a quadratic spline fit ($k \gtrsim 30 \hMpc$; see Appendix~\ref{sec:appendixA2} for further details).}
\label{fig:Pk_ratio}
\end{center}
\end{figure}

Figure~\ref{fig:Pk_ratio} shows the $z=0$ ratio of power spectra from different hydrodynamical simulations with respect to their counterpart DMO simulations. The thin lines indicate the nine different baryonic scenarios in the OWLS simulation suite. For the eight baryonic scenarios without AGN feedback, the common feature is a rapid increase in power on small scales. The power enhancement is due to efficient cooling of gas which eventually leads to formation of galaxies within halos, and further concentrates the DM distribution \citep{Blumenthal86}.
Simulations without SN feedback (NOSN, NOSN\_NOZCOOL) tend to have an even stronger increase in power compared to the reference simulation REF due to the enhanced cooling effect.
When adding AGN feedback to REF, the power is suppressed dramatically, with 1\% reduction for $k \approx 0.3 \hMpc$ and exceeding 10\% for $k \gtrsim 2 \hMpc$ \citep{vanDaalen11}. The suppression of power is due to baryons being pushed outward by the energetic AGN feedback processes.

The thick lines represent power spectra ratio for Eagle, MB2, Illustris and Horizon-AGN simulations. Although they all involve a broad range of astrophysical processes that are believed to be relevant to galaxy formation, the resulting power spectra show significant differences. 
The feedback mechanism in Illustris drastically suppresses the power by 35\% at $k \approx 5 \hMpc$. Eagle reaches its maximum suppression of power of 20\% at $k \approx 20 \hMpc$. A similar trend is also observed in Horizon-AGN, but it reaches its minimum amplitude reduction of 10\% at $k \approx 10 \hMpc$.
Going towards higher $k$, we start to see that the ratio curves bend upward and keep increasing beyond $k$ of $30 \hMpc$. The MB2 power spectrum behaves relatively similar to DMO, but still the baryonic prescription prevents the power spectrum ratio from growing too quickly compared to the OWLS scenarios without AGN feedback, which suffer from severe overcooling effect (e.g. \citealt{Tornatore03, McCarthy11}).   

The input cosmologies ($\pcosim$) for the five simulation suites are different. 
In order to predict matter power spectra with baryonic effects for arbitrary cosmological parameters, we take the power spectrum ratios shown Fig.~\ref{fig:Pk_ratio} and apply the following equation:

\begin{equation}
\label{eq:Pk_ratio}
P_{\delta}^{\rm hydro}(k, z\ |\ \pco) = \frac{  P_{\delta}^{\rm hydro, sim}(k, z\ |\ \pcosim)  }{  P_{\delta}^{\rm DMO, sim}(k, z\ |\ \pcosim)  } P_{\delta}^{\rm theory}(k, z\ |\ \pco) \ ,
\end{equation}
where $P_{\delta}^{\rm hydro, sim}(k, z\ |\ \pcosim)$ denotes the hydrodynamical run from a given simulation; $P_{\delta}^{\rm DMO, sim}(k, z\ |\ \pcosim)$ is the corresponding DMO run; $P_{\delta}^{\rm theory}(k, z\ |\ \pco)$ is the theoretical power spectrum calculated from \textsc{Halofit} \citep{Takahashi12} or \textsc{HMcode} \citep{Mead15}, which are calibrated by DMO simulations. 

Eq.~\eqref{eq:Pk_ratio} illustrates the most important assumption in this work: we assume that baryonic effects on the power spectrum can be represented as a fractional change in the power spectrum, and that this fractional change is independent of cosmology. The cosmology enters our analysis only through the theoretical power spectrum $P_{\delta}^{\rm theory}(k, z\ |\ \pco)$. This is a reasonable assumption. According to \citet{vanDaalen19}, the power spectrum ratio remains more or less the same when varying cosmologies (see their Fig. 6). 



\section{Likelihood Analysis Methodology}
\label{sec:likelihood_sim}
Here we present our methodology in estimating the cosmological constraining power for an LSST-like survey. We start by describing the theoretical models used in the work, our mock observations, the covariance matrix constructed for an LSST-like survey, and finally the likelihood formalism used in estimating the posterior distribution of cosmological parameters. The cosmological model considered in our likelihood simulation is flat $w$CDM, with varying cosmological parameters $\pco=\{\Omegam,\ \sigma_8,\ \Omegab,\ \ns,\ w_0,\ w_{\rm a},\ h\}$.

\subsection{Theoretical Models}
\label{sec:Model}

We rely on two main theoretical models to fit our mock observables in this work.
The first one is the \citet{Takahashi12} version of \textsc{Halofit}.
It adopts empirically-motivated functional forms to characterize the variation of power spectra with cosmology.
Having been calibrated with high-resolution $N$-body simulations, it provides an accurate prediction of the nonlinear matter spectrum with 5\% precision at $k \leq 1 \hMpc$ and 10\% at $1 \leq k \leq 30 \hMpc$ within the redshift range of $0 \leq z \leq 10$. 

The second fitting routine is \textsc{HMcode}, constructed by \citetalias{Mead15}. It utilizes the halo-model formalism to describe the cosmological change of power spectra via physically motivated parameters. \textsc{HMcode} has prescriptions for capturing the impact of baryons on the matter power spectrum via two free parameters: the amplitude of the concentration-mass relation ($A$; see Eq.~(14) in \citetalias{Mead15}), and a halo bloating parameter ($\eta_0$; see Eqs.~(26), (29) in \citetalias{Mead15}) controlling the change of dark matter halo profiles in a halo mass-dependent way to account for different feedback energy levels. When allowing $A$ and $\eta_0$ to vary, it can successfully fit the power spectra from various baryonic scenarios of OWLS (\citetalias{Mead15}). 
When fixing $A=3.13$ and $\eta_0=0.6044$, \textsc{HMcode} functions as a regular DMO-based emulator, which is calibrated with high-resolution N-body simulations to an accuracy of $\approx 5\%$ at $k \leq 10 \hMpc$ for $z \leq 2$. We note that the $\approx 5\%$ discrepancy between the DMO mode of \textsc{HMcode} and \textsc{Halofit} is non-negligible within LSST statistics. We therefore construct two sets of mock observables based on each theoretical model.

\subsection{Mock Observational Data}
\label{sec:data_vector}

We rely on four hydrodynamical simulations: Eagle, MB2, Illustris and Horizon-AGN 
to construct mock observables, 
and investigate the performances of the PCA method \citep[][hereafter \citetalias{Eifler15}]{Eifler15} and the halo model approach \citep[][hereafter \citetalias{Mead15}]{Mead15} on mitigating baryonic effects. These methods will be described in more detail in \S\ref{sec:methods}. For simplicity, besides baryonic effects, our mock data vectors do not include any other source of noise or systematics.

We consider tomographic weak lensing shear power spectra as the summary statistics.  These are defined as: 
\begin{equation}
\label{eq:Cij}
C^{ij} (\ell) = \frac{9H_0^4 \Omega_m^2}{4c^4} \int_0^{\chi_{\rm h}} {\rm d} \chi
 \frac{g^{i}(\chi) g^{j}(\chi)}{a^2(\chi)} P_{\delta} \left(\frac{\ell}{f_K(\chi)},\chi \right) \ .
\end{equation}
Here $C^{ij}(l)$ is the convergence power spectrum for tomographic bin combination $\{i,j\}$ at angular wavenumber $l$, $\chi$ is the comoving distance, $\chi_{\rm h}$ is the comoving horizon distance, $f_K(\chi)$ is the comoving angular diameter distance (set to $\chi$ since we assume a flat Universe), $a(\chi)$ is the scale factor, and $P_{\delta}$ is the 3D matter power spectra.
The lens efficiency in the $i$-th tomographic interval is defined as 
\begin{equation}
\label{eq:lens_efficiency}
g^i(\chi) = \int_{\chi}^{\chi_{\rm h}} {\rm d} \chi^{\prime} n^{i}(\chi^{\prime}) \frac{f_K(\chi^{\prime}-\chi)}{f_K(\chi^{\prime})}
\end{equation}
with $n^{i}(\chi^{\prime}(z))$ being the redshift distribution of source galaxies in tomographic bin $i$. The overall source redshift distribution is parametrized in the form of
\begin{equation}
\label{eq:nz}
n(z) \propto z^{\alpha} \exp \left[ - \left(  \frac{z}{z_0} \right)^\beta \right]\ ,
\end{equation}
where $\alpha = 1.27$, $\beta = 1.02$, and $z_0=0.5$ following Table 2 in \citet{Chang13}, which mimics an LSST cosmic shear source galaxy sample after deblending. 
The number density of source galaxies is 26/${\rm arcmin}^2$.

\begin{figure}
\begin{center}
\includegraphics[width=0.48\textwidth]{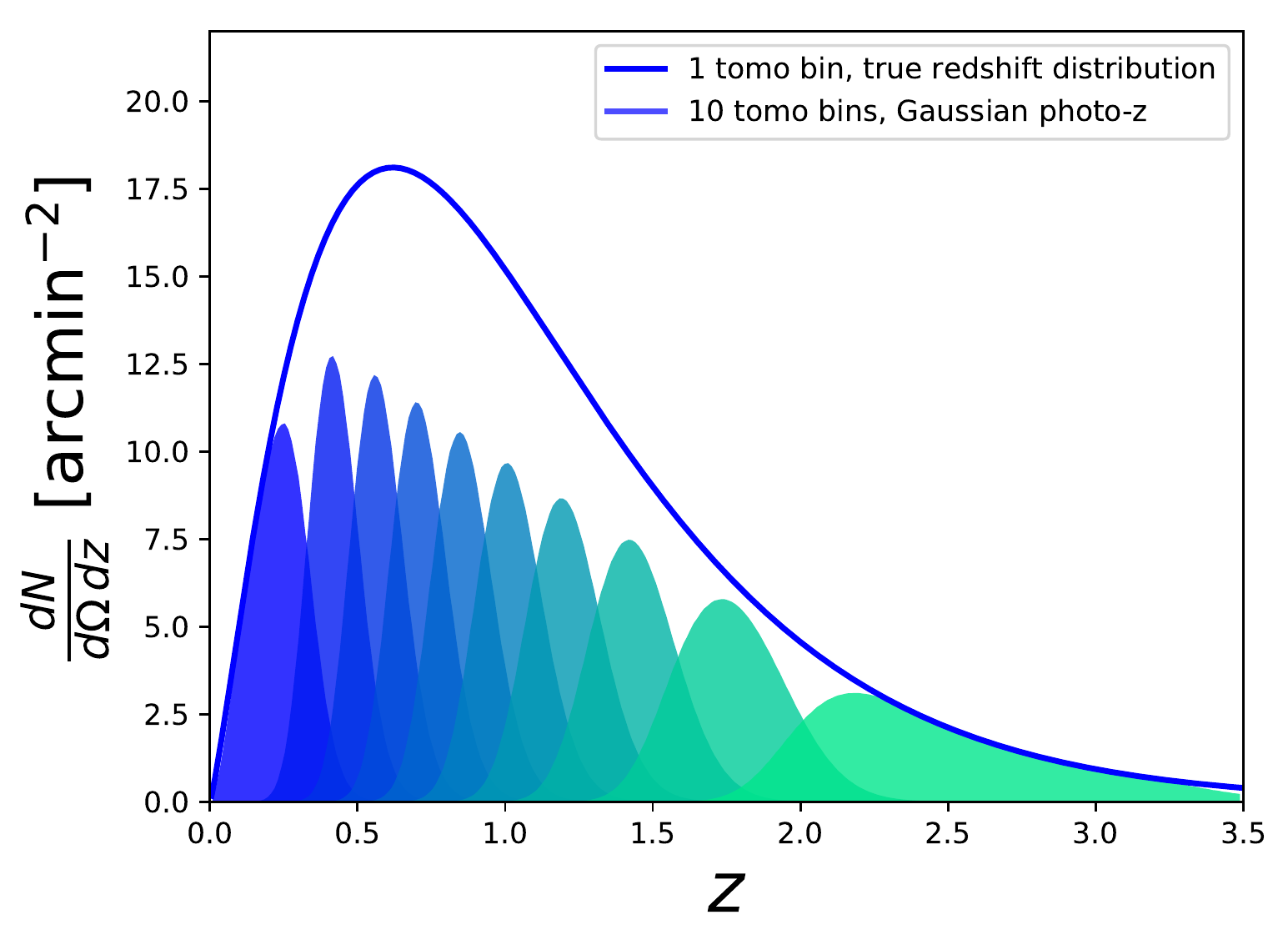}
\caption{The normalized galaxy number density split into ten Gaussian tomographic photo-z bins as shaded regions from blue (low z) to green (high z). For comparison, we show the true underlying redshift distribution as a solid blue line.}
\label{fig:zbins}
\end{center}
\end{figure}
We perform a tomographic analysis by dividing the sources into 10 tomographic bins with equal total number of galaxies in each bin. We also smooth the redshift distribution with a Gaussian kernel to characterize potential photo-z uncertainties. Fig.~\ref{fig:zbins} shows the exact redshift distribution in each bin. 
This results in 55 unique combinations of auto- and cross- correlation shear tomographic power spectra.
For each of the tomographic power spectra, we consider 18 equally spaced logarithmic bins in angular wavenumber $\ell$ ranging from $23\sim2060$. This results in a total of $55\times18=990$ data points in our data vector. 
For the main analysis of this paper, we adopt an upper limit of $\ell_\text{max} \approx 2000$. This limit is driven by the resolution of the hydrodynamical simulations used in this work. We refer readers to Appendix~\ref{sec:appendixA2} for further details on how we extrapolate power spectra to perform the integration to derive $C^{ij} (\ell)$, and how the decision on the  $\ell_\text{max} \approx 2000$ cut is made.

\begin{table}
\caption{Fiducial cosmology, minimum and maximum of the flat prior on the cosmological parameters, and halo-structural parameters in \textsc{HMcode}.} 
\begin{center}
\begin{tabular}{ccl}
\hline
Parameter  		 &     Fiducial   	&   Prior     		 \\  \hline
$\Omega_{\rm m}$	&	0.3156   	& flat (0.05, 0.6) 	\\
$\sigma_8$		&	0.831	& flat (0.5, 1.1) 		\\
$n_{\rm s}$		&	0.9645	& flat (0.84, 1.06) 	\\
$w_0$			& 	-1.0		& flat (-2.1, 0.0)		\\
$w_{\rm a}$		& 	0.0		& flat (-2.6, 2.6)		\\
$\Omega_{\rm b}$	&	0.0049	& flat (0.04, 0.055) 	\\
$h_0$			&	0.6727      & flat (0.4, 0.9)   	\\   \hline
$A$				&	    -		& flat (0.5, 10)		\\
$\eta_0$			&    	    -		& flat (0.1, 1.2)		\\  \hline
 
\label{tb:prior_info}
\end{tabular}
\end{center}
\end{table}

The fiducial cosmology $\pcofid$ of the data vectors is set to be consistent with the Planck 2015  (TT+TE+EE+lowP and assuming $\Lambda$CDM) results \citep{Planck16} as summarized in Table~\ref{tb:prior_info}.

Our mock data vectors for various baryonic physics scenarios are computed with the $P_{\delta}$ term in Eq.~\eqref{eq:Cij} generated from Eq.~\eqref{eq:Pk_ratio}. 
Since \textsc{Halofit} and \textsc{HMcode} (in DMO mode) agree at the level of $\lesssim 5\%$  to $k = 10 \hMpc$, and $\lesssim 10\%$ out to $k \leq 100 \hMpc$ (see Fig.~4 of \citetalias{Mead15}), 
we create two sets of Eagle/MB2/Illustris/Horizon-AGN data vectors, with $P_{\delta}^{\rm theory}(k, z\ |\ \pcofid)$ generated from \textsc{Halofit} or \textsc{HMcode}, and incorporate the baryonic features through the power spectrum ratio. 
Throughout our experiment, when relying on \textsc{Halofit} or \textsc{HMcode} as the theoretical model to perform fitting, we use the same fitting function to generate the mock observational data vectors for the fiducial cosmology.
This way, when comparing the performance of different baryonic mitigation schemes, if one of the methods fails to recover the fiducial cosmological parameters, we can be assured that this failure is purely because of that method's inability to mitigate the modification of the matter power spectrum due to baryonic physics, not because of an inherent discrepancy between the mock data and the DMO matter power spectrum model. 

\begin{figure}
\begin{center}
\includegraphics[width=0.48\textwidth]{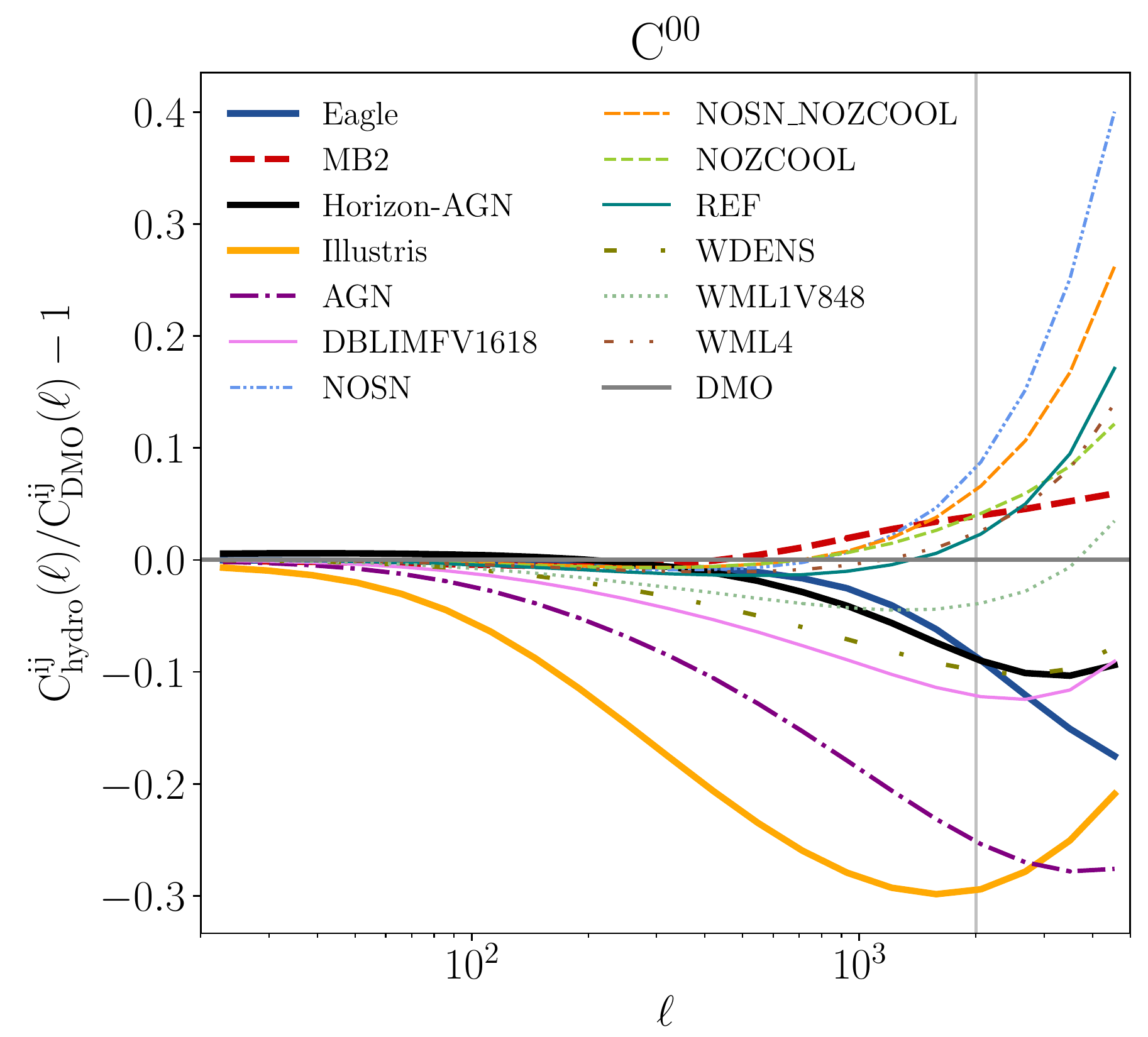}
\caption{The ratio of tomographic shear power spectra of different hydrodynamical simulations with respect to their counterpart DMO simulations for the lowest auto-correlation tomographic bin with the cosmology set at the Planck 2015 result (Table~\ref{tb:prior_info}). The thick lines represent the cases for Eagle/MB2/Illustris/Horizon-AGN simulations, while the thin lines indicate the 9 different baryonic scenarios in OWLS simulation suit. }
\label{fig:Cl_ratio}
\end{center}
\end{figure}

In Fig.~\ref{fig:Cl_ratio} we show the ratio of baryonic to DMO $C^{00}(\ell, \pcofid)$ shear power spectrum for various simulations. The thin lines indicate the nine baryonic scenarios from the OWLS simulation suite. The thick lines represent the Eagle/MB2/Illustris/Horizon-AGN universes, which are the data vectors that we will use for the LSST-like experiment. 
One can see that in this lowest tomographic bin, even for large scales at $\ell \approx 100$, the baryonic scenario of Illustris already causes a deviation from DMO at the 5\% level, with even more severe suppressions at smaller angular scales. 
For higher redshift tomographic bins, the deviations between hydrodynamical and DMO simulations are less severe.
\citet{Semboloni11} showed that a scale cut of $\ell_{\rm max} \approx 500$ would be needed to avoid $w_0$ bias for a Euclid-like survey if the baryonic scenario of our Universe is like OWLS-AGN. When applying the traditional way of mitigating baryonic uncertainty by omitting small scale information, we would need to discard a considerable amount of data before we can rely on DMO-based theoretical model to achieve an unbiased cosmological inference. 

One subtle feature shown in Fig.~\ref{fig:Cl_ratio} is that there is a small but noticeable large-scale excess of power (< 0.4\%) in the Horizon-AGN simulation. 
This is because the power spectrum ratio between hydrodynamical and DMO runs of Horizon-AGN has < 0.1\% excess at large scales (see Fig.~\ref{fig:Pk_ratio}), even though they share the same initial conditions. The true cause of this subtle excess is not clear. After exploring, \citet{Chisari18} concluded that this may originate from the box being too small to reach the linear regime at large scales.  However, the other simulations studied here are similar in size and do not exhibit this feature.

\subsection{Covariance Matrix}
We generate the analytical covariance matrix of tomographic shear power spectra using \textsc{CosmoLike} \citep{Eifler14,Krause17b}.
Briefly, our covariance matrix contains both Gaussian and non-Gaussian parts. 
The Gaussian covariance matrix contains contributions from cosmic variance and shape noise, derived under the assumption that the 4pt-function of the shear field can be expressed in terms of 2pt-functions \citep{Hu04, Takada07}. 
The non-Gaussian part is given by the convergence trispectrum derived using the halo model \citep{Cooray02}, which contains one-, two-, three-, and four-halo terms and a halo sample variance term characterizing the scatter of halo number density due to large-scale density fluctuations \citep{Cooray01, Takada09, Sato09}. The exact equations of our implementation can be found in the appendix of \citet{Krause17b}. 

We assume 18,000 $\rm{deg}^2$ as the survey area in our covariance matrix and adopt the same redshift distribution and source galaxy number density ($26/\rm{arcmin}^2$) as depicted in Fig.~\ref{fig:zbins}. The shape noise is set to be $\sigma_\epsilon=0.26$ in each ellipticity component. 

\subsection{Likelihood Formalism}

Given a data vector $\D$ (at some fiducial cosmology and with baryonic effects from Eagle/MB2/Illustris/Horizon-AGN), one can infer the corresponding posterior probability distribution of cosmological parameters $\pco$ and potential nuisance parameters $\pnu$ via Bayes' theorem:
\begin{equation}
\label{eq:Bayes}
P(\pco, \pnu | \D) \propto L(\D | \pco, \pnu) P_r (\pco, \pnu)\ ,
\end{equation}
where $P_r (\pco, \pnu)$ denotes the prior probability distribution and $L(\D | \pco, \pnu)$ is the likelihood. In this work, we assume a Gaussian likelihood function for the observables,
\begin{equation} 
\label{eq:likelihood}
L(\D | \pco, \pnu) \propto \exp \biggl( -\frac{1}{2} \underbrace{\left[ (\D -\M)^t \, \C^{-1} \, (\D-\M) \right]}_{\chi^2(\pco, \pnu)}  \biggr) \ .
\end{equation}
We further assume that the covariance $\C$ is constant in parameter space for simplicity (but see \citealt{Eifler09, Morrison13} for likelihood analysis with cosmology-dependent covariance matrix). 
As described in \S\ref{sec:Model}, the model vector $\M$ may be derived based on \textsc{Halofit} which is a pure function of cosmology $\M = \M(\pco)$, or it can be a function of some nuisance parameters $\M = \M(\pco, \pnu)$ as well, with factors that are known to affect $\D$ absorbed in $\pnu$. For example, in \textsc{HMcode}, we have $A$ and $\eta_0$ acting as nuisance parameters to account for the baryonic effects (see \S\ref{sec:Model} for details). 
The final posterior distribution on cosmological parameters then can be derived by marginalizing over all other nuisance parameters in the model
\begin{equation}
P(\pco | \D) \propto \int {\rm d} \pnu \  P(\pco, \pnu | \D) \ .
\end{equation}

We use the python \texttt{emcee} package \citep{Foreman-Mackey13}, which relies on the algorithm of \citet{goodman2010ensemble} to sample the parameter space spanned by $\pco$ ($\{\Omegam,\ \sigma_8,\ \Omegab,\ \ns,\ w_0,\ w_{\rm a},\ h_0\}$) as well as $\pnu$ (if needed depending on the model). Altogether, we have conducted $\sim$250 likelihood simulations to present the results for this paper. The MCMC (Markov Chain Monte Carlo) chains contain $\sim$ 200000 to 400000 MCMC steps (after discarding 100000 steps as burn-in phase), depending on the dimension of the parameter space which ranges from 7$\sim$16. 
For simplicity, we assume flat priors for all of our parameters, with their minimum and maximum values summarized in Table~\ref{tb:prior_info}. For likelihood simulations with informative priors based on Planck, we refer readers to \citetalias{Eifler15}. Informative priors help to better constrain $n_s$, $\Omega_b$, and $h$, to which cosmic shear is not very sensitive.

\begin{figure}
\includegraphics[width=0.5\textwidth]{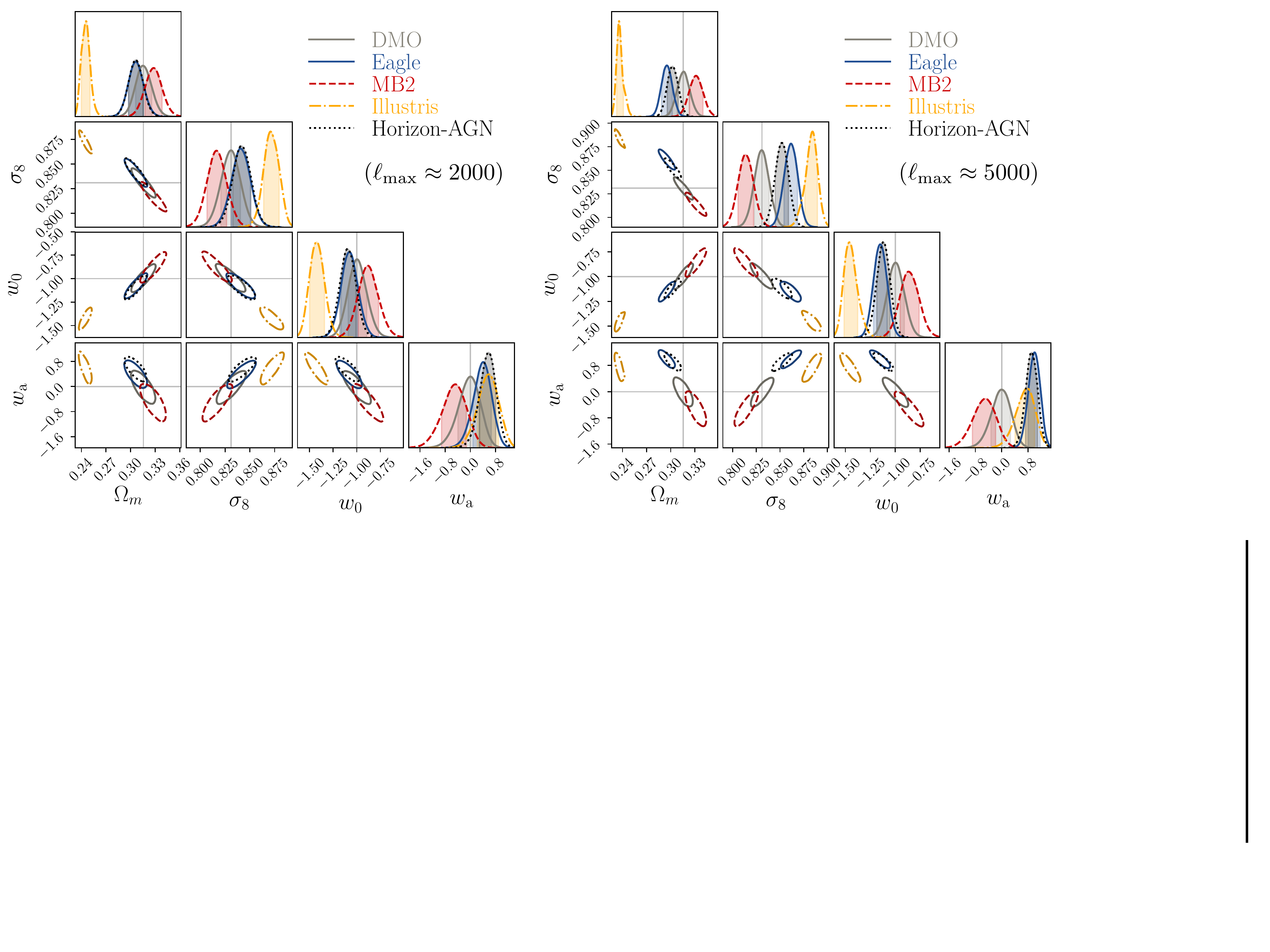}
\caption{Cosmological parameter constraints for an LSST-like weak lensing survey with data vectors generated using various baryonic physics scenarios: pure DM (gray/solid) and the Eagle (blue/solid), MB2 (red/dashed), Illustris (yellow/dot-dashed) and Horizon-AGN (black/dotted) hydrodynamical simulations. In all cases, baryonic physics was ignored during the likelihood analysis, hence providing a worst-case scenario for biases due to baryonic physics. The analyses are carried out assuming non-informative priors on the parameters. Here, and in all such 2D posterior plots below, the contours depict the 68\% confidence levels. \textit{Depending on the intensity of the baryonic feedback, the resulting posterior distributions can be significantly away from the fiducial cosmology (marked in gray lines).} }
\label{fig:posterior_no_trick}
\end{figure}

We will present in \S\ref{sec:methods} on how we implement various baryonic mitigation schemes in the likelihood analysis. But before that,  
in Fig.~\ref{fig:posterior_no_trick} we show the posterior distribution of cosmological parameters derived from our LSST likelihood simulation, when naively applying the \textsc{Halofit} model on fitting the data vectors contaminated with baryonic effects from Eagle/MB2/Horizon-AGN/Illustris simulations. For ease of visualization, we only show posteriors in the subspace of four cosmological parameters out of seven in total. Depending on the intensity of baryonic feedback as reflected in the ratio of hydrodynamical to DMO power spectra shown in Fig.~\ref{fig:Pk_ratio}, the resulting cosmology constraints can be severely biased in the case of Illustris ($2\sigma \sim 13\sigma$ depending on cosmological parameters) or at $1\sigma \sim 2\sigma$ level in the other three cases. 
We note that the degree of bias depends on the $\ell_\text{max}$ used in the analysis. Fig.~\ref{fig:posterior_no_trick} presents the result when applying a cut at $\ell_\text{max} \approx 2000$ on $\D$, which is the default setting in the paper. 
In \S\ref{sec:ell_5000}, we will show how this result changes when extending data vectors to $\ell_\text{max} \approx 5000$.

\vspace{1in}
\section{Methods of mitigating baryonic effects}
\label{sec:methods}

In this section, we describe the methods used to mitigate the impact of baryonic physics on the cosmological parameter estimates from weak lensing. The methods can be classified into two categories: PCA-based methods and the halo-model based approach. We discuss several PCA-based methods that are minor variants of each other in  \S\ref{sec:PCA} to \S\ref{sec:PCA_R}. The halo-model based approach is described in \S\ref{sec:HMcode}.
Throughout the work, we use the nine OWLS simulations as our `training sample' to construct PCs for the PCA-based methods, and use the four  mock data vectors constructed from Eagle/MB2/Illustris/Horizon-AGN simulations as `test sample' to test methods listed in Table~\ref{tb:baryon_methods}.

\begin{table*}
\caption{Summary of baryonic physics mitigation techniques. The first column is the label of each method, which we refer to in the text and plots throughout the work. The second column has simple descriptions that highlight the essential elements of each method. The third column presents the exact $\bm \chi^2$ equations that go into the likelihood analysis. Finally, the last column provides a section number where more information can be found for each method.}
\begin{tabular}{clcc}
\hline 

$\!\!\!\!$\textbf{Method}$\!\!\!\!$ 		&	\textbf{Brief description}	&   \textbf{$\bm \chi^2$ equation}  		& $\!\!\!\!$\makecell[c]{\textbf{Section} \\ \textbf{reference}}			\\  \hline		
A				&	PCA in difference matrix, with exclusion	&   $ [(\D-\M)_{\rm pc, cut}]^{\t}\ \C^{-1}_{\rm pc, cut}\ [(\D-\M)_{\rm pc, cut}] $ 						& \S\ref{subsec:PCA_removal} 	\\  [0.3mm]  \hline
B				&	PCA in difference matrix, with marginalization &   $ [\D-\M_B(\pco, \Q)]^{\t}\ \C^{-1}\ [\D-\M_B(\pco, \Q)] $ 									& \S\ref{subsec:PCA_marginalization_D}		\\ [0.3mm] \hline
C				&	PCA in $\invL$ weighted difference matrix, with exclusion$\!\!\!\!$ 	&   $ [\U_{\rm ch} \P \U_{\rm ch}^{\t} \invL (\D-\M)]^{\t}\ \I\ [\U_{\rm ch} \P \U_{\rm ch}^{\t} \invL (\D-\M)] $  	& \S\ref{sec:Cholesky}				\\ [0.3mm] \hline
D				&	PCA in fractional difference matrix, with marginalization	&   $ [\D-\M_R(\pco, {\bm Q})]^{\t}\ \C^{-1}\ [\D-\M_R(\pco, {\bm Q})] $   			& \S\ref{sec:PCA_R}	\\ [0.3mm] \hline
M				&	Halo model parameter marginalization	&   $\!\!\!\!\!\! [\D-\M_{\rm HMcode}(\pco, A, \eta_0)]^{\t}\ \C^{-1}\ [\D-\M_{\rm HMcode}(\pco, A, \eta_0)] \!\!\!\!\!\! $   		& \S\ref{sec:HMcode}	\\  [0.3mm] \hline
\label{tb:baryon_methods}
\end{tabular}
\end{table*}

\subsection{PCA in Difference Matrix}
\label{sec:PCA}

\subsubsection{Summary of the PCA framework (Method A)}
\label{subsec:PCA_removal}

The original framework for using PCA to mitigate the impact of baryonic physics for weak lensing is described in \citet{Eifler15}. 
The essential idea is that even though hydrodynamical simulations with different baryonic prescriptions predict a range of variations on the matter power spectra (Fig.~\ref{fig:Pk_ratio}), we can still extract the common features of those diversity using PCA, and build an empirical model to mitigate baryonic uncertainty based on these hydrodynamical simulations.
Below we provide a step-by-step description of the PCA framework.

Firstly, we collect the tomographic shear power spectra constructed from the nine OWLS simulations as our training sample, and
label these nine data vectors as $\B_1, ..., \B_9$. 
Next we build a difference matrix ${\bm \Delta}(\pco)$ with dimension of $N_\text{data} \times N_\text{sim} = 990 \times 9$. Each column records the deviation between the baryonic data vector and the DMO model vector $\M$ at any arbitrary cosmology (recomputed for each MCMC step) in terms of their difference
\begin{equation} \label{eq:DiffMatrix}
{\bm \Delta}(\pco) = 
\left[
  \begin{array}{cccc}
    \vertbar & \vertbar &        & \vertbar \\
    \B_1-\M    & \B_2-\M    & \ldots & \B_9-\M    \\
    \vertbar & \vertbar &        & \vertbar 
  \end{array}
\right]_{N_\text{data} \times N_\text{sim}} \ .
\end{equation} 
The left panel of Fig.~\ref{fig:PCmodes_Diff} provides a visualization of the entries of the difference vectors used to construct $\bm \Delta$.
Here notice that both $\B_x(\pco)$ and $\M(\pco)$ are functions of cosmology, and therefore so is ${\bm \Delta}$. We refer readers to Appendix~\ref{sec:appendixA3} for details of how we compute the baryon-contaminated data vectors at different $\pco$.

\begin{figure}
\begin{center}
\includegraphics[width=0.48\textwidth]{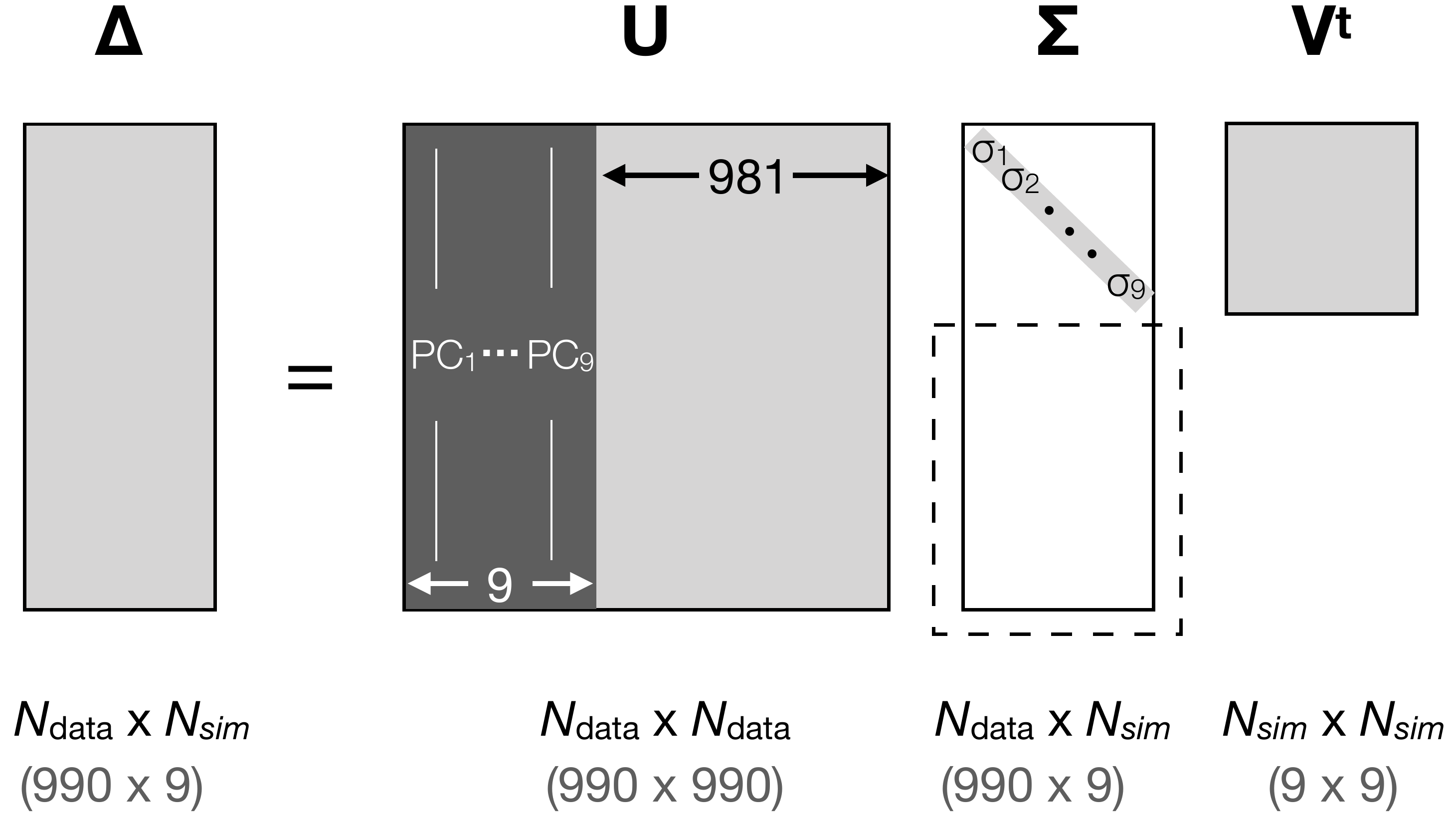}
\caption{We perform singular value decomposition (SVD) on the difference matrix $\bm \Delta$ built based on the 9 baryonic scenarios of OWLS (see Eq.~\eqref{eq:DiffMatrix}). $\U$ is a unitary matrix with columns that form an orthonormal basis set to span the 990-dimensional space of our data vector. Among them, the first 9 PCs of $\U$ form a complete description of the modifications of the data vector due to baryonic physics in the nine OWLS hydro simulations. We will test whether these 9 PCs can also describe the impact of baryonic physics in the Eagle/MB2/Illustris/Horizon-AGN simulations.}
\label{fig:SVD}
\end{center}
\end{figure}

\begin{figure*}
\begin{center}
\includegraphics[width=0.98\textwidth]{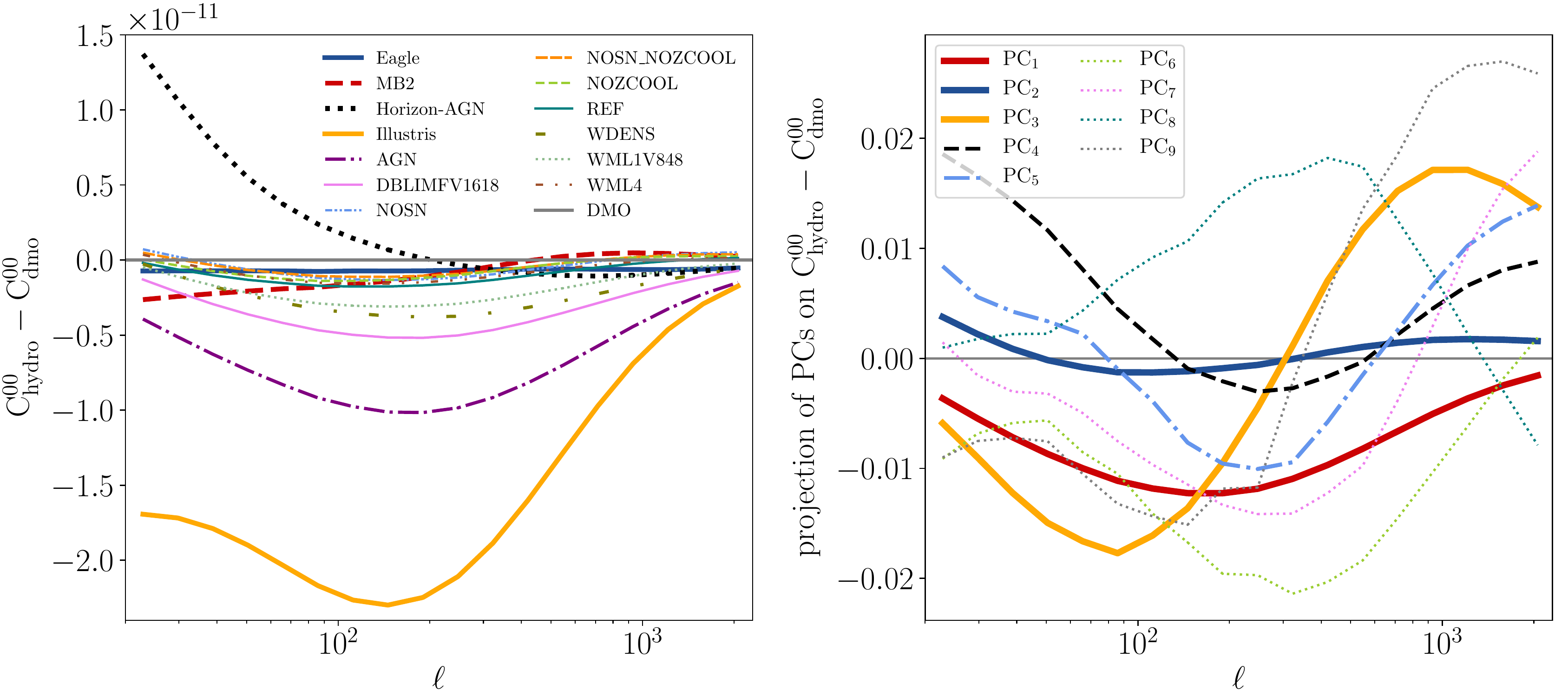}
\caption{Left: The difference vectors, $\B-\M$, from the OWLS simulation set used to construct PCs as input in columns of Eq.~\eqref{eq:DiffMatrix}. The thicker lines indicate the difference vectors for Eagle/MB2/Horizon-AGN/Illustris simulations as our test set. Right: The PC modes constructed from the OWLS simulation set in projection on the difference vector space for the tomographic bin $C^{00}$. \textit{The goal of this work is to check whether these PC modes can flexibly describe the baryonic physics scenarios in the test set hydrodynamical simulations.}}
\label{fig:PCmodes_Diff}
\end{center}
\end{figure*}

The second step is to perform the PCA on the difference matrix, with the goal of identifying the few dominant principle components (PCs) that signify the directions of largest discrepancy between the baryonic and DMO data vectors from the nine OWLS simulations. To find the PCs, we apply the (full) singular value decomposition (SVD) on ${\bm \Delta}$, 
\begin{equation} \label{eq:SVD}
{\bm \Delta} = \U\ \bm \Sigma\ \V^{\t} \ .
\end{equation} 
As shown in Fig.~\ref{fig:SVD}, SVD decomposes $\bm \Delta$ into the product of three matrices. 
Both $\U$ and $\V$ are square unitary matrices with dimensions of $N_{\rm data} \times N_{\rm data}$ $(990 \times 990)$ and $N_{\rm sim} \times N_{\rm sim}$ $(9 \times 9)$ respectively. The upper $N_{\rm sim} \times N_{\rm sim}$ $(9 \times 9)$ block of $\bm \Sigma$ is a diagonal matrix consisting of $N_{\rm sim}$ (9) positive real singular values $\sigma_1 ... \sigma_9$ arranged in descending order, and the remaining $N_{\rm data}-N_{\rm sim}$ (981) rows have only zeros (indicated by the dashed square). 
The $N_{\rm data}$ (990) columns of $\U$ are eigenvectors of $\bm \Delta \bm \Delta^{\t}$, with eigenvalues in the diagonal entries of $\bm \Sigma \bm \Sigma^{\t}$. 

The first 9 eigenvectors constitute a set of orthogonal PCs in order of decreasing importance according to the amount of variation they capture in the different training vectors. The right panel of Fig.~\ref{fig:PCmodes_Diff} shows these 9 PC modes in projection on the $C^{00}$ tomographic bin. The PC modes span a 9-dimensional subspace within the 990 dimensional space which covers entirely the degrees of freedom to explain baryonic uncertainties in the nine OWLS hydro simulations. In other words, any given $\B_x(\pco)-\M(\pco)$, can be described with 9 free parameters via
\begin{equation} \label{eq:PCs_span}
\B_x(\pco)-\M(\pco) = \sum_{n = 1}^{9} Q_n\ {\mathbf {PC}}_n (\pco)\ ,
\end{equation} 
with $Q_n$ being the amplitude of ${\mathbf {PC}}_n$.
The remaining 981 columns of $\U$ are silent orthogonal vectors which extends $\U$ into a unitary matrix. 
With 9 baryonic scenarios as our training sample, we have at most 9 independent PCs to describe modifications to the observables due to baryonic physics.
One of the goals of this work is to understand how effectively the PCA basis can describe baryonic physics scenarios in other more recent hydrodynamical simulations.

The third step is to transform everything to PC basis, and mitigate baryonic uncertainty by excluding PC modes. 
In PC basis, our data and model vectors are defined as
\begin{subequations}
\begin{align}
 \D_{\rm pc} &= \U^{\t} \D \label{eq:Dpc}  \\ 
 \M_{\rm pc} &= \U^{\t} \M \label{eq:Mpc}  \ ,
 \end{align}
\end{subequations}
and the covariance matrix is 
\begin{equation}
\C_{\rm pc} = \U^{\t}\ \C\ \U \ .
\end{equation}
Viewing from PC coordinate, the majority of the baryonic uncertainties between $\D_{\rm pc}$ and $\M_{\rm pc}$ would be absorbed in the first $N$ elements. 
We can then directly cut the data vector $\D_{\rm pc}$ to obtain a shorter vector $\D_{\rm pc, cut}$, and do the same to the model vector $\M_{\rm pc} \rightarrow \M_{\rm pc, cut}$ to avoid modeling challenges on these data points. 

When doing MCMC analysis, we modify the original Eq.~(8) from \citetalias{Eifler15} to properly account for the change of covariance matrix due to loss of information after PC mode removal. We cut the corresponding rows and columns on $\C_{\rm pc}$, and use the corresponding sub-matrix, $\C_{\rm pc, cut}$ to calculate the inverse covariance $\C^{-1}_{\rm pc, cut}$ for $\D_{\rm pc, cut}$. The $\chi^2$ equation can then be written as:
\begin{equation} \label{eq:chi2_new}
\chi'^2(\pco) = (\D-\M)_{\rm pc, cut}^{\t}\ \C^{-1}_{\rm pc, cut}\ (\D-\M)_{\rm pc, cut} \ , 
\end{equation}
and the likelihood equation: 
\begin{equation} 
\label{eq:likelihood_PC_removal}
L(\D | \pco) \propto \exp \biggl( -\frac{1}{2} \underbrace{\left[ (\D-\M)_{\rm pc, cut}^{\t}\ \C^{-1}_{\rm pc, cut}\ (\D-\M)_{\rm pc, cut} \right]}_{\chi'^2(\pco)}  \biggr) \ .
\end{equation}

\subsubsection{The marginalization version of the PCA framework (Method B)}
\label{subsec:PCA_marginalization_D}

We refer to `PC marginalization' as a method that includes (up to nine) amplitudes of PCs as free parameters to parametrize the impact of baryonic physics on the tomographic shear power spectra. As shown in Eq.~\eqref{eq:PCs_span}, the current 9 PCs fully span the baryonic degrees of freedom in the 9 OWLS simulations. We can further check whether they are also effective in describing the impact of baryonic physics on the observables in our test set of hydrodynamic simulations by building a new model with the following parametric form: 
\begin{equation} \label{eq:MD}
\M_B(\pco, \Q ) = \M(\pco) + \sum_{n = 1}^{m} Q_n\ {\mathbf {PC}}_n (\pco)\ .
\end{equation}
where $m \leqq 9$, and $\Q = \{Q_1, Q_2, ... , Q_m \}$ are free parameters in addition to the cosmological parameters. 
The likelihood function for the cosmological parameters can be derived by marginalizing over the amplitude parameters: 
\begin{equation} 
\label{eq:likelihood_PC_mar}
\begin{aligned}
L(\D | \pco) & \propto \int d\Q \  \times    \\
                   &            \exp \biggl( -\frac{1}{2} \left[ (\D-\M_B(\pco, \Q))^{\t}\ \C^{-1}\ (\D-\M_B(\pco, \Q)) \right]  \biggr) \ .
\end{aligned}
\end{equation}

Theoretically, one can prove that the likelihood functions of Eq.~\eqref{eq:likelihood_PC_removal} (method A) and Eq.~\eqref{eq:likelihood_PC_mar} return identical results if the priors on the PC amplitudes are uninformative.
We will provide comparisons of the posterior distributions of $\pco$ in \S\ref{sec:PCex_vs_PCmar} and further comment on both methods there. 




\subsection{Noise-weighted PCA -- Cholesky Decomposition (Method C)}
\label{sec:Cholesky}

As noted at the end of \S2.2 of \citetalias{Eifler15}, performing PCA on the difference matrix $\bm \Delta$ (Eq.~\eqref{eq:DiffMatrix}) is not necessarily the most optimal choice. They suggested an option of conducting the PCA on the `noise'-weighted $\bm \Delta$. As a result of re-weighting, the derived PCs would be more sensitive in accounting for deviations in data vectors due to baryonic physics at well-measured data points, where larger weighting factors are applied. Therefore, when doing PC mode removal, we tend to more effectively remove baryonic physics degrees of freedom that impact better-measured (lower noise) scales, which may more effectively reduce cosmological parameter biases. 

To find the weights, we first decompose our covariance matrix by applying a Cholesky Decomposition
\begin{equation} \label{eq:chy_de}
\C = \L \L^{\t}\ ,
\end{equation}
where $\L$ is a lower triangular matrix with real and positive diagonal entries. We can then weight our $\D$ and $\M$ vectors as
\begin{equation} 
\label{eq:chy_w}
\begin{aligned}
\D_{\rm ch} &= \invL \D  \ ,\\
\M_{\rm ch} &= \invL \M  \ .
\end{aligned}
\end{equation}
After this transformation, our new data vector $\D_{\rm ch}$ has an identity covariance matrix $\mathbbm{1}$, which can be easily proved as follows
\begin{equation} \label{eq:chy_cov}
\begin{aligned}
\C_{\rm ch} &= \mean{ (\D_{\rm ch}-\overline{\D}_{\rm ch}) (\D_{\rm ch}-\overline{\D}_{\rm ch})^{\t} } = \mean{ \invL(\D-\overline{\D}) (\invL(\D-\overline{\D}))^{\t}  } \\
&= \invL \mean{  (\D-\overline{\D}) (\D-\overline{\D})^{\t}  }  (\invL)^{\t}  = \invL  \C^{-1}  (\invL)^{\t}  = \mathbbm{1} \ .
\end{aligned}
\end{equation} 
In other words, after applying Eq.~\eqref{eq:chy_w}, we not only re-weight but also decorrelate the data vector. 

Similar to Eq.~\eqref{eq:DiffMatrix}, we build the new difference matrix as
\begin{equation} \label{eq:DiffMatrix_chy}
\begin{aligned}
{\bm \Delta_{\rm ch}}(\pco) &= 
\left[
  \begin{array}{cccc}
    \vertbar  &        & \vertbar \\
    \B_{\rm 1, ch}-\M_{\rm ch}    & \ldots & \B_{\rm 9, ch}-\M_{\rm ch}    \\
    \vertbar  &        & \vertbar 
  \end{array}
\right]_{N_{\rm data} \times N_{\rm sim}} \\
 & = \invL {\bm \Delta}(\pco) = \U_{\rm ch} \ \bm \Sigma_{\rm ch}\ \V_{\rm ch}^{\t} \ .
 \end{aligned}
\end{equation} 
Here each of the OWLS training data vectors is weighted by $\invL$ as $\B_{\rm x, ch} = \invL \B_{\rm x}$. 
The $\bm \Delta_{\rm ch}$ matrix is equivalent to performing a $\invL$ matrix transformation on $\bm \Delta$ shown in Eq.~\eqref{eq:DiffMatrix}.
We can then apply SVD to derive the PC basis set as stored in the $\U_{\rm ch}(\pco)$ matrix. The first 9 PCs form natural bases to span the weighted difference vector for various baryonic effects
\begin{equation} \label{eq:PCspan_ch}
\B_{\rm ch} - \M_{\rm ch} = \invL (\B - \M)  =  \sum_{n = 1}^{9} Q_n\ {\mathbf {PC}}_n \ .
\end{equation}

\begin{figure}
\begin{center}
\includegraphics[width=0.48\textwidth]{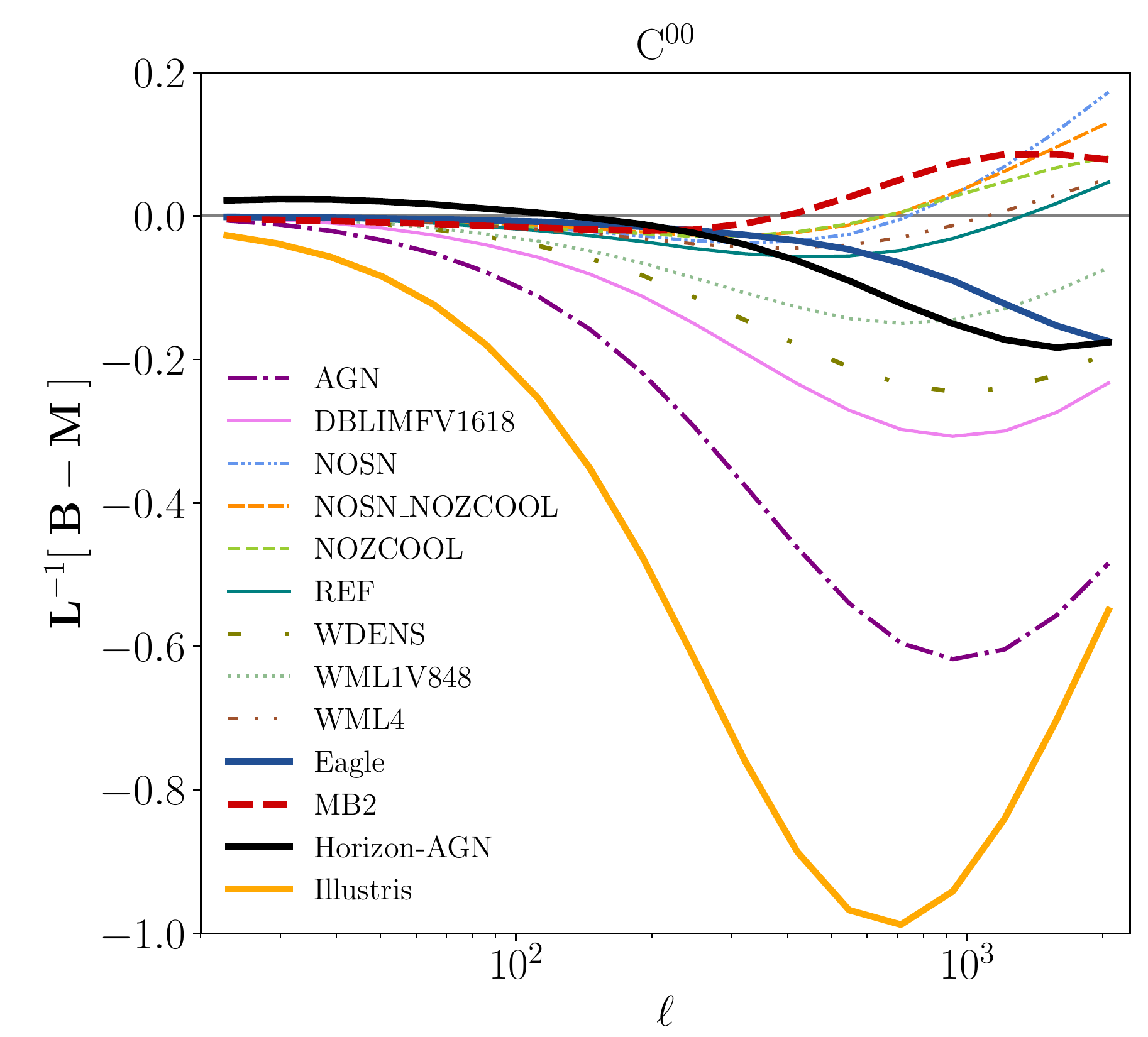}
\caption{The discrepancy between baryon-contaminated data vectors and model in terms of $\B_{\rm ch} - \M_{\rm ch}$ for various hydrodynamical simulations in the lowest tomographic bin. This is similar to the left panel of Fig.~\ref{fig:PCmodes_Diff}, but here shows results for the case when applying Cholesky decomposition on our $\B$ and $\M$ vectors. The nine OWLS baryonic scenarios (thinner lines) compose columns of $\bm \Delta_{\rm ch}$, which are used to build PCs. These PCs are used to span the variation of Eagle/MB2/Illustris/Horizon-AGN simulations in $\D_{\rm ch} - \M_{\rm ch}$ space. \textit{After Cholesky decomposition, the largest data-model inconsistency shifts to smaller scales compared with the upper panel of Fig.~\ref{fig:PCmodes_Diff}, indicating the PCs trained from $\bm \Delta_{\rm ch}$ are more efficient at describing small-scale variations in the matter power spectrum due to baryonic physics compared with performing PCA on $\bm \Delta$.}}
\label{fig:invLDM}
\end{center}
\end{figure}

In Fig.~\ref{fig:invLDM}, we show the $\B_{\rm ch} - \M_{\rm ch} = \invL (\B-\M)$ vectors in our lowest tomographic bin, at $\pcofid$.
The thicker lines represent our four test simulations; the thinner lines are for the nine baryonic scenarios in OWLS, which compose the columns of $\bm \Delta_{\rm ch}$ in Eq.~\eqref{eq:DiffMatrix_chy}. 
Comparing with the left panel of Fig.~\ref{fig:PCmodes_Diff}, one can see that after re-weighting by $\invL$, we more strongly emphasize baryonic fluctuations at smaller scales, so the PCs should also be more effective in accounting for small-scale baryonic features.\footnote{Although we plot $\D_{\rm ch} - \M_{\rm ch}$ vs.\ $\ell$ in Fig.~\ref{fig:invLDM}, we note that actually the new data points are not strictly functions of the original $\ell$ because of the non-zero off-diagonal terms in $\invL$. However, our take-away point from Fig.~\ref{fig:invLDM} still holds due to the fact that our covariance matrix is dominated by Gaussian noise, and thus the off-diagonal terms in $\invL$ are small.}

Similar to \S\ref{subsec:PCA_removal}, to perform the PC mode removal, we transform everything to the PC basis:  
\begin{subequations}
\begin{align}
 \D_{\rm ch, pc} &= \U_{\rm ch}^{\t} \D_{\rm ch} \label{eq:Dchpc}  \\ 
 \M_{\rm ch, pc} &= \U_{\rm ch}^{\t} \M_{\rm ch} \label{eq:Mchpc}  \\ 
 \C_{\rm ch, pc} &= \U_{\rm ch}^{\t}\ \C_{\rm ch}\ \U_{\rm ch} = \mathbbm{1} \ ,
 \end{align}
\end{subequations}
and then cut all the elements from the data and model vectors and the covariance matrix for the PC modes that are to be removed. Here since $\C_{\rm ch}$ is an identity matrix, it and its inverse $\C^{-1}_{\rm ch, pc}$ in the PC basis remain the same after coordinate transformation. 
The final PC mode removal $\chi^2$ equation becomes:
\begin{equation} 
\label{eq:chi2_chy}
\begin{aligned}
\chi^2_{\rm ch}(\pco) &=  (\D_{\rm ch}-\M_{\rm ch})_{\rm pc, cut}^{\t}\ \C^{-1}_{\rm ch, pc\ cut}\ (\D_{\rm ch}-\M_{\rm ch})_{\rm pc, cut} \\
				&=  (\D_{\rm ch}-\M_{\rm ch})_{\rm pc, cut}^{\t}\ (\D_{\rm ch}-\M_{\rm ch})_{\rm pc, cut}
\end{aligned}
\end{equation}


The marginalization version of method C can be viewed as the following. By reorganizing Eq.~\eqref{eq:PCspan_ch}, we can build a baryonic model generator as
\begin{equation} \label{eq:MC}
\M_C(\pco, \Q ) = \M(\pco) + \L \sum_{n = 1}^{m} Q_n\ {\mathbf {PC}}_n (\pco)\ , 
\end{equation}
where $m \leqq 9$. The cosmological parameter-dependence comes in through the DMO model vector, while the amplitudes of PCs are used as higher order correction for baryonic effects.

\subsection{PCA in Fractional Difference Matrix (Method D)}  
\label{sec:PCA_R}
Instead of using the difference matrix $\bm \Delta$ to perform PCA, \citet{Mohammed18} identified PCs based on the fractional difference matrix $\R$ defined as:
\begin{equation} \label{eq:RatioMatrix}
\begin{aligned}
{\R} &= 
\left[
  \begin{array}{cccc}
    \vertbar & \vertbar &        & \vertbar \\
    \frac{\B_1-\M}{\M}    & \frac{\B_2-\M}{\M}    & \ldots & \frac{\B_9-\M}{\M}    \\
    \vertbar & \vertbar &        & \vertbar 
  \end{array}
\right]_{990 \times 9} \\
&= \U_{\rm R} \ \bm \Sigma_{\rm R}\ \V_{\rm R}^{\t} \ .
\end{aligned}
\end{equation}
One fundamental difference between the fractional difference matrix $\R$ and the difference matrices $\bm \Delta$ or $\bm \Delta_{\rm chy}$ is that $\R$ does not depend on cosmology, given our assumption of Eq.~\eqref{eq:Cl_ratio}. 
After the $\U_{\rm R}$ is derived by SVD analysis, a model for the observables with baryonic physics degrees of freedom spanned by OWLS can be built as:
\begin{equation} \label{eq:MR}
\M_R(\pco, \Q ) = \M(\pco)\ \left[ 1 + \sum_{n = 1}^{m} Q_n\ {\mathbf {PC}}_n \right]\ , 
\end{equation}
where $m \leqq 9$, and ${\Q} = \{Q_1, Q_2, ... , Q_m \}$ are free parameters controlling the amplitudes of PCs, and PC$_1$ $\sim$ PC$_9$ are in the first nine columns of $\U_{\rm R}$. 
Similar to the methodology in \S\ref{subsec:PCA_marginalization_D}, the likelihood function for the cosmological parameters can be derived by marginalizing over the amplitude parameters:
\begin{equation} 
\label{eq:likelihood_PC_mar_R}
\begin{aligned}
L(\D | \pco) & \propto \int d\Q \  \times    \\
                   &            \exp \biggl( -\frac{1}{2} \left[ (\D-\M_R(\pco, \Q))^{\t}\ \C^{-1}\ (\D-\M_R(\pco, \Q)) \right]  \biggr) \ .
\end{aligned}
\end{equation}

Similar to the concept mentioned in \S\ref{sec:Cholesky}, performing PCs on the matrix $\R$ can be viewed as putting the weight of $1/\M$ into the PCA analysis. Since $\M$ decreases with increasing $\ell$, and the overall amplitude of $\M$ increases toward higher redshift, after taking its inverse, we upweight data points at smaller scales and lower redshift. 
The fractional difference vectors of OWLS that go into columns of $\R$ are plotted in Fig.~\ref{fig:Cl_ratio}. The PCs derived from $\R$ are expected to be more efficient in accounting for smaller scale and lower redshift variation of the observables due to baryonic physics.

\subsection{HMcode (Method M)} \label{sec:HMcode}

Finally, we compare the above PCA-based methods with the halo model-based approach proposed from \citet{Mead15}, \textsc{HMcode}.
\textsc{HMcode} utilizes two halo profile-related parameters to capture the impact of baryonic physics on the matter power spectrum: the amplitude of the concentration-mass relation ($A$) and a halo bloating parameter ($\eta_0$) controlling the (mass-dependent) change of halo profiles. We refer readers back to \S\ref{sec:Model} for a brief summary of this approach. 

There exists some level of degeneracy between $A$ and $\eta_0$, as shown in Fig.~6 of \citetalias{Mead15}. Thus, when implementing the likelihood analysis, one can either vary both of the parameters, or change only the single parameter $A$ while fixing 
\begin{equation} \label{eq:A_eta0}
\eta_0 = 0.98 - 0.12 A. 
\end{equation} 
For example, \citet{Joudaki17} applied only varying $A$ to marginalize over baryonic physics in CFHTLenS cosmic shear, while \citet{MacCrann17} and \citet{Troxel18} varied both parameters to marginalize over baryonic physics in the Dark Energy Survey (DES).

Equation~\eqref{eq:A_eta0} is derived based on the OWLS simulation suite.
We will test whether it remains valid for the baryonic physics scenarios in Eagle/MB2/Illustris/Horizon-AGN for our forecasted scenario with LSST-like statistical power. 
Also, we will compare the performances of \textsc{HMcode} (marginalization over halo model parameters) with the above PCA-based methods.

\section{Performances of Baryonic Mitigation Techniques}
\label{sec:result}


In this section, we present our simulated likelihood analysis for the different baryonic mitigation schemes listed in Table~\ref{tb:baryon_methods}. We refer readers back to \S~\ref{sec:likelihood_sim} for a description of the simulated likelihood analysis setup.

Ideally, we need a baryonic physics mitigation strategy that can reduce the biases in cosmological parameters due to inaccuracies in theoretical modeling (as demonstrated in Fig.~\ref{fig:posterior_no_trick}) to a level that is much smaller than the statistical uncertainties. In addition, we hope that the increase in statistical errors on cosmological parameters due to the additional nuisance parameters will be as small as possible.
Throughout this section, we will use a criterion of bias $< 0.5\sigma$ (where $\sigma$ represents the marginalized statistical error) for individual cosmological parameters to evaluate whether a method is effective in mitigating the uncertainties due to baryonic physics under various baryonic scenarios. We also compare their performance based on the degradation of cosmological constraining power through the size of the 1D marginalized uncertainties on cosmological parameters. 


\subsection{PC Mode Exclusion versus Marginalizing Over PC Amplitude}
\label{sec:PCex_vs_PCmar}

We start by presenting the results for methods A and B (see Table~\ref{tb:baryon_methods}) with their PCs described using the same difference matrix $\bm \Delta$. 
In Method A, we modify the data vector by excluding the first few PC modes and modify the covariance self-consistently as well. In method B, the data vector and covariance matrix are unmodified, but we introduce free parameters describing the PC amplitudes to marginalize over in the likelihood analysis.

Mathematically, methods A and B are equivalent if no priors are set on the baryonic physics parameters.
From an information perspective, when removing data points (and the corresponding covariance elements), we lose all of the information that can constrain the amplitudes of the excluded PC modes. Thus, this should be equivalent to marginalizing over PC amplitudes with uninformative priors. 

\begin{figure}
\begin{center}
\includegraphics[width=0.5\textwidth]{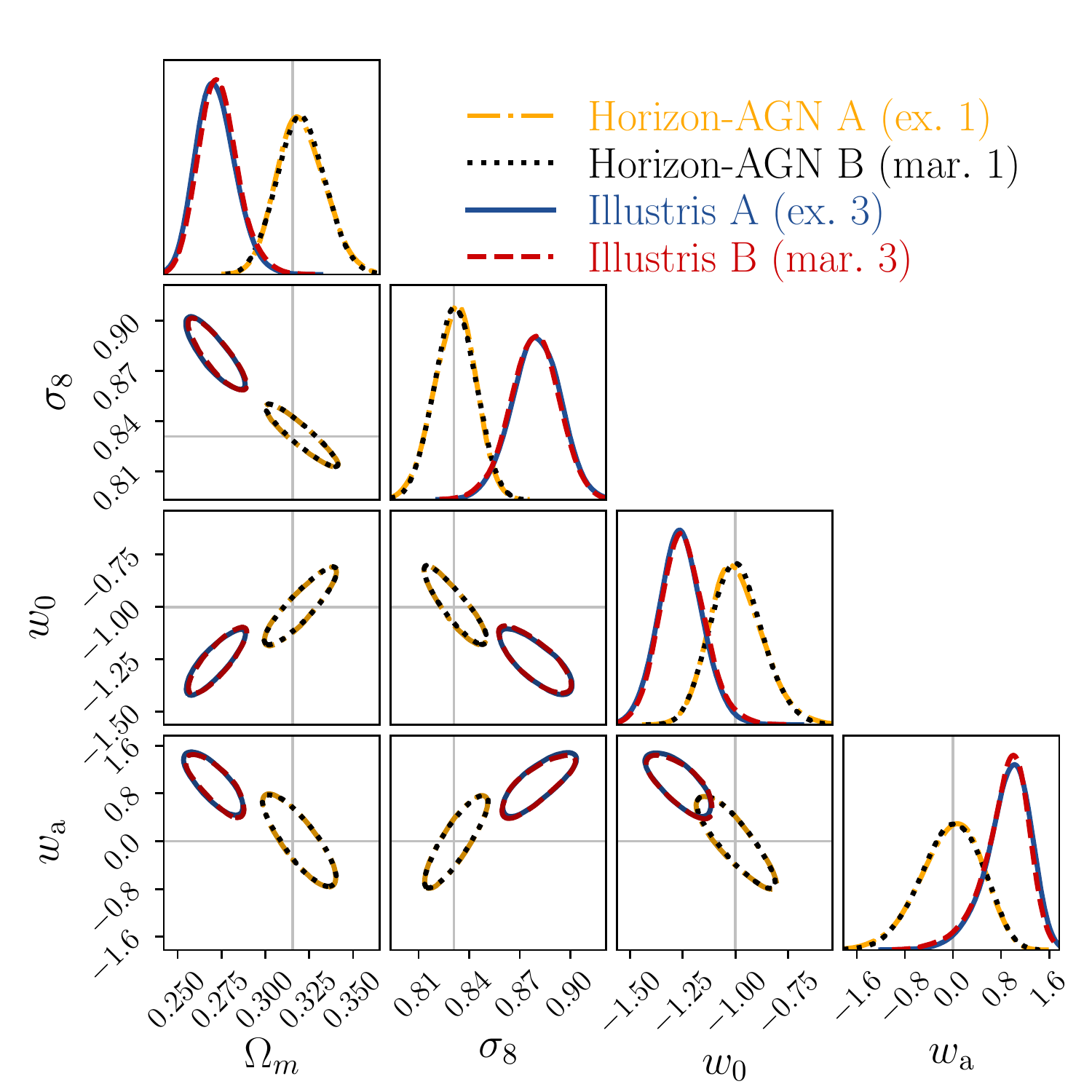}
\caption{Comparison of the posterior distributions of cosmological parameters between baryonic physics mitigation techniques A \& B listed in Table~\ref{tb:baryon_methods}. The yellow dot-dashed and black dotted contours indicate the $1\sigma$ contours of the posterior probability distributions obtained from methods A and B, respectively, for the Horizon-AGN simulation after excluding or marginalizing over the first PC modes. Similarly, the blue solid and red dashed contours indicate the case for Illustris after excluding or marginalizing over 3 PC modes. \textit{The excellent match between the posterior probability distributions for cosmological parameters between methods A and B confirms that the PC exclusion formula shown in Eq.~\eqref{eq:chi2_new} is conceptually equivalent to marginalizing over PC amplitudes.}}
\label{fig:MethodA_vs_B}
\end{center}
\end{figure}

In Fig.~\ref{fig:MethodA_vs_B}, we use simulated likelihood analyses based on Horizon-AGN (yellow dot-dashed \& black dotted) and Illustris (blue solid \& red dashed) to demonstrate the excellent consistency between PC mode exclusion and PC amplitude marginalization. 
For the case of Illustris, large residual biases still exist after performing the baryonic physics mitigation. We will discuss this issue in \S\ref{sec:PCA_vs_HMcode}.
Although not shown, we have also confirmed the consistency between methods A \& B for Eagle and MB2.

In conclusion, we have demonstrated with examples that the PC exclusion formula shown in Eq.~\eqref{eq:chi2_new} gives consistent results as when marginalizing over PC amplitudes with an uninformative prior. 
Method B can provide baryonic information through the constrained PC amplitudes, which can be used as a standard to quantify baryonic effects. So far, we allow the PC amplitudes to vary from ($-\infty$, $\infty$). Reducing the prior ranges on PC amplitudes could potentially increase the constraining power on cosmology if we can develop a consistent way of setting the priors on PC amplitudes, given our knowledge of baryonic physics. The downside of method B is that it requires running longer MCMC chains to ensure convergence due to an increase in the dimensionality of parameter space. 
Therefore if one does not care to learn about baryonic physics, and would simply like to marginalize over it, we recommend method A.



\subsection{Comparison between various PC construction methods}
\label{sec:cp_PCAs}

Here we compare the performances of the PCA-based methods listed Table~\ref{tb:baryon_methods}. We have already shown in \S\ref{sec:PCex_vs_PCmar} that PC mode exclusion (method A) is equivalent to marginalizing over PC amplitudes (method B), so here we only compare methods A, C, and D. 

The fundamental difference between these PCA methods is the way the PCs are constructed from the training simulations, which affects their efficiency in describing how baryonic physics modifies the data vectors on larger or smaller scales. We refer readers back to \S\ref{sec:methods} for more details about this formalism. 
Briefly, when PCs are derived from $\bm \Delta$ (method A; Eq.~\eqref{eq:DiffMatrix}), they are most efficient in describing the difference vector $\D-\M$. 
For PCs trained from $\bm \Delta_{\rm chy}$ (method C; Eq.~\eqref{eq:DiffMatrix_chy}), they are most efficient in describing the noise-weighted difference, $\D_{\rm chy}-\M_{\rm chy} = \invL (\D-\M)$, due to baryonic physics. Finally, PCs trained from $\R$ (method D; Eq.~\eqref{eq:RatioMatrix}) are most efficient in describing variations in the fractional difference $\frac{\D-\M}{\M}$ from baryonic effects. 

\begin{figure*}
\begin{center}
\includegraphics[width=1\textwidth]{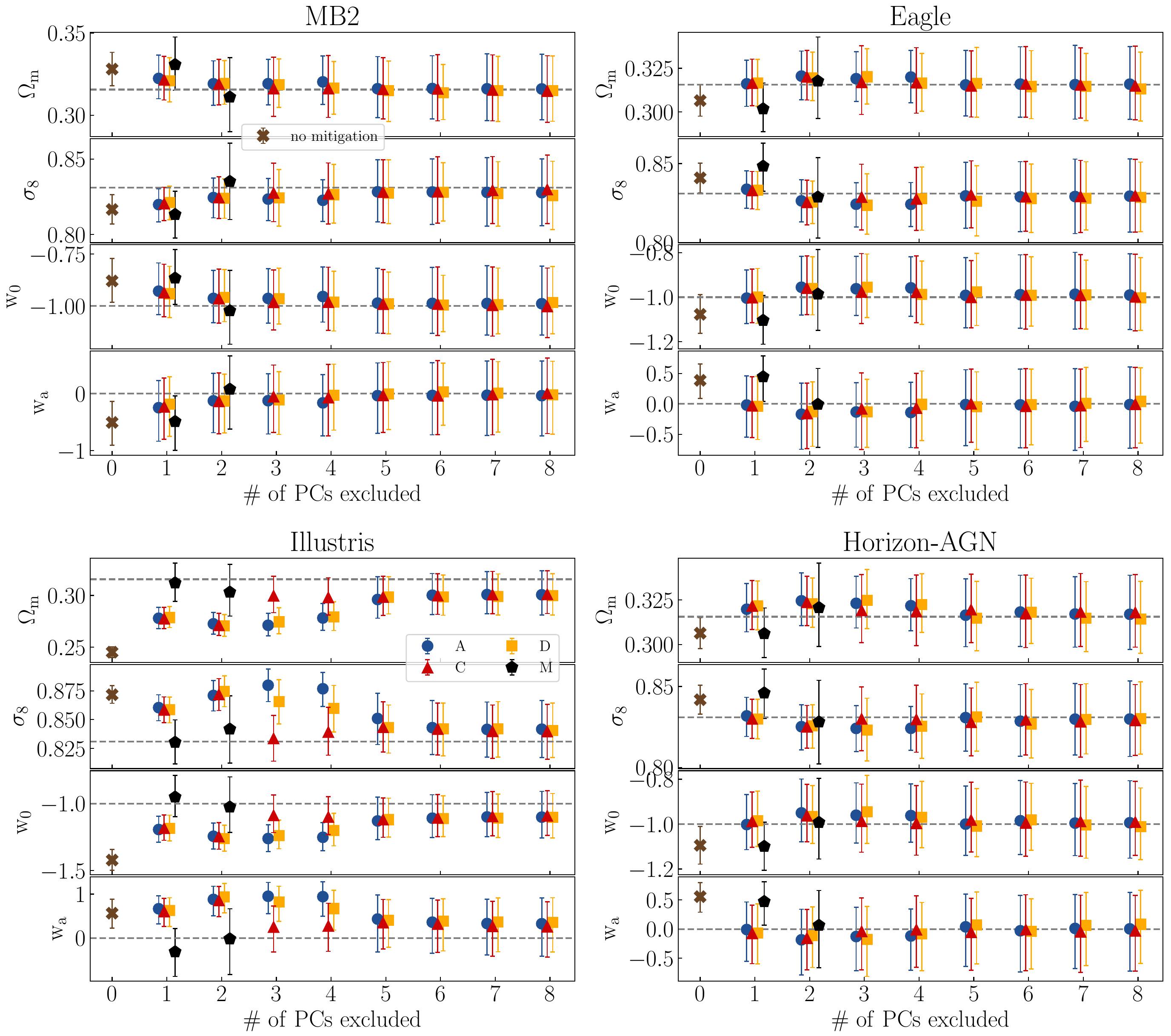}
\caption{Marginalized 1D constraints on cosmological parameters when using different baryonic physics mitigation techniques from Table~\ref{tb:baryon_methods}. Each panel is for a different input data vector based on a different hydrodynamical simulation as explained in the plot title. The gray dashed horizontal lines indicate the fiducial cosmological values. The marker position, the lower and upper error bars indicate the median, the 16th and the 84th percentiles of marginalized 1D posteriors. The brown crosses indicate the results when fitting the data vectors with the DMO-based emulator (\textsc{Halofit}) without applying any baryonic physics mitigation technique. The blue circles, red triangles and yellow squares show the results when applying PCA-based methods A, C, and D respectively, with their positions in the $x-$direction indicating how many PC modes are excluded or numbers of marginalization parameters used when doing the analysis. The black pentagons located at $x=1$ indicate the result when only marginalizing over $A$ in \textsc{HMcode} (with $\eta_0$ fixed via Eq.~\eqref{eq:A_eta0}). The black pentagons located at $x=2$ are the results when marginalizing over both $A$ and $\eta_0$ in \textsc{HMcode}. \textit{For PCA-based methods, we find the 1$\sigma$ posteriors start to enclose the fiducial cosmology after removing 2 PC modes for MB2/Eagle/Horizon-AGN, while excluding 6 PC modes is required for more extreme baryonic scenarios of Illustris. When using \textsc{HMcode} to perform marginalization, except for the Illustris simulation for which marginalizing over $A$ alone is enough, generally it is required to vary both $A$ and $\eta_0$ to mitigate baryonic effects to within 1$\sigma$.} }
\label{fig:pco_1Dist}
\end{center}
\end{figure*}

\begin{figure*}
\begin{center}
\includegraphics[width=1\textwidth]{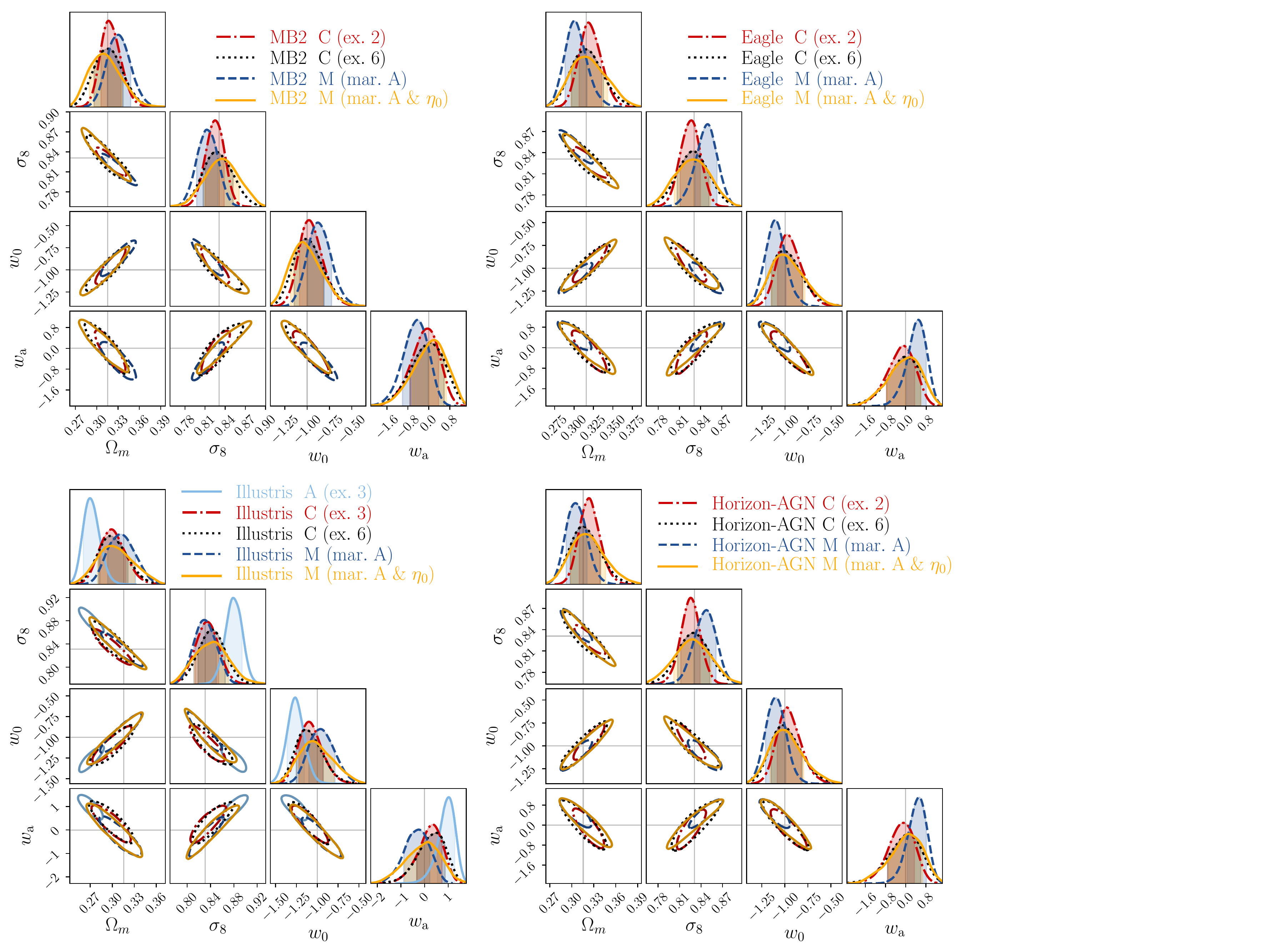}
\caption{The 2D posterior distributions on cosmological parameters for some selected cases shown in Fig.~\ref{fig:pco_1Dist}. Each panel is for a different input data vector based on a different baryonic physics scenario as labeled in the legend. The legend also describes which baryonic mitigation techniques are applied, and how many PC modes are excluded or the \textsc{HMcode} parameters marginalized over.}
\label{fig:posterior4_cp}
\end{center}
\end{figure*}

Figure~\ref{fig:pco_1Dist} shows the median of the marginalized 1D posteriors of cosmological parameters under different baryonic physics mitigation techniques for data vectors derived from our four test simulations. The lower and upper error bars represent for the 16 and 84th quantiles of the 1D marginalized posterior distribution.
The $x-$axes indicate numbers of PC modes excluded or numbers of marginalization parameters used in the analysis. We select some cases from Fig.~\ref{fig:pco_1Dist} and present their 1D and 2D posteriors in Fig.~\ref{fig:posterior4_cp}. 
The brown crosses in Fig.~\ref{fig:pco_1Dist} indicate the case when no baryonic physics mitigation scheme is applied. One can see that the deviations from the fiducial cosmological parameters exceeds 1$\sigma$ for all of our test baryonic scenarios. This is also shown in Fig.~\ref{fig:posterior_no_trick} on the 2D posterior contours. The blue-circle, red-triangle, and yellow-square markers indicate the results of performing baryonic physics mitigation by PCA-based methods A, C, and D, respectively. 
When the modifications of the data vectors due to baryonic physics are relatively weak as in MB2/Eagle/Horizon-AGN, we find that removing up to 2 PC modes is sufficient to marginalize baryonic bias to within $1\sigma$\footnote{The $1\sigma$ criterion is a looser condition than the 0.5$\sigma$ constraint we will later use for defining acceptability; we use this looser condition here as we are just trying to compare the basic behavior of the different PCA methods here. Once we are comparing the best-performing PCA methods with HMCode, we will consistently impose a 0.5$\sigma$ constraint on both.} for the cosmological parameters presented here. For the Illustris simulation, due to its strong baryonic feedback, we need to remove up to 6 PCs for the $1\sigma$ posteriors to include the fiducial cosmological parameters.

\begin{figure*}
\begin{center}
\includegraphics[width=1\textwidth]{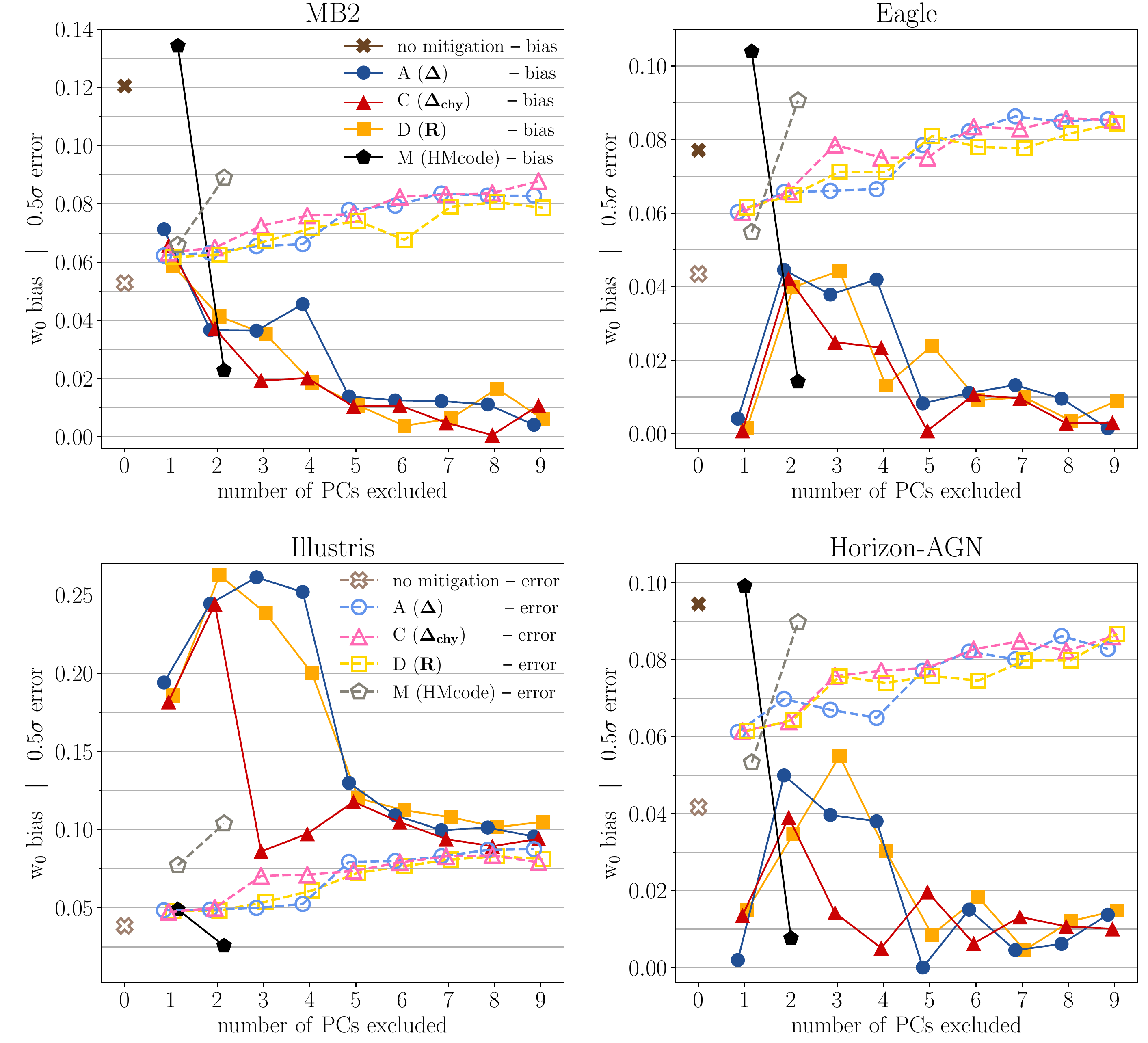}
\caption{The $w_0$ bias and statistical uncertainty under various baryonic physics mitigation techniques listed in Table~\ref{tb:baryon_methods}. The darker colored-filled markers indicate the level of $w_0$ bias, defined as $|w_{\rm 0, best\ fit} - w_{\rm 0, fid}|$. The fainter unfilled markers indicate the $0.5\sigma$ statistical uncertainty, with $1\sigma$ defined as the half difference between the 16th and 84th quantiles of the marginalized 1D $w_0$ posterior distribution. 
We adopt a criterion of residual bias $< 0.5\sigma$ error in this work when determining how many PC modes are required to mitigate biases due to baryonic physics.
\textit{The four main lessons from this plot are that: i) Of various PCA methods, at fixed number of excluded PC modes, the biases of method C are nearly always smaller than methods A and D, indicating method C is the most efficient PCA method. ii) For MB2/Eagle/Horizon-AGN simulations, removing $\geq 2$ PC modes is enough to mitigate baryonic physics-induced bias to 0.5$\sigma$. For the Illustris simulation, all PCA methods fail to pass the bias < 0.5$\sigma$ criteria even after 9 PC modes are removed. iii) No matter which PCA method (A, C or D) is applied, after removing $\geq 6$ PC modes, the statistical errors on $w_0$ converge to similar values. iv) \textsc{HMcode} works particularity well for the Illustris simulation. For MB2/Eagle/Horizon-AGN, marginalizing over both $A$ and $\eta_0$ is required to safely mitigate baryons to 0.5$\sigma$.} 
}
\label{fig:w0_bias_err}
\end{center}
\end{figure*}

\subsubsection{Method C is superior to methods A and D}
\label{sec:C_better_AD}

In Fig.~\ref{fig:w0_bias_err} we plot the $w_0$ bias (in color-filled markers; defined as $|w_{\rm 0, best\ fit} - w_{\rm 0, fid}|$, with $w_{\rm 0, best\ fit}$ being the median value of the marginalized posterior distribution of $w_0$) and the 0.5$\sigma$ error of $w_0$  (in open markers; with $\sigma$ defined as the half difference between the 16th and 84th percentile of the 1D marginalized posterior of $w_0$) for various baryonic physics mitigation schemes. For a method to be effective in mitigating baryonic-induced parameter biases, we require that the bias be below the $0.5\sigma$ errors. 
For all baryonic physics scenarios, we observe that at fixed number of excluded PC modes, the biases of method C (red-solid triangles) are nearly always smaller than methods A (blue-solid circles) and D (yellow-solid squares). If focusing on the lower left panel of Fig.~\ref{fig:posterior4_cp}, using Illustris when removing 3 PC modes as an examples, one can see that the 2D $1\sigma$ posteriors of method C (red dot-dashed curves) enclose the fiducial cosmology, while the posteriors of method A (light blue solid curves) are several $\sigma$ away.
Based on these, we conclude that PCs build from $\bm \Delta_{\rm chy}$ are potentially more effective than others to mitigate baryonic effects. 

To understand why method C performs better, we can go back to the $\chi^2$ equation when both $\D$ and $\M$ are set at $\pcofid$: 
\begin{equation} \label{eq:chy2_bary}
\begin{aligned} 
\mean{\chi_{\rm bary}^2} + \mean{\rm{Noise}}  &= \mean{[\D-\M(\pcofid)]^{\t}\ \C^{-1}\ [\D-\M(\pcofid)]}  \\
           &= \mean{[\D-\M]^{\t}\  {\mathbf L^{-1}}^{\t} \invL \ [\D-\M]} \\
           &= \mean{[\D_{\rm ch}-\M_{\rm ch}]^{\t} \ \mathbbm{1}\ [\D_{\rm ch}-\M_{\rm ch}]} \ .
\end{aligned}
\end{equation}
$\mean{\chi_{\rm bary}^2}$ quantifies the amount of $\chi^2$ caused by baryonic uncertainties. The noise term in our likelihood simulation is zero by construction.
Our goal is to reduce $\mean{\chi_{\rm bary}^2}$ to avoid bias in cosmological parameters due to baryonic physics. 
From Eq.~\eqref{eq:chy2_bary}, one can see that 
when doing PC mode exclusion in $\D_{\rm ch}-\M_{\rm ch}$ (with PCs constructed in $\bm \Delta_{\rm chy}$), there is a direct connection in reducing $\mean{\chi_{\rm bary}^2}$, while when doing PC mode exclusion in $\D-\M$ (with PCs constructed in $\bm \Delta$), the covariance matrix in between makes the reduction of baryonic uncertainties less direct.


\subsubsection{Error bars converge for all PCA methods}
\label{sec:error_converge}
Going back to Fig.~\ref{fig:w0_bias_err}, and focusing on the trend in the 0.5$\sigma$ error bars of $w_0$ shown in open fainter-colored markers. Generally, error bars grow as more PC modes are excluded (see also Fig.~\ref{fig:posterior4_cp} for the growth of error ellipse on 2D posteriors). The size of the error bars varies among the different PCA methods when fewer PC modes are excluded, but eventually converge/saturate to similar error bar sizes when excluding $\gtrsim$ 6 PC modes, independent of how PCs are constructed. This means that the PCs fully absorb the range of matter power spectrum modifications due to baryonic physics across the nine OWLS simulation, characterizing them using 6 dominant degrees of freedom; the last 3 PC modes are subjected to very small singular values ($\sigma$ as depicted in Fig.~\ref{fig:SVD}) such that only a tiny amount of baryonic fluctuation would be projected on them. 
In principle, including more training samples with different features would enrich the PC pool, increasing the number of effective degrees of freedom to characterize other possible baryonic scenarios. 

\subsection{PCA framework versus \textsc{HMcode}}
\label{sec:PCA_vs_HMcode}

We now move to a more detailed comparison of the two main ways to marginalize over baryonic uncertainties, namely the PCA-based methods and the halo-model based approach.
Since we already compared in \S\ref{sec:cp_PCAs} that of all PCA methods listed in Table~\ref{tb:baryon_methods}, method C is more efficient than the other two in mitigating biases in cosmological parameters due to baryonic physics. In the following, we will use method C as a representative for PCA-based methods, and compare it with \textsc{HMcode} (method M).


\subsubsection{Comparison on the effectiveness (criterion: bias $< 0.5\sigma$)}
\label{sec:pass2000}

We begin by discussing the performance of \textsc{HMcode} when using only one ($A$) vs.\ two (both $A$ and $\eta_0$) parameters to marginalize over baryonic physics. When only the parameter $A$ is used, \textsc{HMcode} sets the $\eta_0$ value via Eq.~\eqref{eq:A_eta0}. 
Going back to Fig.~\ref{fig:w0_bias_err}, with $w_0$ as an example, the pentagons at an x-axis value of 1 indicate the bias (black-solid) and $0.5\sigma$ error (gray-open) of only varying $A$ in \textsc{HMcode}. Similarly, the pentagons at an x-axis value of 2 indicate the option for \textsc{HMcode} varying both $A$ and $\eta_0$. 
For the Illustris simulation, both options can successfully mitigate the baryonic bias on $w_0$ to within our 0.5$\sigma$ criterion.  
However, apart from Illustris, for the baryonic scenarios of MB2, Eagle, and Horizon-AGN, we find that varying only $A$ while setting $\eta_0$ following Eq.~\eqref{eq:A_eta0} is not sufficient to mitigate baryonic bias. 
This implies that the current empirical relation described in Eq.~\eqref{eq:A_eta0} may not be precise enough for MB2/Eagle/Horizon-AGN-like data vectors with LSST-like statistical power. We therefore recommend that the extra freedom carried by $\eta_0$ is needed for upcoming weak lensing surveys to effectively mitigate the impact of baryonic physics on cosmological weak lensing measurements.
This is even more true in light of recent findings that indicate that our Universe is not like Illustris, for which the AGN feedback is known to be too strong such that the baryon fractions in massive halos are too low compared with observations \citep{Haider16}.
In Fig.~\ref{fig:OmS8Wa_bias_err} of Appendix~\ref{sec:appendixA4}, we also provide similar bias and error plots for other cosmological parameters: $\Omega_m$, $\sigma_8$, and $w_a$. The same conclusion holds for \textsc{HMcode} on these cosmological parameters (as shown in the filled and open pentagons), except for the $w_a$ constraint for Illustris (Fig.~\ref{fig:OmS8Wa_bias_err}l), where varying only $A$ is not enough to mitigate $w_a$ bias to within 0.5$\sigma$.

For the PCA-based method C, as indicated in red triangles of Fig.~\ref{fig:w0_bias_err} for $w_0$ and Fig.~\ref{fig:OmS8Wa_bias_err} for $\Omega_m$, $\sigma_8$, and $w_a$), we find that removing $\geq 3$ PC modes is sufficient to mitigate baryonic uncertainties to within 0.5$\sigma$ for all cosmological parameters considered here, if our Universe has a baryonic physics scenarios like MB2/Eagle/Horizon-AGN.

For the case of the Illustris simulation, we find that the PCA method fails to mitigate baryonic biases to within 0.5$\sigma$ for $w_0$ and $\Omega_m$ (Fig.~\ref{fig:OmS8Wa_bias_err}d), even after 9 PC modes are removed, but just passes the threshold for $\sigma_8$ (Fig.~\ref{fig:OmS8Wa_bias_err}h) and $w_a$ (Fig.~\ref{fig:OmS8Wa_bias_err}l) after removing 7 PC modes. We note that this is likely not a major concern as the baryonic effects of Illustris are unrealistically large, and the next generation IllustrisTNG hydrodynamical simulation \citep{Pillepich18, Springel18} will address the defects of the old version.

We provide a summary of the results from the above discussion in Table~\ref{tb:pass2000}. In Appendix~\ref{sec:appendixA5}, we further provide the $\chi^2$ values computed at the best-fitted cosmological parameters from various baryon mitigation models.

\begin{table}
\caption{Summary of the effectiveness of baryonic physics mitigation methods in reducing biases to within 0.5$\sigma$ for various cosmological parameters under different baryonic scenarios.  A cosmological parameter is struck out if a mitigation method fails to pass our criterion of bias < 0.5$\sigma$, where $\sigma$ represents the marginalized statistical error (see \S\ref{sec:pass2000} for detail).}
\begin{tabular}{lcc}
\hline 			 		  	&	$\!\!\!\!\!\!\!\!\!\!\!\!$MB2/Eagle/Horizon-AGN$\!\!\!\!\!$ & Illustris	\\  \hline
\textsc{HMcode} ($A$) 			&	all fail	 & $\Omegam$ $\sigma_8$ $w_0$ \cancel{$w_{\rm a}$}  \\
\textsc{HMcode} ($A$, $\eta_0$) 	&	all pass & all pass \\ \hline
PCA (trained by 9 sims)         		& 	all pass &	\cancel{$\Omegam$} $\sigma_8$ \cancel{$w_0$} $w_{\rm a}$ \\ \hline
\label{tb:pass2000}
\end{tabular}
\end{table}

\subsubsection{Comparison on the level of degradation on cosmology}
\label{sec:cp_error}

We now compare the error bars on cosmological parameter constraints between PCA method C and \textsc{HMcode}, on baryonic scenarios of MB2/Eagle/Horizon-AGN, where both methods successfully mitigate the baryonic biases to within $0.5\sigma$. 
The pink open triangles in Fig.~\ref{fig:w0_bias_err} indicate the 0.5$\sigma$ error of $w_0$ under method C, and the gray open pentagons indicate the same for \textsc{HMcode}. Besides $w_0$, other cosmological parameters are also shown in Fig.~\ref{fig:OmS8Wa_bias_err}. 
As discussed in \S\ref{sec:error_converge}, one nice feature of the PCA method is that the error bars converge to a certain level when excluding $\geq 6$ PC modes.
We find that the converged error bars for method C generally are smaller than those for \textsc{HMcode}, even though \textsc{HMcode} only utilizes 2 parameters to marginalize over baryonic physics uncertainties while the PCA method needs 3 parameters to mitigate baryonic effects to 0.5$\sigma$ in the case of MB2/Eagle/Horizon-AGN.
A similar result can be seen from the $1\sigma$ 2D posteriors shown in Fig.~\ref{fig:posterior4_cp}.

\subsubsection{Baryonic feature constraint from \textsc{HMcode}}

\begin{figure}
\begin{center}
\includegraphics[width=0.48\textwidth]{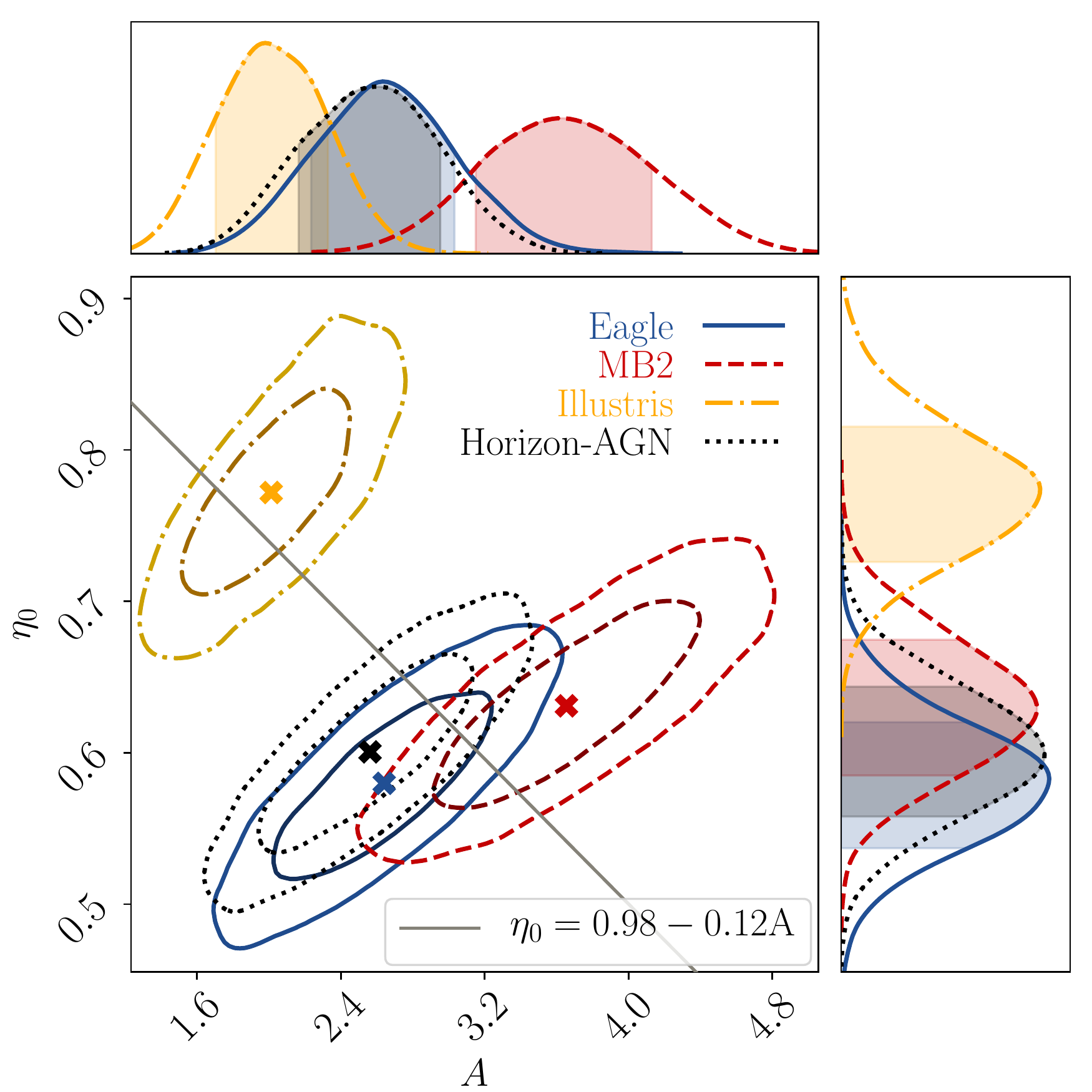}
\caption{The 2D constraints on the \textsc{HMcode} halo structure parameters $A$ and $\eta_0$ from our simulated likelihood analysis for baryonic scenarios of Eagle (blue/solid), MB2 (red/dotted), Illustris (green/dot-dashed), and Horizon-AGN (yellow/dashed). The black line plots the relationship between $A$ and $\eta_0$ that is used to provide single-parameter fit in \textsc{HMcode}. 
Both 68\% and 95\% confidence levels are shown.}
\label{fig:A_eta0}
\end{center}
\end{figure}

In Fig.~\ref{fig:A_eta0}, we plot the 2D posterior distributions on $A$ and $\eta_0$ for various baryonic scenarios in colored contours, along with Eq.~\eqref{eq:A_eta0} shown in the black line. We can see that although relying on this $A$-$\eta_0$ relationship is not effective enough to mitigate baryonic bias in most of baryonic recipes under LSST-like survey, the suggested relationship is still good enough to pass the 68\% contours in all cases. Therefore, instead of following a fixed relationship like Eq.~\eqref{eq:A_eta0} or allowing both $A$ and $\eta_0$ to vary unboundedly, setting an $A$-dependent prior on $\eta_0$ may help recover some cosmological constraining power while still reducing biases in cosmological parameters when using \textsc{HMcode}. 

\begin{figure*}
\begin{center}
\includegraphics[width=0.98\textwidth]{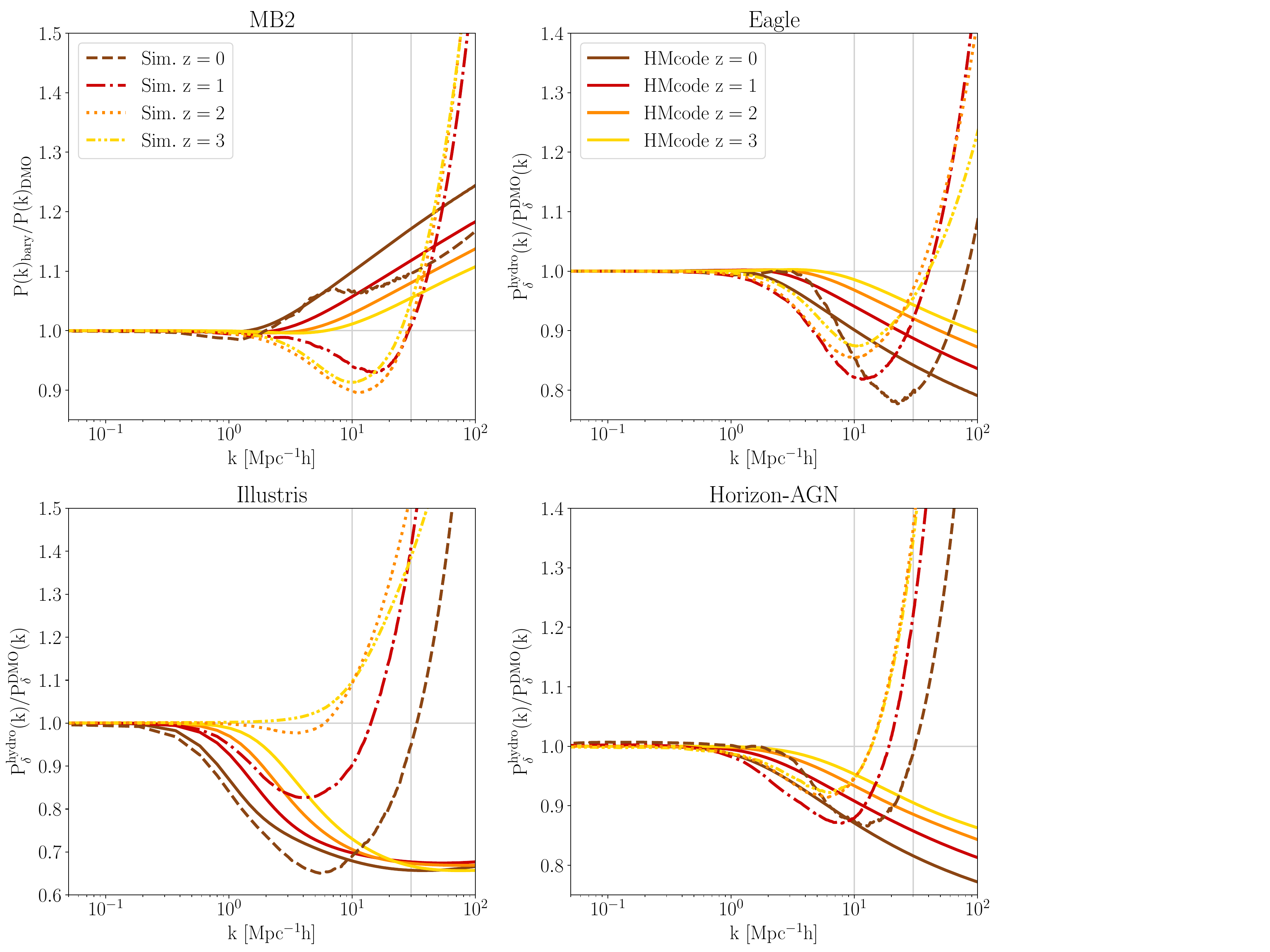}
\caption{Comparisons of power spectra generated from \textsc{HMcode} at the best-fitted $A$ and $\eta_0$ values (solid lines) and power spectra directly derived from hydrodynamical simulations (dotted or dashed lines) at $z=$ 0, 1 , 2, 3.
\textit{The discrepancy indicates that \textsc{HMcode} lacks degrees of freedom to account for the cooling effect at high $k$, and that it is too simplified to capture the complex redshift evolution patterns present in hydrodynamical simulations.}}
\label{fig:cp_Pk_HMvsSim}
\end{center}
\end{figure*}


In Fig.~\ref{fig:cp_Pk_HMvsSim}, we compare the power spectra generated from \textsc{HMcode} at the best-fitted values of $A, \eta_0$ (cross symbols in Fig.~\ref{fig:A_eta0}) to the original power spectra derived directly from the hydrodynamical simulations at redshift $z = 0, 1, 2, 3$.
We note that this is not a fair comparison because the underlying $P_{\delta}(k, z)$ is not constrainable from the projected tomographic power spectra, unless the tomographic bins are fine enough to recover the full 3D information.\footnote{If using \textsc{HMcode} to directly fit the 3D matter power spectrum including baryonic effects at some specific redshift, according to \citetalias{Mead15}, by adjusting $A(z)$, and $\eta_0(z)$, \textsc{HMcode} has enough degrees of freedom to match the baryonic power spectra from the OWLS simulations to $k \gtrsim 10 \hMpc$.} 
Here we simply use these plots to understand the effects of \textsc{HMcode} parameters on $P_{\delta}(k, z)$.
Firstly, \textsc{HMcode} does not have degrees of freedom for the cooling feature of hydro simulations, which leads to a turn-over in the power spectrum ratio at $k \gtrsim 10 \hMpc$. This is expected as according to \citetalias{Mead15}, the halo-model power is accurate to $\approx$ 5\% only for $k \leq 10 \hMpc$ and $z \leq 2$. Because of this limitation, \textsc{HMcode} tends to produce a shallower suppression of power for a given $k$ (when $k \leq 10 \hMpc$) compared with MB2/Eagle/Horizon-AGN, in order to compensate for the lack of cooling prescription at $k > 10 \hMpc$.  
Secondly, the redshift evolution of \textsc{HMcode} power spectra is too monotonic, lacking freedom to capture the complicated evolutionary pattern that generally exists in hydrodynamical simulations. The redshift evolution patterns can be very different for various baryonic scenarios. 
\textsc{HMcode}'s inability to model redshift evolution may be due to the fact that only two nuisance parameters are involved in describing 
the complex scale and redshift dependences of baryonic effects seen in the simulations.  
One straightforward suggestions is to add redshift dependence to $A$ and $\eta_0$. Further development of halo model approaches to account for the modification of the matter power spectrum for $k > 10 \hMpc$ is also needed.
Although the current model does not describe all the complexity of possible modifications of $P(k)$ due to baryonic physics, we can still use \textsc{HMcode} to gain insight into the strength of feedback from the constrained values of $A$ and $\eta_0$. As shown in Fig.~\ref{fig:A_eta0}, the Illustris-like universe tends to have small $A$ and large $\eta_0$.

\subsection{Pushing to even smaller angular scales: $\ell_{\rm max}$ of 5000}
\label{sec:ell_5000}

\begin{table}
\caption{Similar to Table~\ref{tb:pass2000}, but now for the likelihood simulations with mock observables pushing to $\ell_{\rm max} \approx 5000$.}
\begin{tabular}{lccc}
\hline 			 		  	&	$\!\!\!\!$MB2/Eagle$\!\!\!\!$ 						& $\!\!\!\!$Horizon-AGN 					& Illustris	\\  \hline
\textsc{HMcode} ($A$) 			&	all fail	& all fail & all fail  \\
\textsc{HMcode} ($A$, $\eta_0$) 	&	all pass & $\!\!\!\!$$\Omegam$ $\sigma_8$ $w_0$ \cancel{$w_a$} & \cancel{$\Omegam$} \cancel{$\sigma_8$} $w_0$ $w_a$ \\ \hline
PCA (trained by 9 sims)$\!\!\!\!$         		& 	all pass &	all pass & all fail \\ 
PCA (trained by 12 sims)$\!\!\!\!$ 	&	& 	& all fail \\ \hline
\label{tb:pass5000}
\end{tabular}
\end{table}

Until now, all elements of our analysis have been based on mock tomographic shear data vectors with $\ell_{\rm max} \approx 2000$, which is a conservative choice under the limitation that we lack accurate power spectra at $k > 30 \hMpc$. The $\ell_{\rm max} \approx 2000$ cut assures that various extrapolation curves on $P_{\delta}(k)$ ratio out to $k > 30 \hMpc$ would not cause significant change on the resulting $C^{ij}(\ell)$ data vector (see Appendix~\ref{sec:appendixA2} for details on how the scale cut limit is determined). 

To further test the limits of the proposed baryonic mitigation techniques, we generate mock $C^{ij}(\ell)$ data vectors with $\ell_{\rm max} \approx 5000$ (based on the quadratic  extrapolation trends derived by fitting the $P_{\delta}(k)$ ratio in $k \in [10, 30] \hMpc$, see the red curve in Fig.~\ref{fig:Pkratio_extra} as a demonstration), and then perform the same simulated likelihood analyses with mitigation techniques described in \S\ref{sec:likelihood_sim} and \S\ref{sec:methods}. The only difference is that we append 3 extra data points that with equal logarithmic spacing in $\ell \in [2060, 5000]$ to the original data vector $\D$ in each tomographic bin. The new length of $\D$ is thus extended to $55 \times (18+3) = 1155$ data points (see \S\ref{sec:data_vector} for the original format of $\D$). The covariance matrix is also updated accordingly.

\begin{figure}
\includegraphics[width=0.5\textwidth]{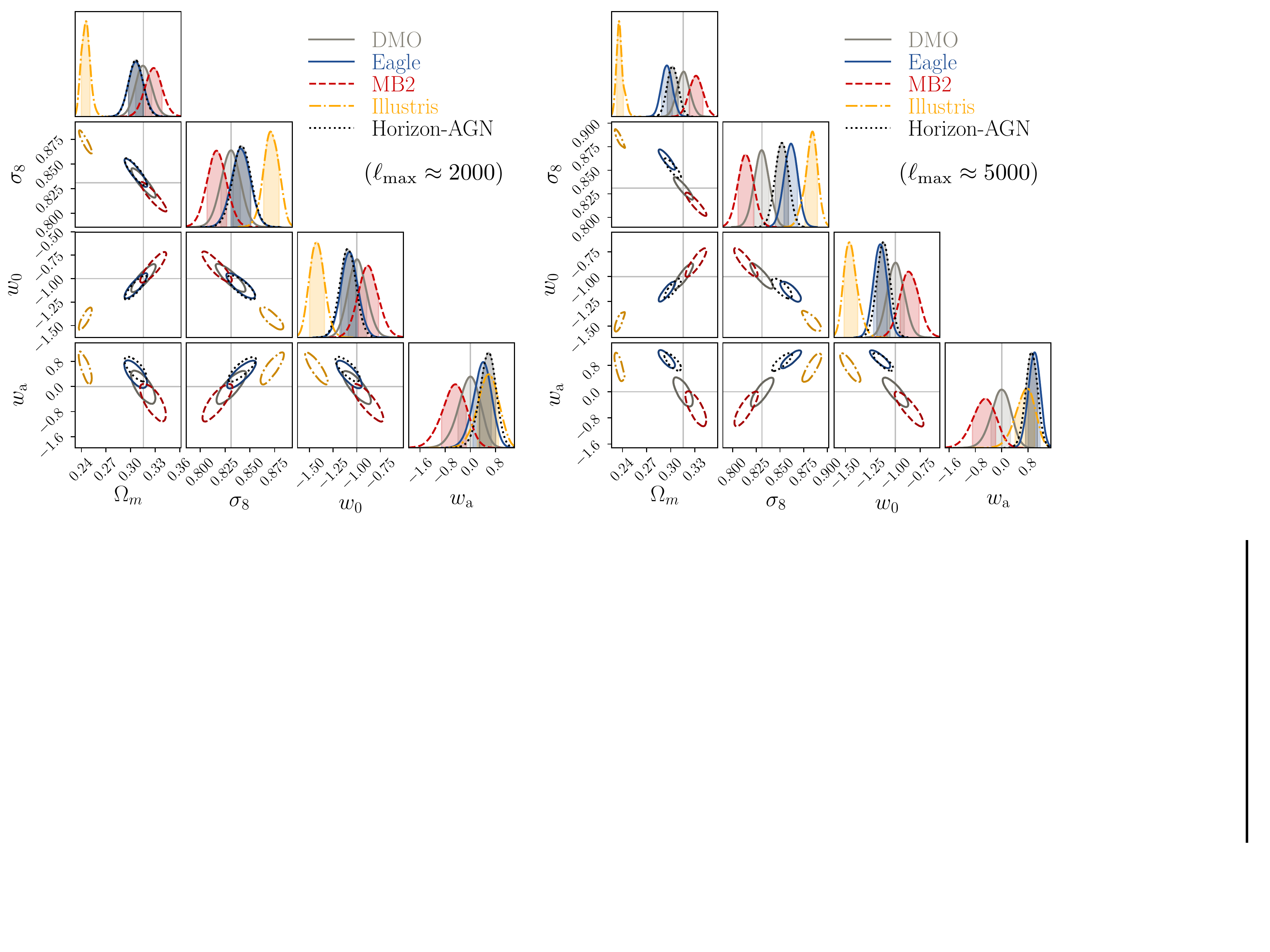}
\caption{Similar to Fig.~\ref{fig:posterior_no_trick}, but for the cases when pushing our mock observables toward $\ell_{\rm max} \approx 5000$.}
\label{fig:posterior_no_trick5000}
\end{figure}

The dark gray contours in Fig.~\ref{fig:posterior_no_trick5000} indicate the 2D posterior distributions of the cosmological parameters, when no baryonic physics mitigation technique is applied. Compared with the similar plot shown in Fig.~\ref{fig:posterior_no_trick}, but for $\ell_{\rm max} \approx 2000$, the biases on cosmological parameters increases to $2\sigma \sim 19 \sigma$ for the various cosmological parameters in an Illustris-like universe, and around $1.5\sigma \sim 6 \sigma$ for the other cases.
This amount of bias is consistent with Fig.~5 of \citetalias{Eifler15}, who showed the posterior distributions for $\ell_\text{max} \sim 5000$ for the OWLS baryonic physics scenarios for an LSST-like likelihood simulations.

\begin{figure*}
\begin{center}
\includegraphics[width=1\textwidth]{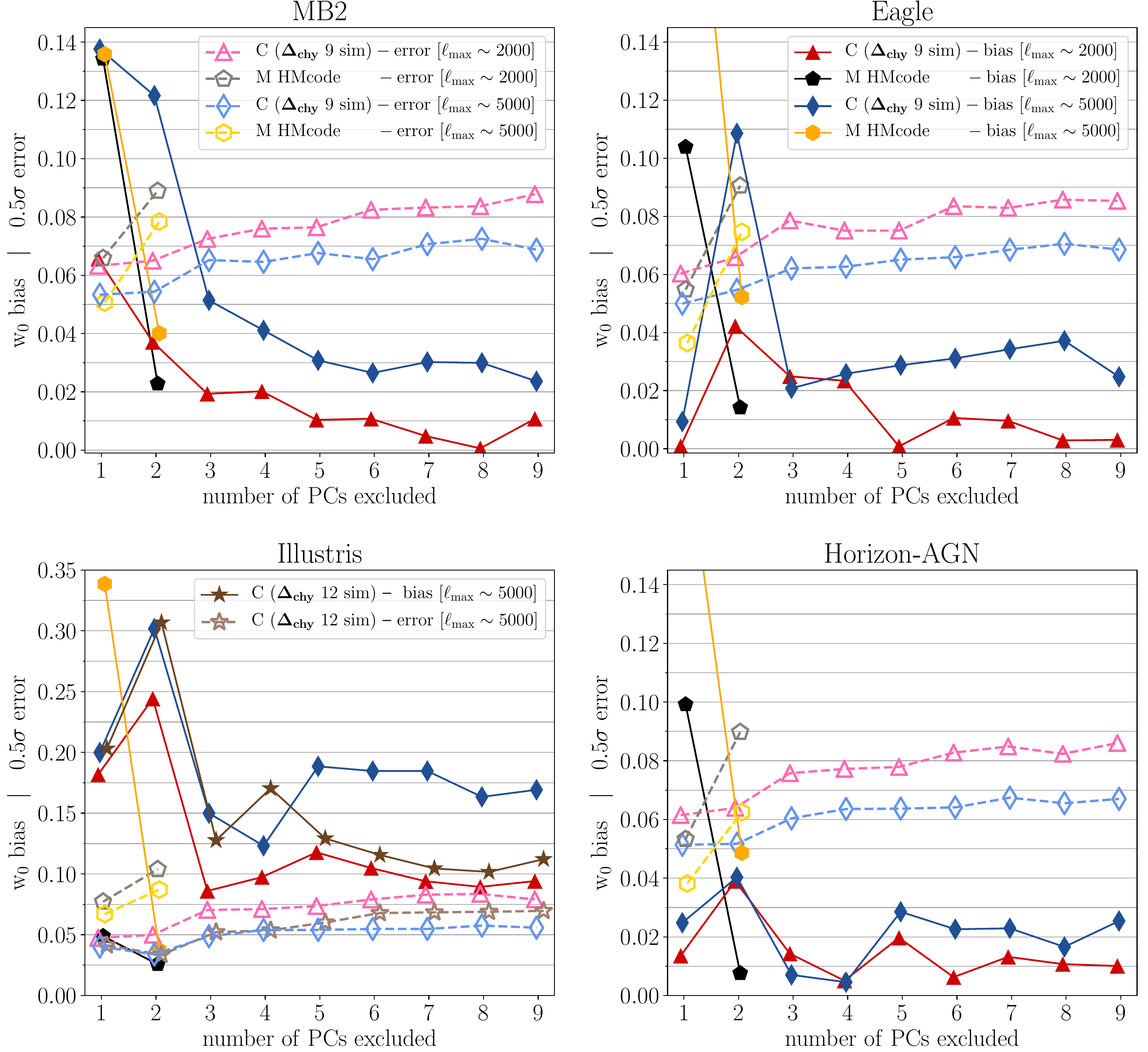}
\caption{The $w_0$ bias and uncertainty for the baryonic physics mitigation methods C (as our representative for PCA-based method) and M (halo model-based method/\textsc{HMcode}) (see Table~\ref{tb:baryon_methods} for details). The darker colored-filled markers indicate the level of $w_0$ bias, defined as $|w_{\rm 0, best\ fit} - w_{\rm 0, fid}|$.
The fainter colored-open markers indicate the 0.5$\sigma$ $w_0$ uncertainty, with 1$\sigma$ defined as the half difference between the 16th and 84th percentile of the marginalized 1D $w_0$ posterior distribution. 
\textit{The four key results on this plot are:  
i) the PCA method (blue diamonds) mitigates bias to within 0.5$\sigma$ for the milder baryonic physics scenarios -- MB2, Eagle, and Horizon-AGN -- after excluding more than 3 PC modes, but fails for the Illustris scenario.
ii) When marginalizing over both $A$ and $\eta_0$, \textsc{HMcode} (yellow hexagons) mitigates $w_0$ to within 0.5$\sigma$ for all baryonic scenarios.
iii) For baryonic scenarios of MB2 and Eagle, the cases for which both methods work, the error bars for the PCA method (light blue open diamonds) converge to smaller values compared with \textsc{HMcode} (yellow open hexagons).
iv) Including more small-scale data in the analysis reduces the statistical error to $\sim 20\%$ for PCA method (light blue open diamonds v.s. pink open triangles) and to about $12\%\sim30\%$ for \textsc{HMcode} (yellow open hexagons v.s. gray open pentagons).
v) Including more training simulations in PCA improves reducing the $w_0$ bias induced by neglecting baryonic effects to $\sim 15\%$ for the case of Illustris (brown solid stars v.s. blue solid diamonds), although the improvement on residual bias is not reaching our criterion of < 0.5$\sigma$.}
}
\label{fig:w0_bias_err_ell5000}
\end{center}
\end{figure*}

Since we showed in \S\ref{sec:cp_PCAs} that method C is the most efficient of the PCA-based methods, we only run simulated likelihood analyses with PCA-based method C, compared with method M using \textsc{HMcode} for $\ell_{\rm max} \approx 5000$. 
In Fig.~\ref{fig:w0_bias_err_ell5000}, we plot the marginalized $w_0$ bias (color-filled symbols) and 0.5$\sigma$ $w_0$ uncertainty (open symbols) as a function of the number of excluded PC modes in method C (blue diamonds) and \textsc{HMcode} (yellow hexagons). (The red triangles and black pentagons are simply copies of the data points shown in Fig.~\ref{fig:w0_bias_err}, to enable easier comparison of results with $\ell_{\rm max}$ of 2000 versus 5000.) The bias and error plots for $\Omegam$, $\sigma_8$ and $w_a$ are also provided in Fig.~\ref{fig:OmS8Wa_bias_err}.

Similar to \S\ref{sec:PCA_vs_HMcode}, we rely on the bias < $0.5\sigma$ criterion to validate the effectiveness of baryonic physics mitigation methods, with the results summarized in Table~\ref{tb:pass5000}. First of all, for \textsc{HMcode}, varying only $A$ is not sufficient to mitigate the bias to within $0.5\sigma$ for the Illustris simulation, which \textsc{HMcode} is particularly good at describing. Both $A$ and $\eta_0$ must be varied to meet our criterion for MB2 and Eagle. For Horizon-AGN and Illustris, \textsc{HMcode} works well for some cosmological parameters, while it fails for the others. For the PCA method, it still works for baryonic scenarios of MB2/Eagle/Horizon-AGN when pushing to $\ell_{\rm max} \approx 5000$, but continues to fail to meet our criterion for the Illustris scenario.

In terms of degradation on cosmological parameter constraints after marginalization, for the cases of MB2 and Eagle, the scenarios in which both PCA and \textsc{HMcode} succeed in mitigating the bias to within $0.5\sigma$, we see that PCA method yields smaller converged error bars (light blue open diamonds) compared with \textsc{HMcode} using 2 parameters (yellow open hexagons) to do marginalization.

Does extending the data vectors to $\ell_{\rm max} \approx 5000$ help to better constrain cosmological parameters compared with $\ell_{\rm max} \approx 2000$? 
As shown in Fig.~\ref{fig:w0_bias_err}, for the PCA method, we observe that the converged $w_0$ errors for the $\ell_{\rm max} \approx 5000$ cases (light blue open diamonds) are smaller by $\sim 20\%$ compared with the errors for $\ell_{\rm max} \approx 2000$ (pink open triangles). For the cases of \textsc{HMcode} when varying both $A$ and $\eta_0$, the $w_0$ errors reduce by $\sim 12\%$ for MB2, $\sim 18\%$ for Eagle, and $\sim 30\%$ for Horizon-AGN, after extending data points to  $\ell_{\rm max} \approx 5000$ (yellow open hexagons) from $\ell_{\rm max} \approx 2000$ (gray open pentagons). This means that we do benefit from additional constraining power when including more small-scale data in the analysis, if the baryonic physics effect in our Universe is near the physics implemented in Eagle/MB2/Horizon-AGN.

\subsection{Including more AGN prescriptions in the training set}
\label{subsec:training_set}

The reason why the PCA method fails to mitigate the impact of baryonic physics on the matter power spectrum in Illustris is that the PCs built from the current training set do not capture the strong variation with $k$ to explain its intense feedback feature. As also discussed in \citet{Mohammed18}, it is better to have a training set that comprises adequately exotic but reasonable models.
Of the nine training OWLS simulations, only the OWLS-AGN contains an AGN feedback prescription, and we rely on this single AGN model to explain Illustris.
However, this shortcoming can be fixed by incorporating more training simulations into the PCA, so that the resulting PCs will include more degrees of freedom to explain the broader range of outcomes due to baryonic physics. 

We try to address the above Illustris problem by including the baryonic scenarios of MB2/Eagle/Horizon-AGN in our training set, and then build a ${\bm \Delta_{\rm ch}}$ matrix with 12 columns to extend the capability of the derived PCs. 
In the bottom left panel of Fig.~\ref{fig:w0_bias_err_ell5000}, we plot the marginalized $w_0$ bias (brown filled stars) and error (light brown open stars) for Illustris simulation with PCs trained from the 12 baryonic scenarios, and a scale cut at $\ell_{\rm max} \approx 5000$. (The results for other cosmological parameters can be found in the last column of Fig.~\ref{fig:OmS8Wa_bias_err} as well.)
With this expanded training set, the PCA method now reduces the $w_0$ bias from 1.5$\sigma$ (blue filled diamonds) to 0.8$\sigma$ (brown filled stars), which is an improvement but still does not enable us to meet our criterion of bias $< 0.5\sigma$.  

The error bars when using 12 simulations in the training set converge after removing $\geq 6$ PC modes.  Notice that the converged errors become bigger when the PCs are trained from 12 simulations rather than just the 9 OWLS simulations. By including more simulations to construct the PCs, we also enlarge the range of baryonic uncertainties, which is a trade-off to ensure a more effective removal of biases due to baryonic physics for a broad range of baryonic physics scenarios. 
However, we can also imagine trying to rely on external information from independent observations 
to rule out baryonic scenarios that fail to describe our Universe. 
By carefully controlling the uncertainty range of the training set, we could potentially improve the cosmological constraining power after mitigation.
 


\section{Summary and Discussion}
\label{sec:summary}

We have explored the two major approaches to mitigate uncertainties in cosmic shear tomographic power spectra due to baryonic physics, with the goal of understanding their performance on cosmological constraints for the upcoming LSST survey. 
The first approach is the PCA-based analysis proposed by \citetalias{Eifler15}. 
Based on a set of training hydrodynamical simulations with various baryonic prescriptions (spanned by OWLS in this work), a difference matrix (Eq.~\eqref{eq:DiffMatrix}) is computed. Its columns are filled with difference vectors between these hydro and DMO simulations. PCA is then performed on the difference matrix to find dominant PC modes that can be used to model baryonic effects in other hydro simulations. 
The second approach is the halo model-based method coded in the package of \textsc{HMcode} by \citetalias{Mead15}, which utilizes two halo structural parameters ($A$ and $\eta_0$) related to the halo concentration-mass relation to marginalize over baryonic uncertainties. 

We examine the basics of the PCA formalism and provide a modification to properly account for the change of covariance matrix after removal of PC modes. Under the new formalism, we demonstrate that PC mode removal is equivalent to marginalization over PC amplitudes (see \S\ref{sec:PCex_vs_PCmar}). 
Instead of difference matrices, we also investigate PCA on other kinds of matrix forms with their columns filled with the fractional difference  (Eq.~\eqref{eq:RatioMatrix}) or noise-weighted difference vectors (Eq.~\eqref{eq:DiffMatrix_chy}) to quantify deviations in the matter power spectrum due to baryonic physics. 
The derived PC bases from different matrices vary in their efficiency in explaining baryon fluctuations at different angular scales. 
Difference matrix PCs can more effectively account for large scale baryonic fluctuations, fractional difference matrix PCs are more effective at describing the small scale fluctuations, and noise-weighted difference matrix PCs most effectively describe the scales at which the S/N is maximal. 
We find that performing PCA on the noise-weighted difference matrix, with the weighting factor derived via performing Cholesky decomposition on the covariance matrix (\S\ref{sec:Cholesky}), is the most efficient way to mitigate the impact of baryonic physics on inferred cosmological parameters (\S\ref{sec:cp_PCAs}). Therefore, for future application on real data, we recommend  applying the noise-weighted PCA technique.
It should be noted that except for method D, the current PCs are wiggling slightly in their directions at each MCMC step, when cosmology changes. 
If we would like to quantify baryon physics via PC amplitudes, we hope the constrained PC amplitudes are subjected to a fixed set of PCs. A more complete design of PCA algorithm therefore would be an iteration process. We will first use the current setting to find the best-fitted cosmology, and once the $\bm p_{\rm co, best\ fit}$ is determined, we will fix PC basis at $\bm p_{\rm co, best\ fit}$, and constrain the posteriors of PC amplitudes subjected to this PC set.

We apply both the PCA and \textsc{HMcode} techniques on mock shear tomographic data vectors ($C^{ij}(\ell)$) with baryonic physics scenarios of MB2/Eagle/Illustris/Horizon-AGN.  
We test whether these mitigation techniques can reduce the bias in cosmological parameters induced by neglecting baryonic effects to within 0.5$\sigma$.
With a scale cut at $\ell_{\rm max} \approx 2000$, and for milder baryonic physics scenarios like MB2/Eagle/Horizon-AGN, both methods succeed in mitigating the impact of baryonic effects on the inferred cosmological parameters. For the PCA method, we find that excluding 3 PC modes is sufficient to mitigate the bias to within 0.5$\sigma$ for $\Omegam, \sigma_8, w_0$ and $w_{\rm a}$.
For \textsc{HMcode}, we find that it is safer to vary both $A$ and $\eta_0$ when performing marginalization, rather than varying only one of them and having the other follow the suggested relation in \citetalias{Mead15}, at least at the level of LSST statistical power.
For the Illustris scenario, only \textsc{HMcode} is sufficient to mitigate the bias to within 0.5$\sigma$. The PCA method fails to pass our criterion even after removing 9 PC modes. 
With a more aggressive $\ell_{\rm max}$ of 5000, the PCA methods still work for MB2/Eagle/Horizon-AGN, but fail for Illustris. \textsc{HMcode} remains sufficient for MB2/Eagle after marginalizing over 2 parameters but only works partially on some of the cosmological parameters for Horizon-AGN and Illustris, as summarized in Table~\ref{tb:pass5000}.


We found that \textsc{HMcode} is most effective at mitigating the impact of baryonic physics for a strong feedback scenario like Illustris, because \textsc{HMcode} is designed to describe the impact of baryonic physics on the matter power spectrum for $k \leq 10 \invhMpc$, where the main feature is the suppression of power due to feedback (Fig.~\ref{fig:Pk_ratio}). 
\textsc{HMcode} and halo model-based approaches in general have the advantage over PCA that they have cosmology dependence built in.
Although the current version of \textsc{HMcode} lacks the complexity to fully describe various baryonic scenarios (Fig.~\ref{fig:cp_Pk_HMvsSim}), it provides a good summary of the level of feedback strength through two nuisance parameters (Fig.~\ref{fig:A_eta0}). Future improvements of the halo model to smaller scales of $k \geq 10 \invhMpc$, as well as adding parameters to allow additional freedom in the redshift evolution of baryonic physics effects, may constrain halo structural parameters and baryonic power spectra ratio curve together with cosmology. Exploring the prior ranges on halo model parameters also help to improve cosmological parameter constraint. For example, we can use the posterior constraints from realistic hydrodynamical simulations as shown in Fig.~\ref{fig:A_eta0} to narrow down the allowed ranges of $A$ and $\eta_0$. Joint constraints from galaxy-galaxy lensing together with cosmic shear may also provide additional information from the data itself on the halo structure parameters \citep{Zentner08}.

There are several advantages of the PCA method. 
Firstly, it successfully mitigates quite general baryonic fluctuations and complex redshift evolution patterns,
when collecting several representative training hydrodynamical simulations to conduct PCA.
The complex baryonic behaviors as well as the redshift evolution would then be naturally absorbed in only a few dominant PC modes, and we can use the amplitudes of these PC modes to perform marginalization (or, equivalently, PC mode exclusion).
Secondly, the PCA method efficiently accounts for baryonic uncertainties without losing too much cosmological constraining power. As discussed in \S\ref{sec:cp_error}, whenever both methods are successful in removing baryonic bias, the error bars are generally smaller for PCA methods compared with the errors of \textsc{HMcode}. 
It is quite important to note that even if we do not know in advance how many PC modes must be excluded to safely remove baryonic bias in our Universe, excluding all effective PC modes does not unacceptably increase the errors, which saturate at a certain limit.
The maximum number of effective PC modes one can remove is equal to the total number of training simulations used in the PCA (see \S\ref{subsec:PCA_removal} for detail). 
Finally, the PCA method has significant flexibility to make adjustments as our knowledge of baryonic physics improves. 
For example, in \S\ref{sec:ell_5000} we tried to improve the bias mitigation of the Illustris simulation by including more realistic baryonic scenarios with AGN prescriptions in our training set, which enriches the space of possible baryonic uncertainties that the PCs can describe.
After the inclusion of MB2/Eagle/Horizon-AGN as well as the original 9 OWLS scenarios in our training set, we can further decrease the residual cosmological parameter bias compared with the results when only using the 9 OWLS simulations as training set. The cost is that we lose some constraining power. The flexibility of the PCA framework makes it easy to adjust the model based on changes in our knowledge of baryonic physics, and allows us to regulate errors by controlling the input training simulations in the PCA. 

There are several aspects regarding the PCA framework that we do not explore within this work. 
Firstly, our training hydro simulations are all run under the flat $\Lambda$CDM model, and we assume that the baryonic fluctuations, as quantified in terms of power spectrum ratios between hydrodynamical and DMO simulations, remain fixed when cosmology changes. In reality, baryonic and cosmological
effects vary jointly. 
Currently, there is no easy way to investigate this assumption, but future fast hydrodynamical simulations under development would be an ideal tool to systematically study this issue. 
Secondly, we adopted a power law extrapolation scheme for $P_{\delta}(k)$ ratio at $k \geq 30 \invhMpc$ (see Appendix~\ref{sec:appendixA2}). 
The most relevant physics that governs the high $k$ behavior is the cooling and inner stellar density profile of galaxies. 
Current large-volume cosmological hydrodynamical simulations lack the resolution to resolve the physics of galaxy formation to galaxy centers. 
We rely on sub-grid models of feedback to avoid the overcooling of gas and to mitigate the differences between the observed and simulated galaxies, but discrepancies still exist \citep{Stinson10, Bottrell17, Furlong17}. This implies that $P_{\delta}(k)$ in the high $k$ regime is still highly uncertain.
Does the $P_{\delta}(k)$ ratio continue the trend of increasing monotonically? Or should it reach a saturation point at some high $k$ regime? 
How to properly propagate the uncertainties of the poorly understood small scale $P_{\delta}(k)$ ratio into the errors of integrated $C^{ij}(\ell)$ which in turn affects the derived PCs? 
These questions require higher resolution hydrodynamical simulations to further address. 
Finally, we have briefly demonstrated in \S\ref{subsec:training_set} that depending on the training simulation set, the derived PCs carry different abilities to mitigate baryonic effects, and differ in the final constraining power. It would be worthwhile to systematically investigate various possible combinations of the training simulations, to find a most effective set of PCs that are able to span a wide enough range of baryonic uncertainties but with less degradation on constraining power.

In future extensions of this work, we will apply the PCA framework to a configuration-space tomographic shear analysis on real data 
to constrain the baryonic feature of our Universe and compare it with hydrodynamical simulations. 
We aim to develop a consistent way of quantifying priors of PC amplitudes, which would provide us with more constraining power on cosmological parameters by shrinking the allowed range of baryonic physics modifications of the matter power spectrum. We will also develop a PCA tool for joint analysis of galaxy-galaxy lensing and galaxy clustering observables. The full 3$\times$2-point analysis then can be self-consistently analyzed within the PCA framework to increase the constraining power on cosmology while safely marginalizing over baryonic physics.

\section*{Acknowledgements}
We thank Alex Hall for reviewing this paper and providing useful suggestions to improve the manuscript. We thank Sukhdeep Singh, Fran\c{c}ois Lanusse, Qirong Zhu, Arya Farahi, Hy Trac, Phil Bull, Tiziana Di Matteo, Alexander Mead, Irshad Mohammed, Ananth Tenneti for many constructive discussions and feedback. RM and HH are supported by the Department of Energy Cosmic Frontier program, grant DE-SC0010118. 
Part of the research was carried out at the Jet Propulsion Laboratory, California Institute of Technology, under a contract with the National Aeronautics and Space Administration and is supported by NASA ROSES ATP 16-ATP16-0084 grant.



\bibliographystyle{mnras}
\bibliography{reference} 



\appendix

\section{Power Spectrum Computation}
\label{sec:appendixA1}

Here we describe the practical implementation of our power spectrum computation from the simulation snapshots.

\subsection{The power spectrum estimator}

The matter density field in the Universe can be quantified via the overdensity $\delta({\bf x})$, defined as $\delta({\bf x}) = \frac{\rho({\bf x}) - \bar{\rho}}{\bar{\rho}}$, 
where $\rho({\bf x})$ specifies the density function at position ${\bf x}$ and $\bar{\rho}$ is the global mean density. We first estimate $\delta({\bf x})$ on a uniform grid of 1024 cells across a side of the simulation box with the particle deposition step carried out via Nearest Grid Point (NGP) assignment. Our estimator is
\begin{equation}
\label{eq:deltax}
\hat{\delta}({\bf x})  = \delta({\bf x}) \ast W({\bf x}) = \int_{\rm V_{box}} d {\bf x'} \delta({\bf x'}) W({\bf x-x'}) \ .
\end{equation}
Here the mass assignment function can be described by $W({\bf x}) = \prod_{i} W(x_i)$, with
\begin{equation}
\label{eq:Wx}
 W(x_i)=\begin{cases} 
               1/ \Delta L  & \mbox{for}~~~~|x_i| < \Delta L /2\\
               0  &  \mbox{else}
               \end{cases} \ ,
\end{equation}
where the grid cell side length  $\Delta L = {\rm L_{box}}/1024$, and the index $i$ is the axies label of the Cartesian coordinate system.
We then perform a discrete Fourier transform on $\hat{\delta}({\bf x})$ to derive its Fourier transformation pair $\hat{\delta}({\bf k})$: 
\begin{equation}
\hat{\delta}({\bf k}) = \delta({\bf k}) W({\bf k}) \ .
\end{equation}
After the Fourier transform, the convolution operation in Eq.~\eqref{eq:deltax} becomes a simple product, 
with $W({\bf k})$ being the Fourier space mass assignment window function: 
\begin{equation}
\label{eq:Wk} 
 W({\bf k}) =  \prod_{i} W(k_i)= \prod_{i} \frac{\sin{( \frac{k_i \Delta L }{2} )}}{ (\frac{k_i \Delta L }{2})}  \ ,
\end{equation}
where $k_i$ ($i = x, y, z$) is the $i$-th component of $\bf k$.
The Fourier transformation pair of $\delta({\bf x})$ then can be computed by
\begin{equation}
\label{eq:deltak}
\delta({\bf k}) = \frac{\hat{\delta}({\bf k})} {W({\bf k})} \ . 
\end{equation} 
We choose the convention of Fourier transform as $\delta({\bf k}) = \int  d{\bf x}\ \delta({\bf x}) e^{{-i {\bf k} \cdot {\bf x}}}$. Under this convention, the power spectrum can be estimated by averaging over all modes $\bf k$ with a length of $k$: 
\begin{equation}
\label{eq:Pk}
\hat{P}_{\delta}(k) = \frac{1}{\rm V_{box}} \left< |\delta({\bf k})|^2 \right>_{k = |\bf k|} \ , 
\end{equation} 
with ${\rm V_{box}}$ being the box size of simulation. 

The raw estimation of $\hat{P}_{\delta}(k)$ above is known to be affected by discreteness effects, which contributes into the power through a constant amplitude called shot noise
\begin{equation}
\label{eq:Pshot}
P_{\rm shot} = {\rm V_{box}} / N_{\rm eff} \ .
\end{equation} 
Here $N_{\rm eff}$ is the effective number of particles, which accounts for their difference in mass: 
\begin{equation}
\label{eq:Neff}
N_{\rm eff} = \frac{ ( \Sigma_{i}^{N} m_i )^2 }{ \Sigma_{i}^{N} m_i^{2} }  \ , 
\end{equation} 
where $m_i$ is the individual particle mass, and $N$ is the total number of particles. For DMO simulations where all tracer particles have equal mass, $N_{\rm eff} = N$. 
The final power spectrum $P_{\delta}(k)$ used throughout this work is derived after subtraction of the shot noise term:
\begin{equation}
\label{eq:Pk_final}
P_{\delta}(k) = \hat{P}_{\delta}(k) - P_{\rm shot}  \ .
\end{equation} 

\subsection{The accuracy of power spectrum} \label{sec:var_Pk}

On large scales, due to the limited size of simulation boxes, the statistical uncertainties of the estimated matter power spectra are dominated by cosmic variance. 
The contribution of cosmic variance on $P_{\delta}(k)$ can be estimated by \citep{Takada13}: 
\begin{equation}
\label{eq:sigma2_Pk} 
 \sigma^2(k) =  2 \frac{P_{\delta}^2(k)}{N_{\rm modes}(k)}\ .
\end{equation}
$N_{\rm modes}$ is total number of modes available in the bin range [$k-\Delta k/2$, $k+\Delta k/2$]: 
\begin{equation} \label{eq:Nmodes} 
\begin{aligned} 
N_{\rm modes}(k) &= \frac{1}{(2 \pi)^3} V_{\rm box} \int_{k-\Delta k/2}^{k+\Delta k/2} 4 \pi k^2\ {\rm d}k \\
                           & \approx \frac{1}{2 \pi^2} V_{\rm box} k^2 \Delta k \ .
\end{aligned} 
\end{equation}

The small-scale error of power spectra is mostly caused by the simulation resolution. For DMO simulations, \citet{Heitmann10} pointed out that the $P_{\delta}(k)$ values of different resolutions are within 1\% agreement for $k < k_{\rm Ny}/2$, where the Nyquist wavenumber is set by the inter-particle separation on the initial grid: 
\begin{equation}
\label{eq:kNy}
k_{\rm Ny} = \frac{\pi N_{\rm p}}{\rm L_{box}} \ ,
\end{equation} 
with $N_{\rm p}$ being the cube-root of the total number of particles used in simulations. 
\textit{For smaller scales at $k > k_{\rm Ny}/2$, we expect a suppression of power for scales around $k \sim k_{\rm Ny}$ followed by a steep rise of power at even larger $k$} (see Fig.~8 of \citealt{Heitmann10} and Fig.~A3 of \citealt{vanDaalen11}). According to \citet{Heitmann10}, the suppression of power is due to discreteness effects in sampling small scale fluctuations. When fewer low-mass halos are resolved, $P_{\delta}(k)$ is suppressed at small scales. The steep rise of power is believed to be caused by incorrect shot noise subtraction. Shot noise should be scale-dependent at small scales rather than a simple constant as in Eq.~\eqref{eq:Pshot}. 


For hydrodynamical simulations, the convergence properties are more difficult to systematically quantify due to the interplay between resolution effects and galaxy formation physics.
For example, as the resolution increases, more lower mass halos are resolved, leading to an increased power of SN feedback. Subgrid parameters regulating SN feedback then must be modified to account for this effect in order to match the observables like galaxy stellar mass function. 
Typically subgrid prescriptions are designed to reach some level of convergence with the variation of resolution. 
However, the functionality of such self-regulation is rather limited. Eagle is currently the only hydro simulation with its subgrid model parameters re-calibrated to match observations when the resolution is changed (see Fig.~7 of \citealt{Schaye15}). In the left-hand panel of Fig.~A2 in \citet{vanDaalen11}, they showed a convergence comparison between the power spectra of OWLS-REF hydro simulations with $k_{\rm Ny}$ of 16 (the current OWLS resolution used in this work) and 32 respectively. The two baryonic power spectra agree to within $\sim 10\%$ out to $k \approx 40 \invhMpc$. We note that this statement only applies to OWLS-REF. There is no general rule on the behavior of convergence for hydro simulations, given that the galaxy observations are not yet converged, and that the subgrid prescriptions are all different.

\section{Power Spectrum Ratio}
\label{sec:appendixA2}

The power spectrum ratio between hydrodynamical and DMO simulations is an important quantity in this work. We rely on it in Eq.~\eqref{eq:Pk_ratio} to derive mock observables at different cosmology, as well as to build difference matrices to perform PCA. Here we discuss the validity of our estimates of this ratio over the range of scales used throughout this work, and describe how we perform extrapolation to smaller scales than those that are well-described in the simulation. 


\subsection{Discussion on the convergence of the power spectrum ratio}
\label{sec:appendixA2_1}

The ratios of matter power spectra that we use in this work are accurate to $k \lesssim 30 \invhMpc$ for Eagle/MB2/Illustris/Horizon-AGN and of $k \lesssim 10 \invhMpc$ for OWLS. Below we will justify this claim.

We have discussed the statistical uncertainty of $P_{\delta}(k)$ due to cosmic variance in \S\ref{sec:var_Pk}.
Based on the first order error propagation, ${\rm Var}[\frac{X}{Y}] = \frac{\overline{X}^2}{\overline{Y}^2} \left(\frac{\sigma^2_{X}}{\overline{X}^2} + \frac{\sigma^2_Y}{\overline{Y}^2} - 2 \frac{{\rm Cov}[X,Y]}{\overline{X}\ \overline{Y}}\right)$, the cosmic variance contribution to the uncertainty in the power spectrum ratio is
\begin{equation} \label{eq:Var_Pkratio} 
\begin{aligned} 
{\rm Var}\left[ \frac{P_{\rm \delta, hydro}(k)}{P_{\rm \delta, DMO}(k)}\right] & = \frac{1}{P_{\rm \delta, DMO}^2(k)} \sigma^2_{\rm hydro}(k) \\
& + \frac{P_{\rm \delta, hydro}^2(k)}{P_{\rm \delta, DMO}^4(k)} \sigma^2_{\rm DMO}(k)\\
&- 2 \frac{P_{\rm \delta, hydro}(k)}{P_{\rm \delta, DMO}^3(k)} {\rm Cov}\left[P_{\rm \delta, hydro}(k),P_{\rm \delta, DMO}(k)\right]\ ,
\end{aligned} 
\end{equation}
where the variance of the power spectrum $\sigma^2_{\rm hydro/DMO}(k)$ is expressed in Eq.~\eqref{eq:sigma2_Pk}.


The hydrodynamical and DMO runs are set at exactly the same initial conditions, and baryonic effects are negligible on large scales. When $P_{\rm \delta, hydro}(k)\approx P_{\rm \delta, DMO}(k)$ and ${\rm Cov}\left[P_{\rm \delta, hydro}(k),P_{\rm \delta, DMO}(k)\right] \approx 1$, as is the case at small $k$, we expect the variance of the power spectrum ratio in Eq.~\eqref{eq:Var_Pkratio} to approach zero.


On small scales, as discussed in \S\ref{sec:var_Pk}, $P_{\delta}(k)$ would achieve $1\%$ convergence out to $k \approx k_{\rm Ny}/2$.  Hence, we expect the uncertainty in the power spectrum ratio due to limited simulation resolution to be $< 2\%$ to $k \approx k_{\rm Ny}/2$, where $k_{\rm Ny}/2$ is $\sim$ 30$\invhMpc$ for Eagle/MB2/Illustris/Horizon-AGN and $\sim$ 10$\invhMpc$ for OWLS.

For the cosmic variance contribution on small scales, we can derive an upper limit by setting ${\rm Cov}\left[P_{\rm \delta, hydro}(k),P_{\rm \delta, DMO}(k)\right]$ to zero in Eq.~\eqref{eq:Var_Pkratio}, and with the $\sigma_{\rm hydro/DMO}^2(k)$ estimated using Eqs.~\eqref{eq:sigma2_Pk} and~\eqref{eq:Nmodes}: 
\begin{equation} \label{eq:Var_Pkratio_upper} 
{\rm Var}\left[ \frac{P_{\rm \delta, hydro}(k)}{P_{\rm \delta, DMO}(k)}\right]_{\rm upper} \rightarrow \frac{8\pi^2 }{V_{\rm box} k^2 \Delta k} \frac{P_{\rm \delta, hydro}^2(k)}{P_{\rm \delta, DMO}^2(k)} \ .
\end{equation} 
For $k \approx 10\invhMpc$, $L_{\rm box} = 100 \hMpc$, $\Delta k = 0.1$, and $\frac{P_{\rm \delta, hydro}}{P_{\rm \delta, DMO}} \approx 0.9$, the estimated variance of the power spectrum ratio is $\sim 0.0005$. Thus the 1$\sigma$ uncertainty in the power spectrum ratio due to cosmic variance is expected be $\lesssim 0.3\%$.\footnote{\citet{Chisari18} have estimated the effect of cosmic variance on the power spectrum ratio (see their Fig.~5) by subsampling the Horizon-AGN simulation with a volume that is 8 times smaller then our setting here. So the expected 1$\sigma$ error due to cosmic variance in their case should be on the order of 1\% ($\sqrt 8$ times larger then our setting). The total spread of their subsampled power spectra ratios is consistent with our derivation within 3$\sigma$.}


If we naively derive the power spectrum ratio by using the raw data points of $P_{\delta , \rm DMO}(k)$ and $P_{\delta , \rm hydro}(k)$ at $k > k_{\rm Ny}/2$, the derived ratio will be overestimated due to the underestimation of the denominator of $P_{\delta , \rm DMO}(k)$ at scales of several times $k_{\rm Ny}$, and underestimated toward even higher $k$ due to the overestimation of $P_{\delta , \rm DMO}(k)$ (see \S\ref{sec:var_Pk}).
We will introduce our extrapolation scheme below to avoid the biases.






\subsection{Power Spectrum Ratio Extrapolation Scheme}
\label{sec:appendixA2_2}

For scales below $k < 0.1 \invhMpc$, we simply let the ratio curve asymptotically approach one, which is a justifiable assumption since we know that baryons hardly modify the matter power spectrum on large scales. For small scales, as shown in Fig.~\ref{fig:Pk_ratio}, the power spectrum ratio between hydrodynamical and DMO simulations tends to increase after $k \gtrsim 20 \invhMpc$. This increase is caused by cooling effects in hydrodynamical simulation. 
In order to capture this physical effect, we make use of data points in $k \in [10, 30] \invhMpc$, perform a smooth quadric spline fitting in $\log(k) \mbox{-} \log(P_{\delta \rm, bary }(k) / P_{\delta \rm, DMO }(k))$ space, and extrapolating the fitted trend out to $k > 30 \invhMpc$. Figure~\ref{fig:Pk_ratio} shows the extrapolation curves for all baryonic scenarios.

\begin{figure}
\begin{center}
\includegraphics[width=0.47\textwidth]{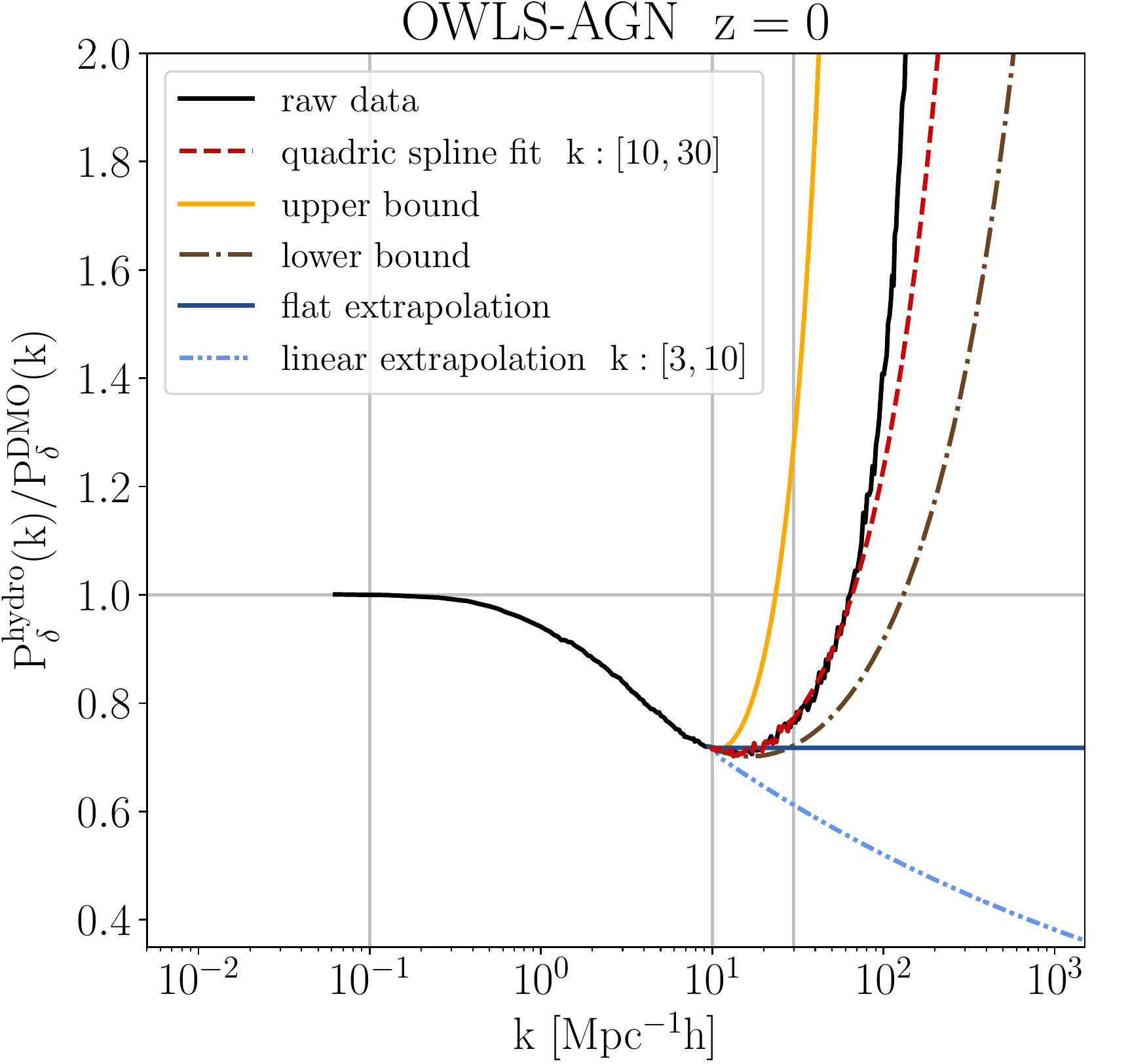}
\caption{Power spectrum ratio between OWLS-AGN and OWLS-DMO at $z=0$, assuming different extrapolation schemes above $k > 10 \invhMpc$. The black line shows the raw data calculated by simply taking ratio from the raw OWLS power spectra from \citet{vanDaalen11}. The red dashed line indicates the extrapolation scheme by extending a quadric spline fitted curve using the raw data points in $k \in [10, 30] \invhMpc$. The yellow and brown dot-dashed lines indicate the two possible upper and lower bounds one may derive, if there is some uncertainty in the raw data points $k \in [10, 30] \invhMpc$. The dark blue line plots a pure flat extrapolation. The light blue dash-dot-dotted line indicates the case of extrapolation when using data points in $k \in [3, 10] \invhMpc$, which are believed to be well-measured.}
\label{fig:Pkratio_extra}
\end{center}
\end{figure}

The above extrapolation scheme holds for Eagle/MB2/IIIustris/Horizon-AGN simulations, where we have reliable power spectrum ratios in the range $k \in [10, 30] \invhMpc$. For the OWLS simulation set, as discussed above, the ratio data is only well-determined to $k < 10 \invhMpc$. 
The black line in Fig.~\ref{fig:Pkratio_extra} indicates the naively derived power spectrum ratio from the raw data of OWLS-AGN and OWLS-DMO. In the case of OWLS-AGN, we do need data points slightly beyond $k$ of $10 \invhMpc$ to capture the transition from suppression to enhancement of power. We therefore still make use of the raw data points in the range of $k \in [10, 30] \invhMpc$, where the uncertainties may $> 2\%$ but not that worse,
to perform extrapolation spline line fitting, and the resulting curve is indicated in the red-dashed curve of Fig.~\ref{fig:Pkratio_extra}. The raw data curve is slightly higher than the extrapolating curve when $k$ is large. This is resulting from the underestimation of the DMO power spectrum as discussed in \S\ref{sec:var_Pk} and \S\ref{sec:appendixA2}. The extrapolation based on fitting data at $k \in [10, 30] \invhMpc$ may exhibit uncertainties in the final extrapolation slope at high $k$. Therefore, we try to explore such uncertainties by setting slightly different extrapolating parameters that are shown as upper (yellow line) and lower (brown dot-dashed) bounds in Fig.~\ref{fig:Pkratio_extra}. Finally, we also check the simplest constant extrapolation scheme as shown in the dark blue line. 
We will explore the effects of different extrapolation schemes on the resulting tomographic shear power spectra later. 

If only using data points around $k < 10 \invhMpc$ to perform smooth extrapolation, one would fail to capture the cooling effect that typically exists in hydrodynamical simulations at high $k$. The light blue dash-dot-dotted line of Fig.~\ref{fig:Pkratio_extra} indicates such an extrapolation based on the data points for $k \in [3, 10] \invhMpc$. 
This sets a very important requirement for our method. If we really want to use the PCA framework to achieve better cosmological parameter constraints by including more small scale information, 
the simulations that are used to build the PC basis must also have high enough resolution to construct a reasonable extrapolation down to the scale that goes into cosmological analysis. This is the reason why we avoid using the \citet{Rudd08} and \citet{Gnedin11} simulations to build our PC basis as in the previous work of \citetalias{Eifler15}. The half-Nyquist wavenumbers of these two simulations are too low to capture the up-turn in the power spectrum ratio, given the angular scales of $\ell \approx 2000$ used in this work.

\begin{figure}
\begin{center}
\includegraphics[width=0.5\textwidth]{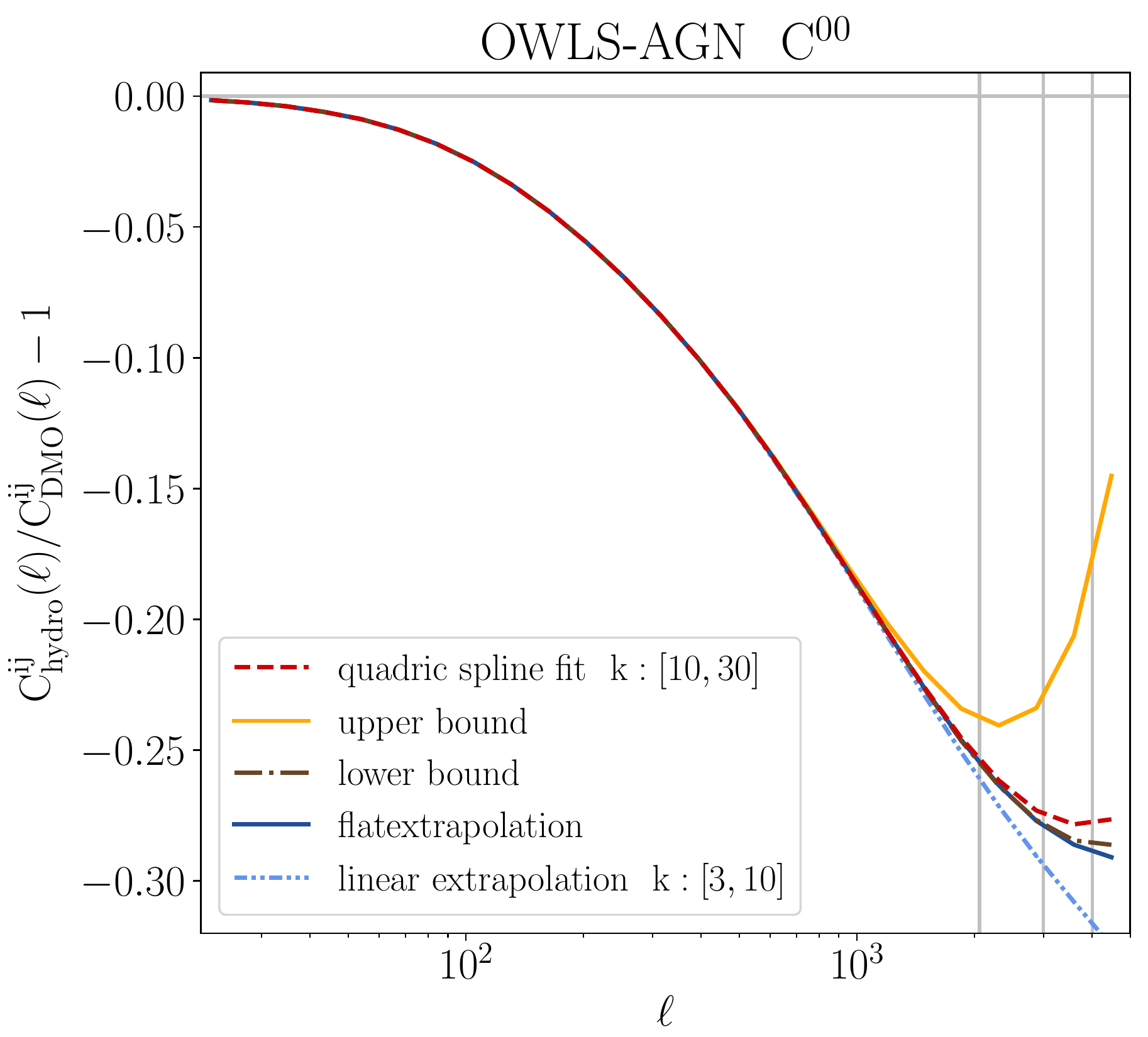}
\caption{The ratio of tomographic shear power spectrum between OWLS-AGN and OWLS-DMO simulations for our lowest tomographic bin. Curves in different colors represent different $P_{\delta}(k)$ ratio extrapolation schemes as colored in Fig.~\ref{fig:Pkratio_extra}. The vertical dashed line indicates the angular scale of $\ell = 2060$, where the cut is made for all of tomographic bins in our data vector. This cut is chosen such that the derived $C^{ij}(\ell)$ curve would not be too sensitive on different $P_{\delta}(k)$ ratio extrapolation schemes.}
\label{fig:Clratio_extra}
\end{center}
\end{figure}

We now justify how the choice of angular scale cut at $\ell \approx 2000$ is made. In Fig.~\ref{fig:Clratio_extra} we present the computed tomographic shear power spectra in our lowest redshift bin, for various extrapolation schemes on the power spectra ratio shown in Fig.~\ref{fig:Pkratio_extra}. The vertical gray line indicates the angular scale cut of $\ell = 2060$ we have adopted. One can see that the $C^{ij}(\ell)$ ratio only differs mildly at this scale. Therefore, although our current extrapolation scheme may lead to a considerable error in the $P_{\delta}(k)$ ratio, after the integration process, such error propagation in $C^{ij}(\ell)$ ratio is estimated to be within 10\% when making an $\ell$ cut at $\approx 2000$. We make a conservative choice of cutting in $\ell$ such that the final result is not too sensitive to our extrapolation scheme. 


\section{Computing baryon-contaminated data vectors at varying cosmology}
\label{sec:appendixA3}

In this Appendix, we describe in detail how we compute baryon-contaminated data vectors as a function of cosmology. This procedure is needed to
build the difference matrix (Eq.~\ref{eq:DiffMatrix}), weighted difference matrix (Eq.~\ref{eq:DiffMatrix_chy}), or ratio matrix (Eq.~\ref{eq:RatioMatrix}) when doing PCA.

\begin{figure}
\begin{center}
\includegraphics[width=0.48\textwidth]{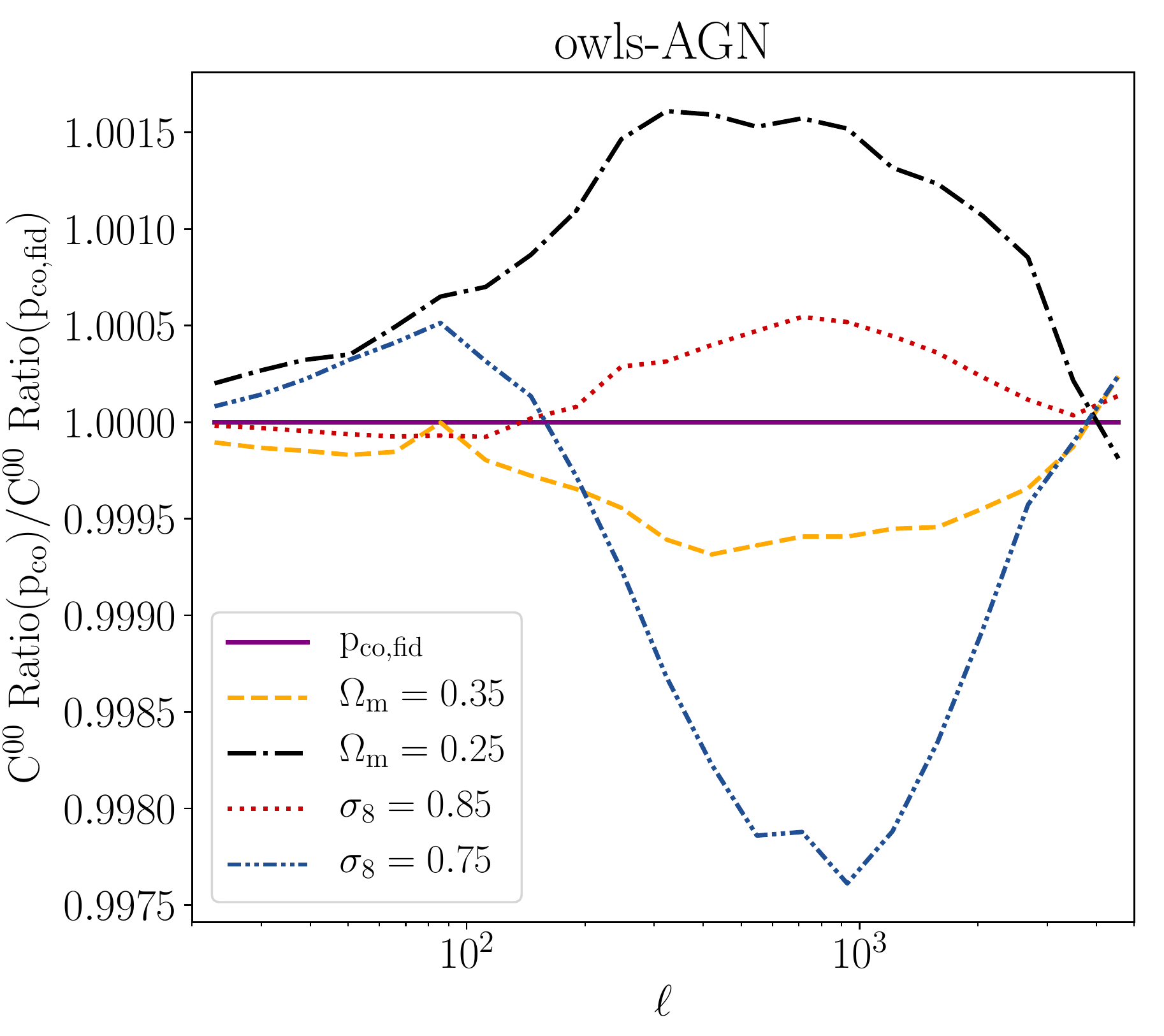}
\caption{The ratio of tomographic power spectra ratio at bin $C^{00}$ between hydrodynamical and DMO simulations evaluated at various $\pco$ v.s. $\pcofid$, i.e, $[\frac{C^{00}_{\rm hydro}(\pco)}{C^{00}_{\rm DMO}(\pco)}] / [\frac{C^{00}_{\rm hydro}(\pcofid)}{C^{00}_{\rm DMO}(\pcofid)}]$. Here different curves indicate changes of $\Omega_m$ or $\sigma_8$ to values shown in the legend, while keeping the remaining cosmological parameters the same as $\pcofid$. The fact that all ratio curves are $\approx 1$ to within 0.25\% indicates the validity of using Eq.~\eqref{eq:Cl_ratio} as our approximation.}
\label{fig:Ratio_Cl_ratio}
\end{center}
\end{figure}

To produce a baryon-contaminated vector $\B_x$ at cosmology $\pco$, in principle we should rely on Eq.~\eqref{eq:Pk_ratio} to generate the matter power spectrum for that cosmology, and integrate it to derive the tomographic shear data vector 
\begin{equation}
\label{eq:Cij_hydro}
\begin{aligned}
C^{ij}_{{\rm hydro, }x}& (\ell |\ \pco)   =  \\
& \frac{9H_0^4 \Omega_m^2}{4c^4} \int_0^{\chi_{\rm h}} {\rm d} \chi
 \frac{g^{i}(\chi) g^{j}(\chi)}{a^2(\chi)} P_{\delta}^{{\rm hydro, }x} \left(\frac{\ell}{f_K(\chi)},\chi |\ \pco \right) \ .
\end{aligned}
\end{equation}
However, to increase the computational speed, we approximate this step by 
\begin{equation} \label{eq:Cl_ratio}
\B_x(\pco) = C^{ij}_{{\rm hydro,} x}(\pco) = \frac{C^{ij}_{{\rm hydro,} x}(\pcofid)}{C^{ij}_{\rm theory}(\pcofid)} C^{ij}_{\rm theory}(\pco)\ ,
\end{equation} 
where $C^{ij}_{{\rm hydro,} x}(\pcofid)$ is pre-computed using Eq.~\eqref{eq:Cij_hydro} setting at $\pcofid$ and stored. $C^{ij}_{\rm theory}(\pco)$ is our model vector $\M(\pco)$ generated from \textsc{Halofit}.
Approximating Eq.~\eqref{eq:Cij_hydro} by Eq.~\eqref{eq:Cl_ratio} avoids the need to integrate nine times when constructing the nine columns of $\bm \Delta(\pco)$/$\bm \Delta_{\rm chy}(\pco)$/$\bm \R(\pco)$ at each MCMC step. 
In using Eq.~\eqref{eq:Cl_ratio}, we basically assume the quantity $[\frac{C^{00}_{\rm hydro}(\pco)}{C^{00}_{\rm DMO}(\pco)}] / [\frac{C^{00}_{\rm hydro}(\pcofid)}{C^{00}_{\rm DMO}(\pcofid)}] \approx 1$ at various $\pco$. To check the validity, we compute all the elements in this quantity using Eq.~\eqref{eq:Cij_hydro} and plot it in Fig.~\ref{fig:Ratio_Cl_ratio}, with $\pco$ set at different values of $\Omega_m$ or $\sigma_8$, while keeping the rest of the cosmological parameters the same as $\pcofid$. As shown, the $C^{00}$ ratio curves are within 0.25\% of 1 for various $\pco$ demonstrated here.

\section{Constraints on cosmological parameters of \lowercase{$\Omegam$, $\sigma_8$}, and \lowercase{$w_a$}}
\label{sec:appendixA4}

In this Appendix, we provide constraints on cosmological parameters of  $\Omegam$, $\sigma_8$, and $w_a$ after applying baryon mitigation techniques using \textsc{HMcode} or various PCA methods. The solid markers indicate the amount of residual bias after mitigation, and the open markers indicate the $0.5\sigma$ errors on the marginalized 1D posteriors. We rely on this plot to check whether a mitigation method can successfully reduce the baryonic physics-induced bias to within  $0.5\sigma$ for different cosmological parameters and baryonic scenarios. The results are briefly summarized in Tables~\ref{tb:pass2000} and~\ref{tb:pass5000}. We refer readers back to discussions at \S\ref{sec:PCA_vs_HMcode} and \S\ref{sec:ell_5000} for details.

\begin{landscape}
\begin{figure}
\begin{center}
\includegraphics[width=1.4\textwidth,keepaspectratio]{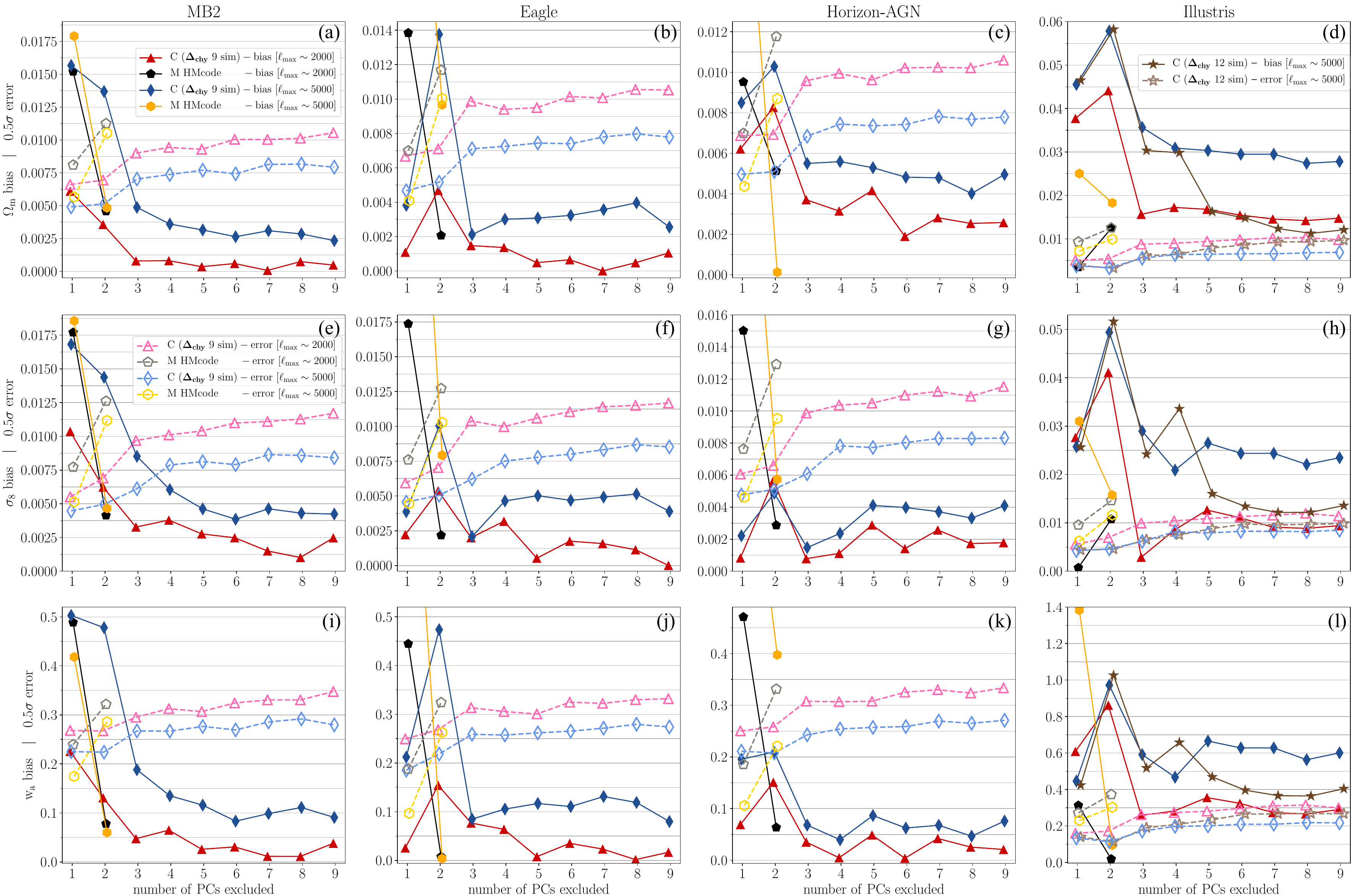}
\caption{Similar to Fig.~\ref{fig:w0_bias_err_ell5000}, but for cosmological parameters of $\Omegam$ (first row), $\sigma_8$ (second row) and $w_a$ (third row), for various baryonic scenarios categorized in each different column.}
\label{fig:OmS8Wa_bias_err}
\end{center}
\end{figure}
\end{landscape}

\section{Goodness of fit for baryon mitigation models}
\label{sec:appendixA5}

\begin{table*}
\caption{Goodness of fit for various baryon mitigation models.}
\begin{tabular}{lccccc}
\hline 		
		 	  	   			  	& Eagle 			& MB2 			& Horizon-AGN 	& Illustris			\\  \hline
no-mitigation at $\pcobest$ ($\pcofid$)	& 2.68 (30.92)	&  0.93 (5.71) 	&  3.85 (103.2) 	& 107.5 (3258) 	\\ \hline
\textsc{HMcode} -- $A$ 				& 2.09			&  0.74 			& 2.34 			& 4.84 			\\ 
\textsc{HMcode} -- $A$, $\eta_0$		& 0.91   			&  0.29 			& 1.69 			& 4.80 			\\ \hline
PCA ex1  at $\pcobest$ ($\pcofid$)		& 1.39  (3.25)	& 0.70 (5.11) 	& 0.75 (6.75) 	& 11.91 (40.0) 	\\
PCA ex2							& 0.41  (0.39)	& 0.37 (0.81) 	& 0.45 (2.72) 	& 9.68 (34.2) 	\\
PCA ex3							& 0.28  (0.24) 	& 0.28 (0.32) 	& 0.26 (2.65) 	& 3.08 (12.0) 	\\
PCA ex4							& 0.15  (0.07) 	& 0.18 (0.30) 	& 0.20 (1.38) 	& 2.06 (4.89) 	\\
PCA ex5							& 0.13  (0.04) 	& 0.14 (0.07) 	& 0.16 (0.64) 	& 1.65 (2.55)		\\
PCA ex6							& 0.09  (0.03) 	& 0.11 (0.06) 	& 0.20 (0.59) 	& 1.58 (2.18) 	\\
PCA ex7							& 0.11  (0.03) 	& 0.12 (0.06) 	& 0.19 (0.54) 	& 1.50 (2.00) 	\\
PCA ex8							& 0.09  (0.03) 	& 0.08 (0.06) 	& 0.15 (0.53) 	& 1.44 (2.00) 	\\
PCA ex9							& 0.07  (0.03) 	& 0.10 (0.05) 	& 0.15 (0.43) 	& 1.37 (2.00)		\\ \hline
\label{tb:chi2}
\end{tabular}
\end{table*}

In this Appendix, we summarize the fitting quality for various baryon mitigation methods. 
In Table~\ref{tb:chi2}, we provide the $\chi^2$ values computed at the best-fitted (fiducial) cosmology for \textsc{HMcode} and the PCA method C when applied on each baryonic scenario for the $\ell_{\rm max} \approx 2000$ likelihood simulations. Here we define our best-fitted parameters to be the median value at the marginalized 1D posterior distribution.

Notice that because our mock data vectors are noiseless, the $\chi^2$ values cannot be used to make statements about overfitting or underfitting based on the reduced $\chi^2$ criterion (i.e.\ we do not expect $\chi^2/\rm{(d.o.f)} \approx 1$). 
However, using the information from the relative $\chi^2$ values ($\Delta \chi^2$) and the relative degrees of freedom ($\Delta$ d.o.f) between two models, we can determine the model complexity needed from the data by performing the Chi-square difference test\footnote{We refer readers to this \href{https://www.psychologie.uzh.ch/dam/jcr:ffffffff-b371-2797-0000-00000fda8f29/chisquare_diff_en.pdf}{link} for more detail about Chi-square difference test.} .

For example, for the Illustris scenario, we see that when comparing the PCA results between excluding 1 PC mode to 5 PC modes, the $\Delta \chi^2 = 11.91 - 1.646 = 10.264$, and the $\Delta$ d.o.f $=4$. The corresponding $p-$value is 0.036, which means that the improvement is marginally statistically significant ($p-$value < 0.05). 
Excluding 6 PC modes does not significantly improve the goodness of fit compared with the result when excluding 5 PC modes ($\Delta \chi^2 = 0.062$, $\Delta$ d.o.f $=1$, $p-$value $=0.8$).


After a few PC modes are excluded, we see that the $\chi^2$ values computed at $\pcobest$ is comparable to that computed at $\pcofid$ for all baryonic scenarios. 
This means that excluding PC modes does not just reduce parameter bias in our simulated likelihood analysis, but the resulting best-fitting model also provides a good fit to the data.


\bsp	
\label{lastpage}
\end{document}